\documentclass[aip,pop,reprint,notitlepage, nofootinbib]{revtex4-1}

\usepackage{anysize}

\usepackage[utf8]{inputenc}

\usepackage{amsmath}
\usepackage{amssymb}
\usepackage{bm}
\usepackage{hyperref}
\usepackage{graphics,graphicx}

\usepackage{array}   % for \newcolumntype macro
\newcolumntype{L}{>{$}l<{$}} % math-mode version of "l" column type
\newcolumntype{C}{>{$}c<{$}} % math-mode version of "l" column type
\newcolumntype{R}{>{$}r<{$}} % math-mode version of "l" column type

\newcommand{\gn}{\bar{g}}
\newcommand{\xn}{\bar{X}}
\newcommand{\ag}{\alpha}
\newcommand{\bg}{\beta}
\newcommand{\g}{\gamma}
\renewcommand{\i}{\alpha}
\renewcommand{\j}{\beta}

\marginsize{0.63in}{0.63in}{0.45in}{0.45in}

\begin{document}
\begin{abstract}
Grad's method is used on the linearized Boltzmann collision operator to derive the most general expressions for the collision coefficients for a multi-component, multi-temperature plasma up to rank-2. In doing so, the collision coefficients then get expressed as series sum of pure coefficients of temperature and mass ratios multiplied by the cross-section dependent Chapman-Cowling integrals. These collisional coefficients are compared to previously obtained coefficients by Zhdanov et al [Zhdanov V.M., Transport processes in multi-component plasma, Taylor and Francis (2002)] for $13N$-moment multi-temperature {  scheme}. First, the differences in coefficients are compared directly, and then the differences in first approximation to viscosity and friction force are compared. For the $13N$-moment multi-temperature coefficients, it is found that they behave reasonably similarly for small temperature differences, but display substantial differences in the coefficients when the temperature differences are high, both for the coefficients and for viscosity and friction force values.
Furthermore, the obtained coefficients are compared to the $21N$-moment single-temperature approximation {  provided by Zhdanov et al}, and it is seen that the differences are higher than the $13N$-moment multi-temperature coefficients, and have substantial differences even in the vicinity of equal temperatures, especially for the viscosity and friction force calculations.

\end{abstract}
 \title{Generalized Collisional Fluid Theory for Multi-Component, Multi-Temperature
Plasma Using The Linearized Boltzmann Collision Operator for Scrape-Off Layer/Edge
Applications}
\author{M.\,Raghunathan}
\author{Y.\,Marandet}
\affiliation{Aix-Marseille Univ., CNRS, PIIM, Marseille, France}
\author{H.\,Bufferand} 
\author{G.\,Ciraolo}
\author{Ph.\,Ghendrih}
\author{P.\,Tamain}
\affiliation{IRFM-CEA, F-13108 Saint-Paul-Lez-Durance, France}
\author{E.\,Serre}
\affiliation{Aix-Marseille Univ., CNRS, M2P2, Marseille, France}
\maketitle

% \section{Introduction}

Power exhaust is a key challenge in next step fusion devices. Reducing the peak heat fluxes on the plasma facing components to tolerable levels, to a large extent, relies on impurity radiation in the boundary layer of the tokamak. The impurity radiation pattern in turn depends on plasma transport, both parallel and perpendicular to the magnetic field. This problem is generally addressed by solving plasma fluid models coupled to kinetic neutrals (e.g. as with the fluid code Soledge2d-EIRENE\cite{bufferand_near_2013}). Generally, the collision terms for the fluid dynamical equations are obtained by averaging over the kinetic equation with the constants of motion.

The moment-averaged collisional term can be determined by the use of different forms of ansatz for the distribution functions entering the collision operator of the kinetic equation. The two major ansatz used are namely, the Chapman-Enskog ansatz\cite{chapman_mathematical_1952} and Grad's Hermite polynomial ansatz\cite{grad_asymptotic_1963}. The Chapman-Enskog method involves decomposing the distribution function in terms of a small parameter expansion, and forming a moment-averaged hierarchy of equations at each order each with own collisional contribution. This method has been the most dominant so far, owing to its quick convergence. Grad's Hermite polynomial ansatz, on the other hand, involves decomposing the distribution function in terms of a series orthogonal tensorial polynomials, leading to a hierarchy of fluid equations for each order of the Hermite polynomial. These polynomials demonstrate mathematical properties which are somewhat easier to manipulate algebraically, however they have no clear rule for convergence. Often the convergence is either checked through brute force methods\cite{struchtrup_macroscopic_2005}, or by direct comparison of terms relevant to the physics in consideration\cite{ferziger_mathematical_1972,zhdanov_transport_2002}. Depending on the complexity of the collision operator, the treatment of the collision terms may get quite cumbersome.

Recently, in plasma physics oriented towards nuclear fusion, there has been a resurgence in the use of Grad's Hermite polynomial ansatz for use in calculating the collisional term, probably because of the improvement of algebraic techniques and availability of computer algebra systems. In recent plasma physics, it has been used to calculate the moments of the kinetic equation and the Landau collision operator\cite{landau_kinetic_1936} expressed in terms of Rosenbluth\cite{rosenbluth_fokker-planck_1957} potentials both for the linearized collision operator\cite{ji_exact_2006} and fully non-linear collision operator\cite{ji_full_2009}, and its extension to magnetized plasmas\cite{ji_framework_2014}. Extension of moments of the Landau collision operator to strong flow cases has also been done\cite{hirvijoki_fluid_2016}, and some closed-form analytic expressions for the involved integrals in the operator were also formulated\cite{pfefferle_exact_2017}. The Hermite polynomial ansatz expressed in terms of the product of Laguerre polynomials and irreducible monomial, has also been used to formulate and study drift-kinetic models\cite{jorge_drift-kinetic_2017} and gyrokinetic models\cite{jorge_nonlinear_2019} for the scrape-off layer\cite{frei_gyrokinetic_2020}, and also has been used to formulate a linear theory of electron plasma waves\cite{jorge_linear_2019}. 

However, the Landau collision operator is only valid for warm plasmas where the weak coupling conditions apply\cite{balescu1997statistical}. The generalization of these operators to different plasma regimes, for example in trying to account for shielding effects as in the Balescu-Lenard collision operator, leads to an increase in mathematical sophistication and a corresponding difficulty in solving\cite{balescu_irreversible_1960}. The alternative is to use the much simpler Boltzmann collision operator. Given that the Landau operator can be thought of the Boltzmann operator under weak coupling limits\cite{balescu1997statistical}, we can expect that the use of the Boltzmann collision operator with the shielded Coulomb potential should provide quantitatively similar effects to the Landau or Balescu-Lenard collision operator as long as the plasma is in local thermodynamic equilibrium\cite{silin_introduction_1971} and does not exhibit large scale fluctuations\cite{klimontovich_kinetic_2013}.  The added advantage of the Boltzmann collision operator is its use of explicit collision cross-sections. 
Any coefficients derived in this manner would have the advantage of being applicable for a wide variety of gas and plasma dynamics merely by using the relevant cross-section for the system in question\cite{capitelli_transport_2013}, for example for ion-ion\cite{kihara_coefficients_1959,liboff_transport_1959,hahn_quantum_1971}, ion-neutral\cite{kihara_transport_1960}, neutral-neutral\cite{chapman_mathematical_1952,monchick_collision_1959,smith_automatic_1964, neufeld_empirical_1972,mason_transport_1954,rainwater_binary_1982}, and charge exchange collisions\cite{helander_fluid_1994,krasheninnikov_edge_2020}.

In the scope of this article, we focus on the derivation of fluid collision coefficients from the linearized Boltzmann collision operator. In the past, two sets of collisional terms, one for any temperature range and the other for temperature range close to plasma common temperature, have been derived and provided\cite{alievskii_1963_transport,yushmanov_diffusion_1980,zhdanov_transport_2002}. However, an explicit derivation process was not provided for the values of the collision coefficients. Therefore, in order to verify the accuracy of the coefficients provided, we firstly rederive the collision operator in terms of partial bracket integrals, and derive the exact values of the partial bracket integrals\cite{chapman_mathematical_1952,rat_transport_2001}. We provide, for the first time, expressions for calculating the general collisional terms up to rank-2, in a manner that can be implemented efficiently in modern computer algebra systems. We also explicitly provide the range of validity of our and the aforementioned coefficients, and clearly delineate the underlying assumptions. This would be useful in clearly defining the simulation parameter range for a number of code packages which have implemented certain versions of the previous collisional coefficients. For example, the previous two sets of coefficients, taken from Ref.\cite{zhdanov_transport_2002}, have been implemented in B2/SOLPS\cite{bergmann_implementation_1996,rozhansky_momentum_2015,sytova_impact_2018}, EDGE2D\cite{fichtmuller_multi-species_1998}, and more recently in Soledge2d-EIRENE\cite{bufferand_2019}, which solves an energy equation for each species { (while other codes mentioned solve only one total energy equation)}.

The article is organized as follows. We first provide a small introduction to the moment-averaged Boltzmann kinetic equation and the corresponding Boltzmann collision operator in Sec.\ref{sec:boltzmann}. We demonstrate its conservation properties in the process. Then, we introduce the Hermite polynomial ansatz and Grad's method, including the expression of the ansatz as a product of the Sonine polynomials and the irreducible tensorial monomial in Sec.\ref{sec:ansatz}. In Sec.\ref{sec:derivation_operator}, we  present the derivation of the most general collision operator in terms of the partial bracket integrals. In Sec.\ref{sec:general_expressions}, we provide the general forms of the partial bracket integrals (with the full derivations in Appendix \ref{sec:bracket_integral_derivation}), alongwith the formulation of the cross-section integrals.  Then we compare our obtained expressions for the collision coefficients with previously derived expressions in Sec.\ref{sec:comparisons} for collisions of various species relevant to fusion, ranging from a deuterium-tritium plasma collisions to heavy impurity collisions such as with tungsten. Finally, in Sec.\ref{sec:intuitive}, we compare calculations of approximate values of physically intuitive quantities such as viscosity and friction force, and provide recommendations on the range of validity for the sets of coefficients. 

\section{The Boltzmann equation for multi-species plasma}
\label{sec:boltzmann} 

The Boltzmann equation for the distribution function for species $\alpha$, $f_\alpha$ in the frame of the peculiar velocity of species $\i$, $\mathbf{c_\alpha}=\mathbf{v_\alpha}-\mathbf{u}$, is given by
\begin{equation}
 \frac{d f_\alpha}{d t}+\mathbf{c_\alpha}.\nabla{f_\alpha}+\frac{1}{m_\alpha}\mathbf{F}^*_\alpha.\nabla_{c_\alpha}{f_\alpha} -c_{\alpha s}\frac{\partial f_\alpha}{\partial c_{\alpha_r}}\frac{\partial u_r}{\partial x_s}= \sum_{\beta} J_{\alpha\beta}, \label{boltzmannc}
\end{equation} 
where the common plasma flow velocity $\mathbf{u}$ is given by
\begin{equation}
 \rho \mathbf{u} = \sum_\alpha \rho_\alpha \mathbf{u_\alpha},\ \rho = \sum_\alpha \rho_\alpha,
\end{equation}
where $\rho$ represents the mass density.The $d/dt$ represents full time derivative given by $d/dt=\partial/\partial t+\mathbf{u}.\nabla$, and where the force term $\mathbf{F_\alpha}$ and $d\mathbf{u}/dt$ are combined to write the relative force in the moving frame given by $\mathbf{F^*_\alpha}=\mathbf{F_\alpha}-m_\alpha d\mathbf{u}/dt$.

The LHS is referred to as the ``free-streaming term'', and the RHS is the collisional contribution between species $\i$ and every other species of the system. The  general ``gain-loss'' type Boltzmann collisional RHS is given by,
\begin{equation}
 J_{\alpha\beta} = \iint (f^\prime_\alpha f^\prime_{1\beta}-f_\alpha f_{1\beta})g\sigma_{\alpha\beta}(g,\chi)d\Omega d\mathbf{c_{1\beta}}, \nonumber
\end{equation}
where $\alpha,1\beta$ refer to species of the two particles colliding, subscript $\prime$ refers to properties after the collision, $g$ is the relative velocity between the colliding particles, $\sigma_{\alpha\beta}$ is the collision cross section, and $\Omega$ is the solid angle in which the collision occurs. For the specific case of multi-species system, it takes the form
\begin{equation}
 J_{\alpha\beta} = \iint \{f_\alpha(\mathbf{c}_\alpha^\prime) f_{1\beta}(\mathbf{c}_{1\beta}^\prime)-f_\alpha(\mathbf{c}_\alpha) f_{1\beta}(\mathbf{c}_{1\beta})\}g\sigma_{\alpha\beta}(g,\chi)d\Omega d\mathbf{c_{1\beta}}, \nonumber
\end{equation}
where the distribution functions for each species are only dependent on the velocity of the species. Such a form is valid for elastic collisions. 

Now, for any quantity $\psi_\alpha$ depending purely on species $\alpha$, one can average over Eq.~\ref{boltzmannc} which attains the following form
\begin{multline}
\frac{d}{dt}n_\alpha \langle\psi_\alpha\rangle+n_\alpha\langle\psi_\alpha\rangle\nabla.\mathbf{u} +\nabla.(n_\alpha\langle\psi_\alpha\mathbf{c_\alpha}\rangle)\\
-n_\alpha\left\{\left\langle\frac{d\psi_\alpha}{dt}\right\rangle +\langle \mathbf{c_\alpha}.\nabla\psi_\alpha\rangle +\frac{1}{m_\alpha}\langle\mathbf{F}^*_\alpha.\nabla_{c_\alpha}{\psi_\alpha}\rangle\right.\\
\left. -\left(\left\langle c_{\alpha s}\frac{\partial \psi_\alpha}{\partial c_{\alpha_r}}\right\rangle\frac{\partial u_r}{\partial x_s}\right) \right\} = R_\alpha, \label{eq:transport}
\end{multline}
where $n_\alpha$ is the number density of the species $\alpha$, and where,
\begin{multline}
 R_\alpha=\sum_{\beta} \int \psi_\alpha J_{\alpha\beta}d\mathbf{v_\alpha}\\
 =\sum_{\beta}\iiint \psi_\alpha(f^\prime_\alpha f^\prime_{1\beta}-f_\alpha f_{1\beta})g\sigma_{\alpha\beta}(g,\chi)d\Omega d\mathbf{c_{\alpha}}d\mathbf{c_{1\beta}}.
\end{multline}
{  For elastic collisions, the moment averaged collision operator can be transformed into
\begin{equation}
 R_\alpha=\sum_{\beta}\iiint (\psi^\prime_\alpha-\psi_\alpha)f_\alpha(\mathbf{c}_\alpha) f_{1\beta}(\mathbf{c}_{1\beta})g\sigma_{\alpha\beta}(g,\chi)d\Omega d\mathbf{c_{\alpha}}d\mathbf{c_{1\beta}},
 \label{eq:boltzmann2}
\end{equation}
since the distribution functions for any given species are purely a function of the species peculiar velocity.} 

By its form, the averaged Boltzmann operator is meant to conserve mass, and by the choice of velocities, it is meant to conserve energy and momentum. One can notice in the averaged collision operator of the form Eq.\,(\ref{eq:boltzmann2}), choosing $\psi_\alpha=m_\i$ leads to a strict conservation of mass for $R_{\i\j}$ for the averaged kinetic equation of each species. Hence, mass is strictly conserved. However, momentum and energy conservation can only be demonstrated over the sum of the averaged right hand sides of the kinetic equations for all species, i.e. $\sum_{\i,\j} R_{\i\j}=0$. In order to demonstrate this, it would be sufficient to show that $R_{\i\j}+R_{\j\i}=0$. This is as follows
\begin{multline}
 R_{\alpha\beta}+R_{\j\i}=\int \psi_\alpha J_{\alpha\beta}d\mathbf{c_{\alpha}}+\int \psi_\beta J_{\j\i}d\mathbf{c_{\j}}\\
 =\iiint (\psi^\prime_\alpha+\psi^\prime_\j-\psi_\alpha-\psi_\j)f_\alpha f_\beta g\sigma_{\alpha\beta}(g,\chi)d\Omega d\mathbf{c_{\alpha}}d\mathbf{c_{1\beta}}\nonumber.
\end{multline}
One can notice that for momentum and energy, the term $\psi^\prime_\alpha+\psi^\prime_\j-\psi_\alpha-\psi_\j$ vanishes as a result of the elastic nature of the collisions. Therefore, we now see that the Boltzmann collision operator is constructed to conserve mass, energy and momentum (and any linear combination of the three, for that matter), for any arbitrary form of the distribution function $f$. This fact is also useful to check the validity of the solutions obtained in the succeeding sections, acting as another check against calculation errors. 
We now proceed to choosing an ansatz in order to expand the collisional term. 

\section{Sonine-Hermite polynomial ansatz and Grad's method}
\label{sec:ansatz}

In this section, we describe the modification of Grad's method\cite{grad_asymptotic_1963} as used by Zhdanov\cite{zhdanov_transport_2002} in his previous papers. In the ansatz for the solution of the Boltzmann equation, it is assumed that the solution $f_\alpha$ is already near thermodynamic equilibrium for species $\alpha$, $f_\alpha^{(0)}$ as follows
\begin{align}
 f_\alpha^{(0)}(\mathbf{c}_\i) &= n_\alpha \left( \frac{m_\alpha}{2\pi kT_\alpha}\right)^{3/2}\exp{\left(-\frac{m_\alpha c_\alpha^2}{2kT_\alpha}\right)}\nonumber\\
 &=n_\alpha\left( \frac{\gamma_\alpha}{2\pi} \right)^{3/2}\exp{\left( -\frac{\gamma_\alpha}{2}c_\alpha^2\right)},
 \label{eq:ansatz}
\end{align}
where $\gamma_\alpha={m_\alpha}/{kT_\alpha}$. In order to solve the Boltzmann equation (\ref{boltzmannc}), Zhdanov and Yushmanov choose an ansatz of the form
\begin{equation}
f_\alpha(\mathbf{c}_\i) = f_\alpha^{(0)}(\mathbf{c}_\i)\sum_{m,n} 2^{2n}m_\alpha^{-2}\gamma_\alpha^{2n+m} \tau_{mn} b^{mn}_{\alpha r_1\ldots r_m}G^{mn}_{\alpha r_1\ldots r_m},\label{eq:ansatz}
\end{equation}
where
\begin{multline}
 G_\alpha^{mn}(\mathbf{c_\alpha},\gamma_\alpha) = (-1)^n n! m_\alpha\gamma_\alpha^{-(n+m/2)}\\
 \times S^n_{m+1/2}\left(\frac{\gamma_\alpha}{2}\mathbf{c}^2_\alpha\right)P^{(m)}(\gamma_\alpha^{1/2}\mathbf{c_\alpha}).
 \label{eq:sonine-hermite}
\end{multline}
Here, $S^n_{m+1/2}$ are the Sonine polynomials, given by,
\begin{equation}
 S^n_{m+1/2}\left(\frac{\gamma_\alpha}{2}\mathbf{c}^2_\alpha\right) = \sum_{p=0}^n \left(-\frac{\gamma_\alpha}{2}\mathbf{c}^2_\alpha\right)^p\frac{(m+n+1/2)!}{p!(n-p)!(m+p+1/2)!},\nonumber
\end{equation}
where the first few $S^n_{m+1/2}$ are
\begin{equation}
  S^0_{m+1/2}\left(\frac{\gamma_\alpha}{2}\mathbf{c}^2_\alpha\right)=1,\ S^1_{m+1/2}\left(\frac{\gamma_\alpha}{2}\mathbf{c}^2_\alpha\right) = m+\frac{3}{2}-\frac{\gamma_\alpha}{2}\mathbf{c}^2_\alpha. \nonumber
\end{equation}

Further, $P^{(m)}$ are the irreducible projection of the tensorial monomial $\mathbf{c}_\alpha^m=c_{\alpha r_1}\ldots c_{\alpha r_m}$, derived by the  following recurrence relation
\begin{equation}
 P^{(m+1)}(\gamma_\alpha^{1/2}\mathbf{c_\alpha}) = \gamma_\alpha^{1/2}\mathbf{c_\alpha} P^{(m)} -\gamma_\alpha^{1/2}\frac{c_\alpha^2}{2m+1}\frac{\partial P^{(m)}}{\partial \mathbf{c_\alpha}},\nonumber
\end{equation}
with $P^{(0)}=1$. This expression is a sum of an outer product and the gradient with respect to the first-rank tensorial monomial $\gamma_\alpha^{1/2}\mathbf{c_\alpha}$. The first few $ P^{m+1}(\gamma_\alpha^{1/2}\mathbf{c_\alpha})$ are given by
\begin{multline}
  P^{(0)}(\gamma_\alpha^{1/2}\mathbf{c_\alpha})=1,\  P^{(1)}(\gamma_\alpha^{1/2}\mathbf{c_\alpha})=\gamma_\alpha^{1/2}\mathbf{c_\alpha},\\
  P^{(2)}(\gamma_\alpha^{1/2}\mathbf{c_\alpha})=\gamma_\alpha\mathbf{c_\alpha}\mathbf{c_\alpha}-\frac{1}{3}\gamma_\alpha U c_\alpha^2.\nonumber
\end{multline}
As one can observe, each $P^{(m)}$ is a rank-$m$ irreducible tensor. We don't need to calculate any more than rank-2 for the scope of the current work.
The constant $\tau_{mn}$ arises as a result of internal contractions between $b_\alpha^{mn}$ and $G_\alpha^{mn}$, and is given by
\begin{equation}
 \tau_{mn}= \frac{(2m+1)!(m+n)!}{n!(m!)^2(2m+2n+1)!}. \nonumber
\end{equation}
The forms mentioned in Refs.~\cite{grad_asymptotic_1963} and \cite{zhdanov_transport_2002} are cosmetically different because of the choice to use full factorial representations of functions and because of summing over full indices rather than over half-indices, but they are exactly the same. The coefficients $b^{mn}_\alpha$ are calculated as
\begin{equation}
 n_\alpha b^{mn}_\alpha = \int G_\alpha^{mn} f_\alpha d\mathbf{c}_\alpha.
\end{equation}
Some values of $G_\alpha^{mn}$ are as follows
\begin{align}
 G^{00}_\alpha &= m_\alpha,\ G^{10}_\alpha = m_\alpha \mathbf{c}_\alpha, \ G^{01}_\alpha=\frac{m_\alpha}{2}\left( c_\alpha^2-\frac{3}{\gamma_\alpha}\right),\nonumber\\
  G^{11}_\alpha &= \frac{m_\alpha}{2}\mathbf{c}_\alpha \left( c_\alpha^2-
 \frac{5}{\gamma_\alpha}\right),\ G^{20}_\alpha=m_\alpha\left(\mathbf{c}_\alpha\mathbf{c}_\alpha-\frac{1}{3}Uc_\alpha^2\right)\nonumber\\
 G^{12}_\alpha &=\frac{m_\alpha}{4}\mathbf{c}_\alpha(c_\alpha^4-14\gamma_\alpha^{-1}c_\alpha^2+35\gamma_\alpha^{-2}),\nonumber\\
 G^{21}_\alpha&=\frac{m_\alpha}{2}(c_\alpha^2-7\gamma_\alpha^{-1})\left(\mathbf{c}_\alpha\mathbf{c}_\alpha-\frac{1}{3}Uc_\alpha^2\right),\nonumber
\end{align}
and the corresponding $b^{mn}_\alpha$ are given by
\begin{align}
 n_\alpha b^{00}_\alpha &= \rho_\alpha,\ n_\alpha b^{10}_\alpha = \rho_\alpha \mathbf{w}_\alpha,\ n_\alpha b^{01}_\alpha = 0,\nonumber\\
 n_\alpha b^{11}_\alpha &= \mathbf{h}_\alpha,\ n_\alpha b^{20}_\alpha = \pi_\alpha,\nonumber\\
 n_\alpha b^{12}_\alpha &= \mathbf{r}_\alpha,\ n_\alpha b^{21}_\alpha = \sigma_\alpha.\nonumber
\end{align}
Here, $b^{00}_\alpha,\ b^{01}_\alpha$ and $b^{10}_\alpha$ represent the intuitive hydrodynamical moments density, diffusion velocity and temperature $\rho_\alpha$, $\mathbf{w}_\alpha=\mathbf{u}_\i-\mathbf{u}$, and $T_\alpha$. the higher moments $b^{11}_\alpha$ and $b^{20}_\alpha$ represent the thermodynamically privileged moments (as per Balescu's nomenclature\cite{balescu_transport_1988}), the heat flux $\mathbf{h}_\alpha$ and the divergence free pressure-stress tensor $\pi_\alpha$, which are privileged because they contribute to the entropy. The higher-order moments $b^{12}_\alpha$ and $b^{21}_\alpha$ are non-privileged moments $\mathbf{r}_\alpha$ and $\sigma_\alpha$, which do not have a clear physical meaning, however which may contribute to the accuracy of moment equations in terms of representing the Boltzmann equation. As one can notice, these are all moments of ranks less than 2. Moments of rank-0 are scalar, like density $\rho_\i$ and temperature $T_\i$, are constitute $N$ variables each. Moments rank-1 are vectorial moments, like momentum $m_\i\mathbf{w}_\i$ and heat-flux $\mathbf{h}_\i$, and contribute $3N$ variables each. Moments of rank-2 are tensorial in nature, like the stress-tensor $\pi_\i$ and $\sigma_\i$, and contribute $5N$ variables each (and not $9N$, since they are symmetric and traceless). In principle, one can construct a $5N$-moment system with just the hydrodynamical moments, a $13N$-system with including the thermodynamically privileged moments, and a $21N$-system including $\mathbf{r}_\i$ and $\sigma_\i$.

Furthermore, the Sonine-Hermite polynomials are chosen by Eq.\,(\ref{eq:sonine-hermite}) so as to form the moments in the most physically intuitive manner possible. They are related to the regular Sonine-Hermite polynomials in Ref.\,\onlinecite{ grad_asymptotic_1963}, as follows
\begin{equation}
 H^{mn}_{\i}(\xi\leftarrow\gamma_\i^{1/2}\mathbf{c}_\i)=\frac{1}{m_\i}2^n\gamma_\i^{n+m/2}G^{mn}_\i(\mathbf{c}_\i,\gamma_\i), \nonumber
\end{equation}
which then reduce to the Hermite polynomials defined in Refs.\,\onlinecite{grad_note_1949, grad_principles_1958} by the following relation
\begin{equation}
 H^{m}_{\i}(\xi)=\sum_{n=0}^{[m/2]}H^{(m-2n)n}_{\i}(\xi). \nonumber
\end{equation}

One limitation to note, is that the choice of a zero function $f_\i^{(0)}$, defined at the common flow of the plasma, is valid for any difference of temperatures among different species, but limits the solution to requiring flow velocities of all species being approximately the same when the number of moments retained is finite. In principle, if one retained infinite moments, then the solution space provided by Eq.\,(\ref{eq:ansatz}) would be the same as one provided by choosing an $f_\i^{(0)}$ defined at the individual species' flow velocity\cite{suchy_collision_1996}. However, since we truncate this series at a very low number of moments, the solution spaces no longer overlap. However, the assumption of flows being close to each other is valid for SOL/edge of tokamaks, since the exit velocities of all species are close to the sound speed $c_s$\cite{stangeby_plasma_2000}. In general, one must keep in mind the general ordering of the diffusion velocities as follows
\begin{equation}
 |\mathbf{w}_\i|\ll\left(\frac{kT_\i}{m_\i}\right)^{1/2}.
\end{equation}

One can now introduce $\psi_\alpha=G^{mn}_\alpha$ in the averaged Boltzmann equation (\ref{eq:transport}), in order to compute the moments, obtaining an infinite hierarchy of transport equations.

Generally, in the LHS of the hierarchy of balance equations for the moments obtained in this manner, the $k^{th}$ moment equation contains both the $(k-1)^{th}$ and $(k+1)^{th}$ moments. Hence, one has only $k$ equations for $k+1$ variables. In order to provide a  closure, in Grad's method, one truncates at a moment $k$, such that moments higher than $k$ are calculated using the expansion for $f_\alpha$ truncated at the $k^\mathrm{th}$ term, which approximates the higher moments in terms of the lower ones. This closes the set of equations obtained. Illustration of Grad's closure and also Zhdanov closure are out of the scope of the current article, and will be demonstrated in an upcoming article. In this article, we illustrate the development and solution of the RHS of the equation, i.e. the moment averaged Boltzmann collisional operator, and compare it to previously obtained values by Zhdanov et al\cite{zhdanov_transport_2002,alievskii_1963_transport,yushmanov_diffusion_1980}.

\section{Derivation of the right hand side of the Boltzmann equation}
\label{sec:derivation_operator}

In the Boltzmann collision integral $J_{\alpha\beta}$, it is possible to choose a distribution function which takes the form
\begin{equation}
 f_\alpha = f^{(0)}_\alpha (1+\Phi_\alpha), \nonumber
\end{equation}
which essentially represents the ansatz as a perturbed Maxwellian. The moment-averaged collision operator Eq.\,(\ref{eq:boltzmann2}) can be written as
\begin{multline}
 R_{\alpha\beta}=\int \psi_\alpha J_{\alpha\beta}d\mathbf{c_{\alpha}}\\
 \approx\iiint f^{(0)}_\alpha f^{(0)}_\beta(\psi^\prime_\alpha-\psi_\alpha)(1+\Phi_\alpha+\Phi_\beta)g\sigma_{\alpha\beta}(g,\chi)d\Omega d\mathbf{c_{\alpha}}d\mathbf{c_{1\beta}}, \nonumber
\end{multline}
on neglecting the $\Phi\Phi$ terms. This linearization of the collision operator makes it bilinear, i.e.\,they satisfy the relations
\begin{align}
 J(f_\alpha,f_\beta+f_\gamma)&=J(f_\alpha,f_\beta)+J(f_\alpha,f_\gamma),\nonumber\\
 J(f_\alpha+f_\gamma,f_\beta)&=J(f_\alpha,f_\beta)+J(f_\gamma,f_\beta)\nonumber\\ 
 J(kf_\alpha,lf_\beta)&=klJ(f_\alpha,f_\beta),  \nonumber
 \end{align}
and correspondingly the moment-average $R(\psi,f_\alpha,f_\beta)$ is trilinear.
\begin{align}
 R(\psi,f_\alpha,f_\beta+f_\gamma)&=R(\psi,f_\alpha,f_\beta)+R(\psi,f_\alpha,f_\gamma),\nonumber\\
 R(\psi,f_\alpha+f_\gamma,f_\beta)&=R(\psi,f_\alpha,f_\beta)+R(\psi,f_\gamma,f_\beta),\nonumber\\
 R(\psi+\eta,f_\alpha,f_\beta)&=R(\psi,f_\alpha,f_\beta)+R(\eta,f_\alpha,f_\beta),\nonumber\\
 R(j\psi,kf_\alpha,lf_\beta)&=jklR(\psi,f_\alpha,f_\beta).  
\end{align}
This allows us to decompose the moment-average into sums of smaller terms, which is useful analytically. This also is similar to the properties exhibited by some other linearized operators such as the linearized Landau operator\cite{balescu_transport_1988,helander_collisional_2005}. On substituting the Sonine-Hermite polynomial ansatz from Eq.\,(\ref{eq:ansatz}) for the distribution functions $f$, and set the moment $\psi=G^{mn}_{\i}$ from Eq.\,(\ref{eq:sonine-hermite}), we obtain
\begin{multline}
 R^{mnkl}_{\alpha\beta} = \iiint f^{(0)}_\alpha f^{(0)}_\beta \{G^{mn}_\alpha(\mathbf{c}_\alpha^\prime)-G^{mn}_\alpha(\mathbf{c}_\alpha)\}\\
 \times\{1+
 2^{2l}\gamma_\alpha^{2l+k}m_\alpha^{-2}\tau_{kl}G_\alpha^{kl}(\mathbf{c}_\alpha)b^{kl}_\alpha\\
 +2^{2l}\gamma_\beta^{2l+k}m_\beta^{-2}\tau_{kl}G_\beta^{kl}(\mathbf{c}_\beta)b^{kl}_\beta
 \}\\
 \times g\sigma_{\alpha\beta}(g,\chi)d\Omega d\mathbf{c_{\alpha}}d\mathbf{c_{1\beta}}, \nonumber
\end{multline} 
where $R^{mnkl}_{\alpha\beta}$ represents the part of $R^{mn}_{\alpha\beta}$ averaging over the $kl$ term of the ansatz Eq.\,(\ref{eq:ansatz}).  Noting that $G^{kl}b^{kl}$ is an inner product, we now substitute the definition of $G^{mn}$, and use the following integral identity\cite{ji_exact_2006,weinert_spherical_1980}
\begin{multline}
 \int P^{(m)}(P^{(k)}:W)G(v)d\mathbf{v}\\
 =\frac{W}{2m+1}\delta_{km}\int P^{(m)}:P^{(m)}G(v)d\mathbf{v}, \label{eq:pmidentity}
\end{multline}
where $W$ is symmetric and traceless tensor of rank $k$ not a function of $\mathbf{v}$. Furthermore, we define a ``bracket'' integrals of the following form
\begin{equation}
 n_\alpha n_\beta [F,G]=\iiint  f^{(0)}_\alpha f^{(0)}_\beta G(F^\prime-F)g\sigma_{\alpha\beta}(g,\chi)d\Omega d\mathbf{c_{\alpha}}d\mathbf{c_{1\beta}},
\end{equation}
through which we can contract over index $k$ in $R^{mnkl}_{\alpha\beta}$ and write it as $R^{mnl}_{\alpha\beta}$, such that 
\begin{equation}
 R_{\alpha\beta}^{mnl}=(1-\delta_{m0}\delta_{l0})(A_{\alpha\beta}^{mnl}b^{ml}_\alpha+B_{\alpha\beta}^{mnl}b^{ml}_\beta)+\delta_{m0}\delta_{l0}C_{\alpha\beta}^{mnl},
 \label{eq:collision_second_form}
\end{equation}
where $A_{\alpha\beta}^{mnl}$, $B_{\alpha\beta}^{mnl}$ and $C_{\alpha\beta}^{mnl}$ are given by
\begin{align}
 A_{\alpha\beta}^{mnl}&=  Q_{\alpha\beta}^{mnl}  
 \gamma_\alpha^{l-n}\times\nonumber\\
 &\left[S^n_{m+1/2}\left(W_\alpha^2\right)P^{(m)}(\mathbf{W}_\alpha),S^l_{m+1/2}(W_\alpha^2)P^{(m)}(\mathbf{W}_\alpha)\right]\nonumber\\
B_{\alpha\beta}^{mnl}&=Q_{\alpha\beta}^{mnl}\frac{\gamma_\beta^{l+m/2}}{\gamma_\alpha^{n+m/2}}\frac{m_\alpha}{m_\beta}\times\nonumber\\
 &\left[S^n_{m+1/2}(W_\alpha^2)P^{(m)}(\mathbf{W}_\alpha),S^l_{m+1/2}(W_\beta^2)P^{(m)}(\mathbf{W}_\beta)\right]\nonumber\\
 C^{mnl}_{\alpha\beta} &= (-1)^n n! \gamma_\alpha^{-(n+m/2)}m_\alpha n_\i n_\j\left[S^n_{1/2}(W_\alpha^2),1\right],
 \label{eq:AmnBmn1}
\end{align}
where
\begin{equation}
 Q_{\alpha\beta}^{mnl}=(-1)^{n+l}2^{2l+m} \frac{(2m)!(m+l)!n!}{(m!)^2 (2m+2l+1)!}n_\alpha n_\beta.
 \label{eq:qmnl}
\end{equation}
This moment-averaged collision operator is valid for any difference of masses or temperatures of the colliding species.

{  Having now derived our expressions for the moment-averaged collision operator, we now illustrate a slightly different expression derived originally in the appendix of Ref.\,\onlinecite{zhdanov_transport_2002} by Zhdanov, which brings with it some additional assumptions.}  The collision operator is linearized by choosing a distribution function defined at the common temperature of the plasma $T=\sum_\alpha n_\alpha T_\alpha/\sum_\alpha n_\alpha$, such that  $f^{\prime(0)}_\alpha f^{\prime(0)}_\beta=f^{(0)}_\alpha f^{(0)}_\beta$, leading to 
\begin{equation}
 J_{\alpha\beta} = \iint f^{(0)}_\alpha f^{(0)}_\beta(\Phi^\prime_\alpha+\Phi^\prime_\beta-\Phi_\alpha-\Phi_\beta)g\sigma_{\alpha\beta}(g,\chi)d\Omega d\mathbf{c_{1\beta}}. \nonumber
\end{equation}
on neglecting the squared $\Phi\Phi$ terms. However, in such a form of the collision operator, Zhdanov et al assume that the temperatures of each species is close to the common temperature, i.e. $|T-T_\i|\ll T$. One could term this as the linearized Boltzmann collision integral under quasi thermodynamic equilibrium conditions. {  Such a scheme is well-suited when the masses of the colliding species are of the same order, thus both species possessing a collisional relaxation timescale of the same order as well, leading to their distribution functions being close enough to a Maxwellian defined at the common temperature $T$. However, one needs to be careful when the masses of the species are at different orders, leading to different relaxation timescales, such as in the case of heavy impurities colliding with the plasma fuel species, where relaxation timescales may be different.}

On integrating over a moment $\psi_\alpha=G^{mn}_\i$ to obtain the moment averaged collision integral using the ansatz Eqs.\,(\ref{eq:ansatz}) for the distribution function, { and on noting that $b_\i^{\prime kl}=b_\i^{kl}$ for a given species $\i$ by construction since the form of $f_\i$ remains the same}, we can obtain  
\begin{equation}
R_{\alpha\beta}^{mnl}=A_{\alpha\beta}^{mnl}b^{ml}_\alpha+B_{\alpha\beta}^{mnl}b^{ml}_\beta,
\label{eq:collision_first_form}
\end{equation}
where $A_{\alpha\beta}^{mnl}$ and $B_{\alpha\beta}^{mnl}$ are given by
\begin{align}
 A_{\alpha\beta}^{mnl}&= Q_{\alpha\beta}^{mnl}   
 \gamma_\alpha^{l-n}\times\nonumber\\
 &\left[S^l_{m+1/2}(W_\alpha^2)P^{(m)}(\mathbf{W}_\alpha),S^n_{m+1/2}\left(W_\alpha^2\right)P^{(m)}(\mathbf{W}_\alpha)\right]\nonumber\\
 B_{\alpha\beta}^{mnl}&=Q_{\alpha\beta}^{mnl}\frac{\gamma_\beta^{l+m/2}}{\gamma_\alpha^{n+m/2}}\frac{m_\alpha}{m_\beta}\times\nonumber\\
 &\left[S^l_{m+1/2}(W_\beta^2)P^{(m)}(\mathbf{W}_\beta),S^n_{m+1/2}(W_\alpha^2)P^{(m)}(\mathbf{W}_\alpha)\right]
 \label{eq:singletempamnlbmnl}
\end{align}
where $Q_{\alpha\beta}^{mnl}$ has the same expression as Eq.\,(\ref{eq:qmnl}). This is also the collision operator employed for deriving the $21N$-moment single-temperature collision coefficients in Ref.\,\onlinecite{yushmanov_diffusion_1980}. Note, however, that the $A_{\alpha\beta}^{mnl},B_{\alpha\beta}^{mnl}$ here correspond to the $A_{\alpha\beta}^{mln},B_{\alpha\beta}^{mln}$ of Eq.\,(\ref{eq:AmnBmn1}).

The limit of the moment averaged collision operator (\ref{eq:collision_first_form}) is that it is only valid for quasi thermodynamic equilibrium conditions with the species temperatures being close to the plasma common temperature. However, the one in Eq.\,(\ref{eq:collision_second_form}) has no such assumption. Therefore, each individual moment averaged term of Eq.\,(\ref{eq:collision_first_form}) will be less accurate than Eq.\,(\ref{eq:collision_second_form}) for increasing difference in the temperatures of the colliding species.

{ It is worth discussing what the assumption of $b^{\prime kl}=b^{kl}$ means. Since $b^{kl}$ enter the distribution functions, this assumption essentially implies no change in the physical quantity in the pre-collision and post-collision distribution functions.  For example $b_\i^{01}=\rho_\i\mathbf{w}_\i$ is the diffusion velocity of the species $\i$, and assuming that  $b_\i^{01}= b_\i^{\prime 01}$ implies that the diffusion velocities of the pre-collision and post-collision distributions remain the same for the same species $\i$, not changing in the timescale of the collision. This would also ensure that the distribution function $f$ has the same form for a given species $\i$ for all four pre-collision and post-collision distributions, as necessitated by the Boltzmann equation. This is a reasonable assumption when the duration of the collision is very small as in the case of short-range forces, e.g.\,in case of rigid-sphere collisions where the collision only lasts an instant, or for weakly-coupled long-range forces such as Coulomb potential, where again the small-angle collision duration is very small. However, for long-range interaction potentials with strong coupling, one has to be careful that the collision duration is much smaller than the timescale of the system evolution, and in general for such an interaction potential the Boltzmann collision operator is anyhow not appropriate.  }

To use either expression of the moment-averaged collision operator, we need to calculate these bracket integrals for the required $(m,n)$ values in order to derive the desired forces. Because of Eq.\,(\ref{eq:pmidentity}), the only brackets that survive in the moment-averaged linearized collision operator are the ones possessing the same rank, which significantly reduces the number of terms one must calculate. Certain methods for deriving these bracket integrals are provided in Refs.\cite{chapman_mathematical_1952}, \cite{ferziger_mathematical_1972} and \cite{rat_transport_2001} for $m=1,2$, as they are well suited enough within the scope of the $21N$-moment scheme. In the next section we indicate the general expressions for the bracket integrals obtained by following these methods for a case of species $\alpha,\beta$ possessing different masses and different temperatures.

\section{General expressions for the bracket integrals}
\label{sec:general_expressions}

 The general expressions for the rank-$m$ bracket integrals take the following general forms { 
 \begin{multline}
 \left[S^p_{3/2}(W_\i^2)P^{(m)}(\mathbf{W}_\i),S^q_{3/2}(W_\j^2)P^{(m)}(\mathbf{W}_\j)\right]\\
 \sim\sum_{rl} \frac{A^{pqrl,m}_{\i\j}}{k_{\i\j}^{r+3/2}}\Omega_{\i\j}^{lr},\nonumber
 \end{multline}
 and
 \begin{multline}
  \left[S^p_{3/2}(W_\alpha^2)P^{(m)}(\mathbf{W}_\alpha),S^q_{3/2}(W_\alpha^2)P^{(m)}(\mathbf{W}_\alpha)\right]\\
  \sim\sum_{rl} \frac{A^{pqrl,m}_{\i\i}}{k_{\i\j}^{r+3/2}}\Omega_{\i\j}^{lr}.\nonumber
 \end{multline}
The coefficients $A^{pqrl,m}_{\i\j}$, $A^{pqrl,m}_{\i\i}$ are functions of mass and temperature ratios of the species $\i$ and $\j$. The terms $\Omega_{\i\j}^{lr}$ are the effective cross-section moment integrals of Chapman and Cowling (henceforth referred to as the ``Chapman-Cowling integrals'') which are dependent on the potential of interaction between species $\i$ and $\j$. The factor $k_{\i\j}$ is an arbitrary function of the masses and temperatures of the colliding species, which can be chosen freely as long as $k_{\i\j}>0$. The choice also affects the forms of $A^{pqrl,m}_{\i\j}$,$A^{pqrl,m}_{\i\i}$ and $\Omega_{\i\j}^{lr}$ which depend on $k_{\i\j}$ individually, such that, in principle, the overall bracket integral values do not depend on $k_{\i\j}$.} 
The exact derivations the generalized coefficients $A^{pqrl,m}_{\i\j}$ and $A^{pqrl,m}_{\i\i}$ for different bracket integrals up to rank-2, with all steps supplied for verification purposes, are provided in Appendix \ref{sec:bracket_integral_derivation}. For the purpose of our work, up to rank-2 suffices, as in the context of the Boltzmann collision operator, only the moments up to rank-2 have been considered in previous works. The solution for higher-order bracket integrals is out of the scope of the current article, and will be reserved for a future manuscript.

The ``reversed'' bracket integrals $\left[S^p_{3/2}(W_\j^2)P^{(m)}(\mathbf{W}_\j),S^q_{3/2}(W_\i^2)P^{(m)}(\mathbf{W}_\i)\right]$ required for the second form of the collision operator, and $\left[S^p_{3/2}(W_\beta^2)P^{(m)}(\mathbf{W}_\beta),S^q_{3/2}(W_\beta^2)P^{(m)}(\mathbf{W}_\beta),\right]$ required for verifying conservation properties, can be obtained by transforming $\i\rightleftharpoons\j$ in the above expressions provided. Since the two types bracket integrals are calculated independently of each other, any mistake in calculating them would lead to non-conservation of momentum or energy. Hence, demonstrating conservation of momentum and energy with the obtained quantities is an adequate testament to their accuracy (see Appendix \ref{sec:bracket_values}). Furthermore, the expressions for the bracket integrals are very amenable to being implemented in computer algebra systems, as they are just composed of sums and products which any computer algebra system should be able to perform. We implement the expressions for the bracket integrals and the moment-averaged collision term in Mathematica\cite{mathematica}.

\label{sec:cross_sections}

{  The Chapman-Cowling integral is written in the following form for our case
\begin{align}
 \Omega_{\i\j}^{lr} =& \left(\frac{\pi}{d_{\i\j}}\right)^{1/2}\int_0^\infty \exp(-\zeta^2)   \zeta^{2r+3} \phi^{(l)}_{\i\j} d\zeta,\\
 \phi^{(l)}_{\i\j}=&\int^\infty_0 (1-\cos^l{\chi}) \sigma_{\alpha\beta}(g,\chi)\sin{\chi}d\chi,
\end{align}
where $\zeta=d_{\i\j}^{1/2}g$ and where the factor $d_{\i\j}$ is related to $k_{\i\j}$ by the following relation
\begin{equation}
 d_{\i\j}=k_{\i\j}\left\{\mu_{\i\j}^2\left(\frac{\gamma_\i}{2m_\i^2}+\frac{\gamma_\j}{2m_\j^2}\right)\right\}. \nonumber
\end{equation}
Choosing $k_{\i\j}$ fixes $d_{\i\j}$ and vice versa, and note therefore that $\Omega_{\i\j}^{lr}$ is a functional of $d_{\i\j}$ and the effective cross-section $\phi^{(l)}_{\i\j}$.}
Essentially we are now left with calculating the effective cross sections $\phi^{(l)}_{\i\j}$, which in turn depend on the physics of the particle-potential interaction. Therefore, the choice of potential is crucial to calculating these effective cross sections accurately. We essentially have a choice between pure Coulomb potential or the shielded Coulomb potential, as these are the only ones that apply to fully ionized plasmas in the fusion domain. However, there are always integrability and convergence issues with the potentials used in these calculations. For example, the pure Coulomb potential diverges at the limit of low collision angles with high impact parameters which constitute the majority of collisions in a hot plasma. This is usually mitigated by choosing a cutoff for the integral at the Debye length radius, which then leads to a converged integral. For the shielded Coulomb potential, for high energy and low angle interactions, it may lead to an issue where some forms of the integrals obtained do not converge at high impact parameter limits. Some physical approximations, such as ignoring large angle collisions for the shielded Coulomb potential part of the integration, are used to express the integral approximately in forms that converge. From a modelling point of view, it is worth keeping in mind that these two potentials will offer slightly different collisional coefficients, which can provide a range of values which could be useful for comparison with experiments. 

In general, the integration of the effective cross-sections with the shielded Coulomb potential is not so simple, and often has to be manually done for different values of $l$, with unique approximations applied at each value. However, it is possible to find the asymptotic values of the cross-sections through a perturbation method\cite{kihara_coefficients_1959}.  {  For our case, the asymptotic form of $\Omega$-integral for the shielded Coulomb potential looks like\footnote{A previous version of this article had a typographical error in the expression for $ \Omega^{lr}_{\i\j,sh}(d_{\i\j})$. The factor inside the Coulomb logarithm should be 1 and not 4.}
\begin{multline}
 \Omega^{lr}_{\i\j,sh}(d_{\i\j})= l\Gamma(r)\left(\frac{\pi}{d_{\i\j}}\right)^{1/2}\frac{\Delta_{\i\j}^2}{4}\left(\frac{2kTd_{\i\j}}{\mu_{\i\j}}\right)^2\\
 \times\left\{\ln\left(\frac{\lambda_D}{\Delta_{\i\j}} \frac{\mu_{\i\j}}{2kTd_{\i\j}}\right)+A_r-C_l -2\ln\gamma\right\},
 \label{eq:kihara_formula}
\end{multline}
}where
\begin{align}
 C_l&=\left\{\begin{tabular}{ll}
             $(1+\frac{1}{3}+\frac{1}{5}+\ldots+\frac{1}{l})-\frac{1}{2l},$ & for odd $l$\\
             $(1+\frac{1}{3}+\frac{1}{5}+\ldots+\frac{1}{l-1}),$ & for even $l$
             \end{tabular}\right.\nonumber\\
 A_r&=1+\frac{1}{2}+\frac{1}{3}+\ldots+\frac{1}{r-1},\ A_{1}=0,\nonumber 
\end{align}
where $T$ is the plasma common temperature given by $T=(1/n)\sum_\i n_\i T_\i$, where $\Delta_{\i\j}$ is the mean distance of closest approach (also called the ``particle diameter''), $\lambda_D$ is the Debye length given by
\begin{equation}
 \lambda_D^{-2}=\sum_\alpha \frac{ n_\i Z_\i^2 e^2}{\epsilon_0 kT_\i},\nonumber
\end{equation}
and  $\gamma$ is the Euler-Mascheroni constant. This formula has the advantage of being easily implementable in computer algebra softwares. {  However, it is worth noting that this expression leads to the Chapman-Cowling integral for the shielded Coulomb potential (and hence also the bracket integrals) evaluating to different values depending on the chosen value of $d_{\i\j}$. This arises from two assumptions in the calculation of the cross-section from the shielded Coulomb potential in Refs.\,\onlinecite{kihara_coefficients_1959} and \onlinecite{liboff_transport_1959}. The first being the non-dimensionalizing of the impact parameter with respect to the particle diameter and the Debye length, which then  vanishes under the asymptotic lower limit of the collision integral, leaving the final result the same for any choice of the Debye length or the particle diameter. The second is that the model of collisions is essentially a collision model of one particle being deflected by one potential. The cross section integrals $\phi^{(l)}$, found in this manner depend only on the relative velocity $g$ other than the potential of interaction. However, the relative velocity $g$ in the Chapman-Cowling is scaled by a factor $d_{\i\j}^{1/2}$, and hence the value of $\phi^{(l)}$ will also need to be transformed to the scaled value of $\zeta=d_{\i\j}^{1/2}\g$, leading to the additional factors of ${2kTd_{\i\j}}/{\mu_{\i\j}}$ seen in Eq.\,(\ref{eq:kihara_formula}). For our coefficients, we choose $d_{\i\j}$ in such a manner that $d_{\i\j}={\mu_{\i\j}}/{2kT}$, in order to agree with the two aforementioned calculations of the cross-sections for the shielded Coulomb potential in Refs.\,\onlinecite{kihara_coefficients_1959} and \onlinecite{liboff_transport_1959}. }
Having $d_{\i\j}$ in such a manner also allows us to immediately use cross-section values from numerical and empirical databases such as AMJUEL\cite{reiter2000data}, LXCat\cite{pancheshnyi2012lxcat} and ADAS\cite{summers2011atomic}, because the cross-sections in these databases are often expressed as a polynomial of ${\mu_{\i\j}g^2}/{2kT}$.

Some attention also needs to be paid to the chosen value of $\Delta_{\i\j}$. Following Liboff\cite{liboff_transport_1959}, Kihara et al\cite{kihara_coefficients_1959} and Hahn et al\cite{hahn_quantum_1971}, one may choose it as follows
\begin{equation}
 \Delta_{\i\j}=\frac{|Z_\i Z_\j| e^2}{4\pi\epsilon_0 kT},\label{kihara_closest_distance}
\end{equation}
where $T$ is the common temperature of the plasma. Such a value of $\Delta_{\i\j}$ is also termed the ``particle diameter'' in the literature (where $\pi\Delta_{\i\j}^2$ would be the collision cross-section for the rigid-sphere collision case\cite{chapman_mathematical_1952}). Following Zhdanov\cite{zhdanov_transport_2002}, one could also choose it as the average inverse impact parameter  $\langle 1/b_0\rangle^{-1}$ over the distributions as follows
\begin{equation}
 \Delta_{\i\j}=\frac{|Z_\i Z_\j| e^2 \gamma_{\i\j}}{12\pi\epsilon_0 \mu_{\i\j}},\label{zhdanov_closest_distance}
\end{equation}
where $\gamma_{\i\j}$ is given by $\gamma_{\i\j}=\gamma_\i\gamma_\j/(\gamma_\i+\gamma_\j)$. (which for equal temperatures makes the logarithmic term in the expression equal to the Coulomb logarithm $\ln\Lambda_{\i\j}$). Whichever is chosen has to be used consistently\footnote{What is interesting is that using Zhdanov's calculation of the mean distance of closest approach would effectively add a factor of $\ln4$ to the value of the Coulomb logarithm. If one decided to retain the Kihara-Liboff values, then the value of $\Omega^{lr}_{\i\j,sh}(d_{\i\j})$, neglecting $A_r,C_l$ would be roughly
\begin{equation}
 \Omega^{lr}_{\i\j,sh}(d_{\i\j}) \sim\ln\left(\frac{4}{\gamma^2}\frac{\pi\epsilon_0}{ |Z_\i Z_\j| e^2} \frac{\mu_{\i\j}}{\gamma_{\i\j}}\lambda_D\right).\nonumber
\end{equation}
But follwing Zhdanov's method of calculation from Chapter 1 of Ref.\,\onlinecite{zhdanov_transport_2002}, one can obtain
\begin{equation}
 \Omega^{lr}_{\i\j,sh}(d_{\i\j}) \sim\ln\left(12\frac{\pi\epsilon_0}{ |Z_\i Z_\j| e^2} \frac{\mu_{\i\j}}{\gamma_{\i\j}}\lambda_D\right).\nonumber
\end{equation}
It is quite a coincidence that $4/\gamma^2\approx12$, and that, by using two different approaches, one essentially comes up with a very similar numerical factor! Nonetheless, this still leaves open the question of which mean distance of closest approach to use, as it is somewhat arbitrary.}. In Ref.\,\onlinecite{zhdanov_transport_2002} by Zhdanov, the cross-section has been treated by using Eq.\,(\ref{zhdanov_closest_distance}) for the Coulomb logarithm, but Eq.\,(\ref{kihara_closest_distance}) for the mean distance of closest approach outside the logarithm. This is essentially the same approximation as in Rosenbluth et al\cite{rosenbluth_fokker-planck_1957}, but which has non-negligible effect on the $\Omega^{lr}_{\i\j}$ for high values of $(l,r)$ (since neither $A_n$ nor $C_n$ converge for large $n$, and can become larger than the log term for a large enough value of $n$). This still remains a reasonable approximation however, because we are only concerned with a small number of lower-order moments. 

One can also, following the interpolated formula provided above, extrapolate results obtained by following Liboff's procedure\cite{liboff_transport_1959} for results provided by Bonnefoi\cite{bonnefoi_thesis_1975}, and one will find the same expression as Eq.\,(\ref{eq:kihara_formula}), but with $-2\gamma$ in place of the $-2\ln\gamma$.
The cause of this is that, in Liboff's procedure, the general angle of deflection of the collision takes the form of the first-order modified Bessel function of the second kind $K_1$, and one must use approximations on the order of the energy of the interaction to express the integral at any order as the square of $K_1$, as every other power of $K_1$ is non-integrable in the given limits. This leads to the integral having a correction of $A_r-\gamma$ instead of that in Eq.\,(\ref{eq:kihara_formula}). 

One feature to note of these shielded Coulomb potential cross-sections is that they ignore any difference in shielding for the attractive and repulsive potential cases, because the particle diameter is assumed to be positive for both attractive and repulsive cases. Depending on the kind of plasma, this may be relevant\cite{capitelli_transport_2013, dangola_thermodynamic_2008}. However, for the case of fusion plasmas, this may safely be neglected, given that the logarithmic term is dominant for hot plasmas. 

We also mention for the sake of completeness that the original formula for the Chapman-Cowling integrals used by Zhdanov for the values of $\Omega^{lr}_{\i\j}$ in Ref.\,\onlinecite{zhdanov_transport_2002} found using the Coulomb potential with the Debye length cutoff is given by
\begin{equation}
 \Omega^{lr}_{\i\j}=\sqrt{\pi}l(r-1)! \left( \frac{Z_\i Z_\j e^2}{4\pi\epsilon_0} \right)^2 \frac{\ln{\Lambda_{\i\j}}}{\mu_{\i\j}^{1/2} (2kT)^{3/2}},
 \label{eq:zhdanov_omegarl}
\end{equation}
where the plasma common temperature $T$ refers to $T=(1/n)\sum_\i n_\i T_\i$. This follows from Zhdanov's forms of the $\Omega$-integrals, when the temperatures of the components are close to each other and when the Debye length $\lambda_D$ is much smaller than the inverse of the average inversed impact parameter, i.e. 
\begin{equation}
\lambda_D\ll \bar{\left(\frac{1}{b_0} \right)}^{-1},\nonumber
\end{equation}
and where $\Lambda_{\i\j}$ is the Coulomb logarithm given by 
\begin{equation}
 \Lambda_{\i\j}=\frac{12\pi\epsilon_0kT}{Z_\i Z_\j e^2}\lambda_D.\nonumber
\end{equation}
This follows the general result in Eq.\,(\ref{eq:kihara_formula}) to the order of the logarithmic term.

{  We will calculate our cross-sections with Eq.\,(\ref{eq:kihara_formula}) with Eq.\,(\ref{kihara_closest_distance}) and with $d_{\i\j}=\mu_{\i\j}/(2kT)$.} With expressions for the coefficients $A^{pqrl}_{\i\j/\i\i}$ and the cross-sections $\Omega^{lr}_{\i\j,sh}$, we can now proceed to calculate the collisional coefficients and compare them to the ones found in existing literature.

\section{Range of Validity of Zhdanov's values} 
\label{sec:comparisons}

Zhdanov et al have previously derived two sets of coefficients for the values of the collisional coefficients. The first set was derived for a multi-temperature, multi-component plasma without any explicit assumptions on the temperatures of the species (see Appendix \ref{sec:zhdanov13}). These $13N$-moments multi-temperature coefficients were also calculated  using the linearized Boltzmann operator. The derivation method for a few lower order moments is given in Refs.\,\onlinecite{zhdanov_transport_2002} and \onlinecite{zhdanov_influence_2013}. However, to the best of our knowledge, no explicit general derivation scheme method was provided, making it difficult to verify some of the cumbersome higher-order moments, and generate higher-order ones, in case we need larger number of moments.  These coefficients generally take the form
\begin{equation}
 A^{mpq}_{\i\j},B^{mpq}_{\i\j}\sim\sum_{k=0}^2 \Theta_{\i\j}^k \sum_{rl} K_{\i\j,\i}^{m,pqrl} \Omega_{\i\j}^{rl},
 \label{eq:zhdanov13_forms}
\end{equation}
where $K_{\i\j,\i}^{m,pqrl}$ is a term depending solely on the masses, temperatures and densities of the species, $\Omega_{\i\j}^{rl}$ is the Chapman-Cowling effective cross-section moment defined by Eq.\,(\ref{eq:zhdanov_omegarl_proper}), 
and $\Theta_{\i\j}$ is given by
\begin{equation}
 \Theta_{\i\j}=\left(1-\frac{T_\j}{T_\i}\right)\left/\left(1+\frac{m_\j}{m_\i}\right)\right..\nonumber
\end{equation}
{  We can verify that on choosing $d_{\i\j}=\gamma_{\i\j}/2$, our coefficients equal the ones provided by Zhdanov (see Appendix C), thus verifying the coefficients provided in known literature for the chosen value of $d_{\i\j}$. We can now also generate higher-order moments of this multi-temperature form if required.}

The second set of collision coefficients, provided by Zhdanov in Refs.\,\onlinecite{zhdanov_transport_2002} and \onlinecite{yushmanov_diffusion_1980}, are derived using the form of the collision operator given by Eq.\,(\ref{eq:collision_first_form}), for $21N$-moments (see Appendix \ref{sec:zhdanov21}). They take the general form 
\begin{equation}
 A^{mpq}_{\i\j},B^{mpq}_{\i\j}\sim \sum_{rl} L_{\i\j,\i}^{m,pqrl} \Omega_{\i\j}^{rl},
  \label{eq:zhdanov21_forms}
\end{equation}
where $L_{\i\j,\i}^{m,pqrl}$ is a term depending on the masses, temperatures and densities of the species. The bracket integrals are evaluated at the  plasma common temperature $T=\sum_\i n_\i T_\i/\sum_\i n_\i$, however. { There still remains, however, an individual species temperature dependence from the terms multiplying the bracket integral (See Exs.\,(\ref{eq:singletempamnlbmnl})). The values of $\Omega_{\i\j}^{rl}$, however, are proportional to the approximate formula Eq.\,(\ref{eq:zhdanov21Ncrosssections}), which is accurate to the order of the Coulomb logarithm.} 

When comparing our expressions derived with $d_{\i\j}=\mu_{\i\j}/(2kT)$ with the expressions Eqs.\,(\ref{eq:zhdanov13_forms}) and Eqs.\,(\ref{eq:zhdanov21_forms}), we can expect the $13N$-moment multi-temperature coefficients to agree in the case of equal temperatures and the $21N$-moment coefficients to modestly agree, and for both have a range of reasonable agreement in the vicinity of equal temperatures. To understand where the coefficients begin to diverge, we consider some physical situations relevant to SOL/edge physics, i.e. 1.\,a three component plasma intended to be the fusion fuel, with electrons, deuterium (D) and tritium (T), the three being at comparable densities, as D-T fusion is planned to be used in current and future high-Q campaigns, 2.\,a three component plasma with light impurities at significant fraction (10\%) of the main fuel species, i.e.\,electrons, hydrogen (H) and carbon (C), with the carbon in the plasma originating from facing plasma components made of graphite, 3.\,a three component plasma with injected mid-weight impurities with densities at a small fraction (1\%) of the fuel species density, electrons at a small fraction, hydrogen and argon (Ar), often used for controlled experimentation with impurities or for other purposes, and finally 4.\,a three component plasma with a heavy impurity at trace levels (0.001\%), i.e.\,electrons, ions and tungsten (W), where the tungsten usually originates from the walls and divertors made of tungsten. Such a choice of scenarios will help us scan over operationally relevant mass ratios and density ratios, allowing us to focus on the effect of the temperature ratio. The values of masses, charges, densities and temperatures we choose can be found in Table (\ref{table:values}). The temperatures are chosen so as to provide a range of temperature ratio spanning $0.1-2$.

\begin{table}
\centering
 \begin{tabular}{|c||c|c|c|c|}
\hline \rule{0pt}{2ex}
     $\i-\j\rightarrow$      & \text{T-D} & \text{C-H} & \text{Ar-H} & \text{W-H} \\
        \hline\hline
 \rule{0pt}{3ex} $n_\i$   & $10^{19}$  &  $10^{18}$ & $10^{17}$ & $10^{14}$\\
  $Z_\i$   & $+1$       & $+6$       & $+7$      & $+7$\\ 
  $m_\i$   & $3$ amu    & $12$ amu   & $40$ amu  & $184$ amu\\
  $T_\i$   & $100$ eV   & $100$ eV   & $100$ eV  & $100$ eV\\   
  $n_\j$   & $10^{19}$  &  $10^{19}$ & $10^{19}$ & $10^{19}$\\
  $Z_\j$   & $+1$       & $+1$       & $+1$      & $+1$\\ 
  $m_\j$   & $2$ amu    & $1$ amu    & $1$ amu   & $1$ amu\\
  $T_\j$   & $10-200$ eV& $10-200$ eV& $10-200$ eV& $10-200$ eV\\
  \hline
\end{tabular}
\caption{Values of constants used for the different operational cases chosen. At 100eV, the maximum excitation state for higher-Z impurities is around +7, which is why we chose to limit Argon and Tungsten charge state to +7.}
\label{table:values}
\end{table}

Firstly, we begin by looking at coefficients for the D-T case with a physical significance that can be intuitively understood, e.g. the friction force, governed by $A^{100}_{\ij},B^{100}_{\ij}$, the thermal gradient force, which is governed by $A^{110}_{\ij},B^{110}_{\ij}$, the energy exchange term given by $C^{010}_{\ij}$. 
\begin{figure}
 \centering
 \includegraphics[width=\columnwidth]{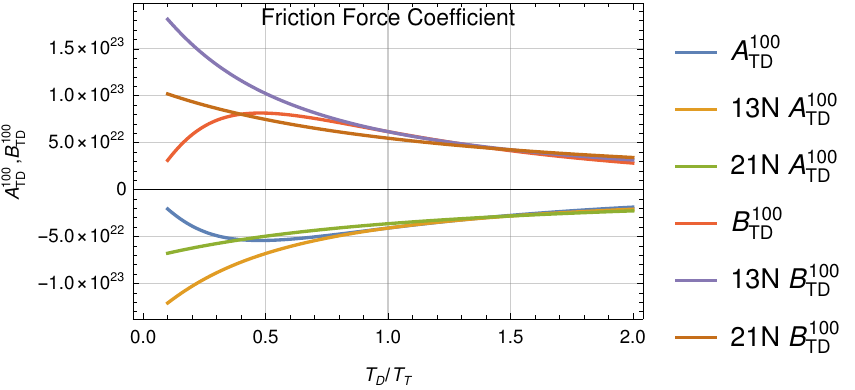}
 \includegraphics[width=\columnwidth]{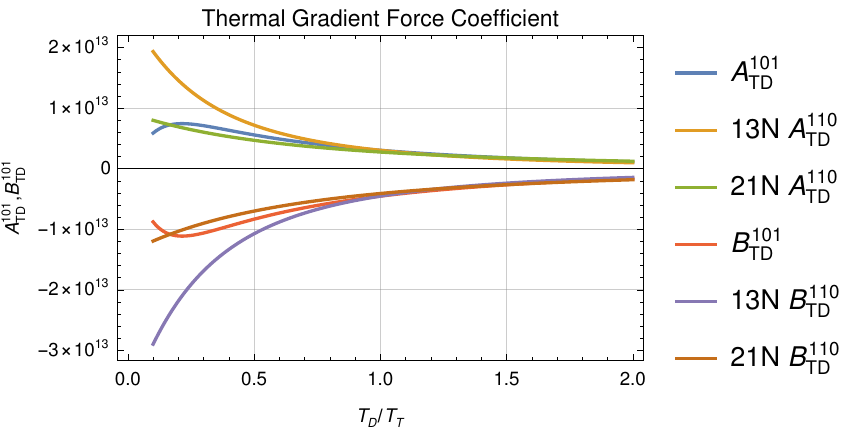}
 \includegraphics[width=\columnwidth]{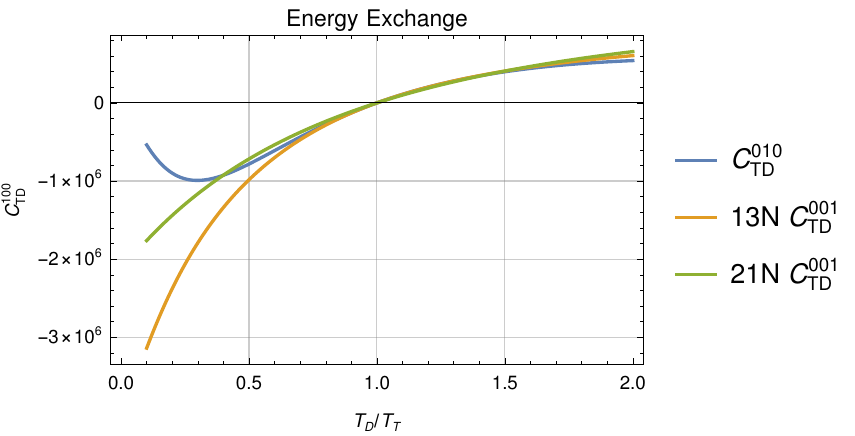}
  \caption{Comparison of physically relevant moments, i.e., from top to bottom, friction force, the thermal gradient force, and the energy exchange, between the two species D and T. The plots are of the coefficient values plotted against the temperature ratio $T_D/T_T$. { In the legends for the plots, first $A^{mnl}_{\i\j}/B^{mnl}_{\i\j}$ refer to our coefficients, the ``$13N$'' indicates Zhdanov's multi-temperature collisional coefficients and similarly the ``$21N$'' refers to Zhdanov's single-temperature collisional coefficients.} }
  \label{fig:intuitive_terms}
\end{figure}
We can immediately notice in Fig.\,(\ref{fig:intuitive_terms}), that generally in the vicinity of equal temperatures $T_D/T_T=1$, the curves for all coefficients tend to follow each other quite closely. In particular, the curves for the $13N$-moment multi-temperature coefficients follow our obtained values much closer than the $21N$-moment single-temperature ones. However, they deviate quite significantly going away from equal temperatures.

Based on these observations, we state that the $13N$-moment multi-temperature coefficients agree more with ours in the vicinity of equal temperature than the  $21N$-moment single-temperature ones. However, to better recommend a range of validity and quantitatively characterize the deviations, we proceed to plot the percentage differences in the coefficients, defined as the absolute value percentage difference of Zhdanov's two sets of coefficients with respect to our coefficients Eqs.\,(\ref{eq:collision_second_form}).

\begin{figure}
\centering
 \includegraphics[width=0.9\columnwidth]{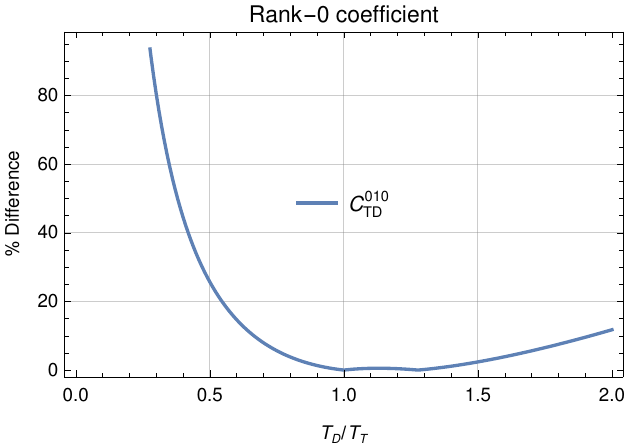}
  \includegraphics[width=0.9\columnwidth]{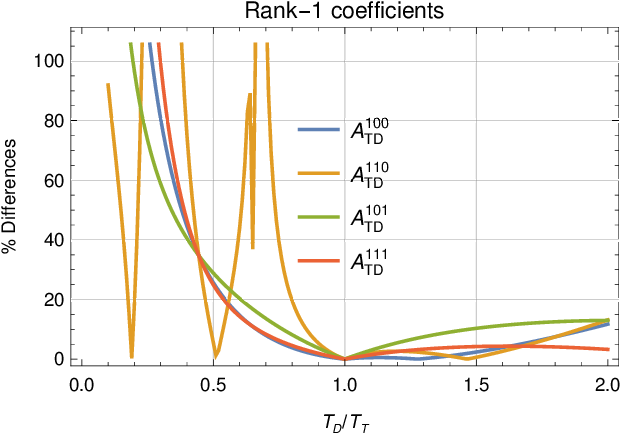}
 \includegraphics[width=0.9\columnwidth]{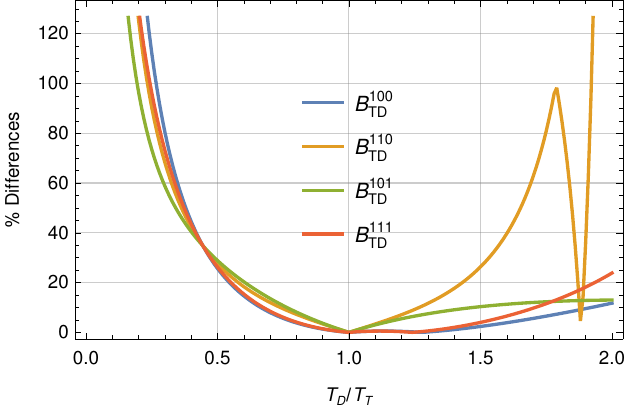}
 \includegraphics[width=0.9\columnwidth]{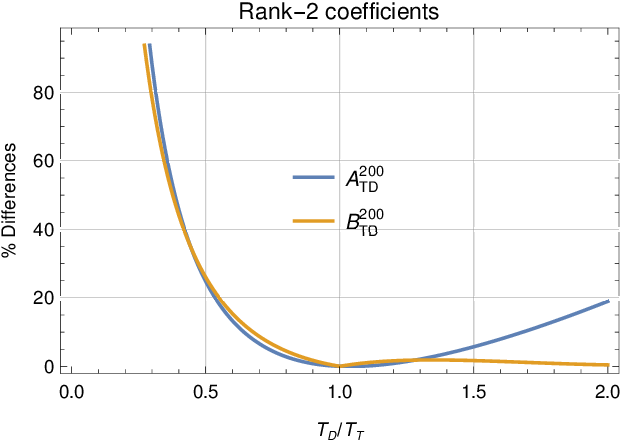}
 \caption{Plots of percentage differences in multi-temperature coefficients calculated by Eqs.\,(\ref{eq:zhdanov13_forms}) and ours calculated from Eq.\,\ref{eq:collision_second_form}. { The top plot showcases the difference for the rank-0 coefficient, the two middle plots shows the differences for the rank-1 coefficients, and the bottom plot shows the differences for the rank-2 coefficients.}}
 \label{fig:zhdanov13_errors}
\end{figure}

The percentage differences in the $13N$-moment multi-temperature coefficients are plotted in Fig.\,(\ref{fig:zhdanov13_errors}). 
We can notice exact agreement at equal temperatures, and that the differences in the coefficients are very low in the vicinity of equal temperatures. However, they seem to deviate rapidly as the temperature ratio decreases below 0.5. In particular, the coefficients related to the heat flux transmission $A^{110}_{\i\j},B^{110}_{\i\j}$ seem to deviate very quickly. This would indicate that at significant temperature differences, the representation of heat flux gains more importance. { In Table (\ref{table:13n}), we showcase the maximum differences in the temperature ratio range of $0.8-1.2$. We can see that all differences for all four physical cases are less than 22\%, with most coefficients having differences less than 6\%.} Furthermore, we can notice that the heavier the impurity becomes, the lower the differences are. This indicates that the $13N$-moment multi-temperature coefficients may be more suitable for simulation the heavier the impurity species being simulated.

\begin{table}
\centering
 \begin{tabular}{|c||c|c|c|c|}
 \hline
 \rule{0pt}{2ex}
 \text{Coefficient} & \text{D-T} & \text{C-H} & \text{Ar-H} & \text{W-H} \\ 
\rule{0pt}{3ex}$C^{010}$& 5.29 & 6.19 & 5.32 & 5.20 \\
$A^{100}$& 5.29 & 6.19 & 5.32 & 5.20 \\
$B^{100}$& 5.29 & 6.19 & 5.32 & 5.20 \\
$A^{101}$& 14.50 & 26.06 & 23.28 & 23.39 \\
$B^{101}$& 14.50 & 26.06 & 23.28 & 23.39 \\
$A^{110}$& 12.51 & 19.77 & 18.27 & 18.37 \\
$B^{110}$& 42.95 & 14.89 & 2.18 & 2.48 \\
$A^{111}$& 5.73 & 6.92 & 6.19 & 6.11 \\
$B^{111}$& 6.77 & 13.36 & 12.15 & 12.20 \\
$A^{200}$& 6.65 & 8.60 & 7.50 & 7.39 \\
$B^{200}$& 3.24 & 3.25 & 2.75 & 2.65 \\
\hline
\end{tabular}
\caption{Table of maximum percentage differences in the $13N$-moment multi-temperature coefficients in the range $T_\i/T_\j=0.8-1.2$. It can be noticed that the differences remain reasonably low for small temperature differences.}
\label{table:13n}
\end{table}

\begin{figure}
 \centering
 \includegraphics[width=0.9\columnwidth]{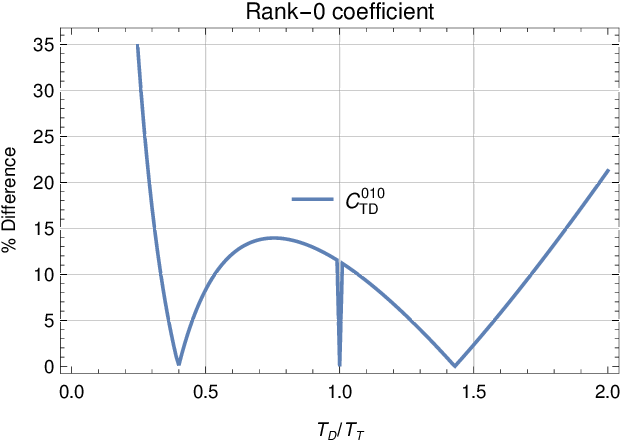}
  \caption{Plots of percentage differences in multi-temperature coefficient $C^{210}_{TD}$ calculated by Eqs.\,(\ref{eq:zhdanov21_forms}) and ours calculated from Eq.\,\ref{eq:collision_second_form}. { The spike near unity temperature ratio is because of our collision coefficient changing signs, but the error being the absolute value of the relative difference.}}
 \label{fig:zhdanov21_errors0}
\end{figure}

\begin{figure}
 \centering
 \includegraphics[width=0.9\columnwidth]{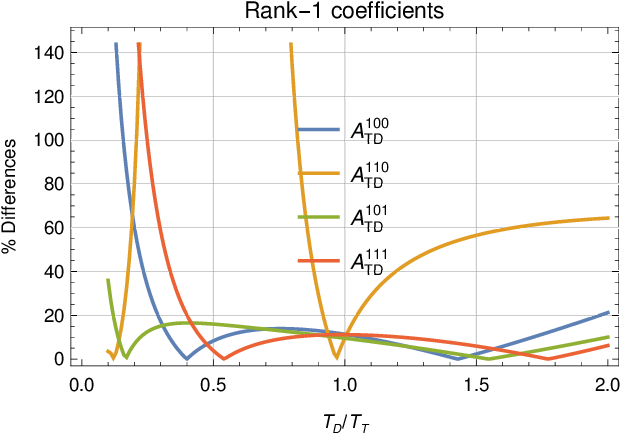}
 \includegraphics[width=0.9\columnwidth]{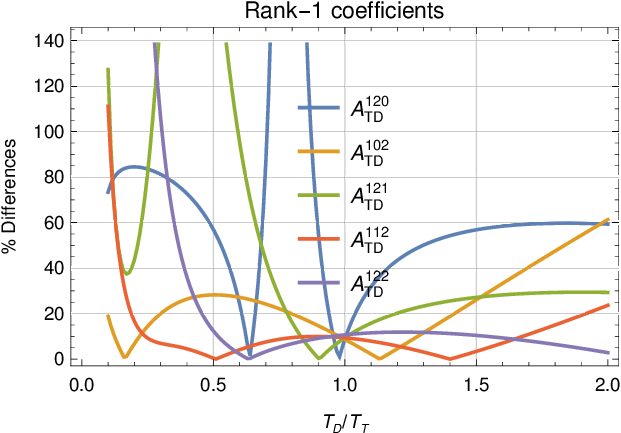}
 \includegraphics[width=0.9\columnwidth]{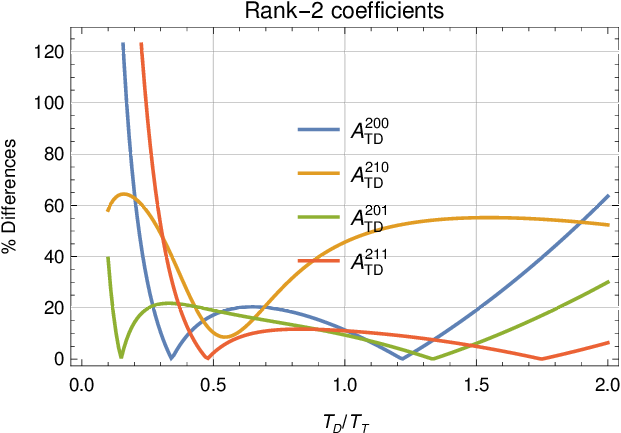}
 \caption{Plots of percentage differences in multi-temperature coefficient $A^{mnl}_{TD}$ calculated by Eqs.\,(\ref{eq:zhdanov21_forms}) and ours calculated from Eq.\,\ref{eq:collision_second_form}. { The top and the middle plots showcase the differences for the rank-1 coefficients, and the bottom plot shows the differences for the rank-2 coefficients.}}
 \label{fig:zhdanov21_errorsA}
\end{figure}

\begin{figure}
 \centering
 \includegraphics[width=0.9\columnwidth]{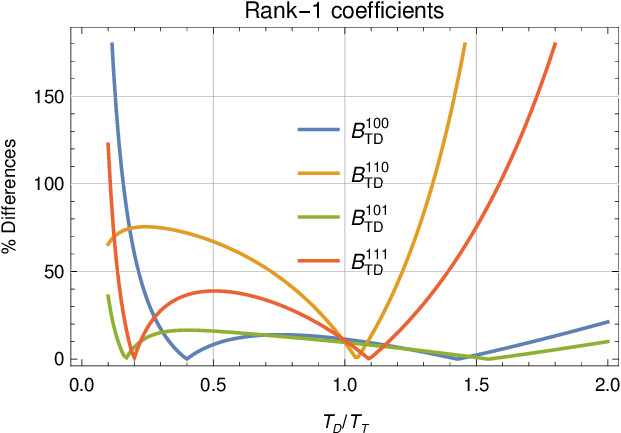}
 \includegraphics[width=0.9\columnwidth]{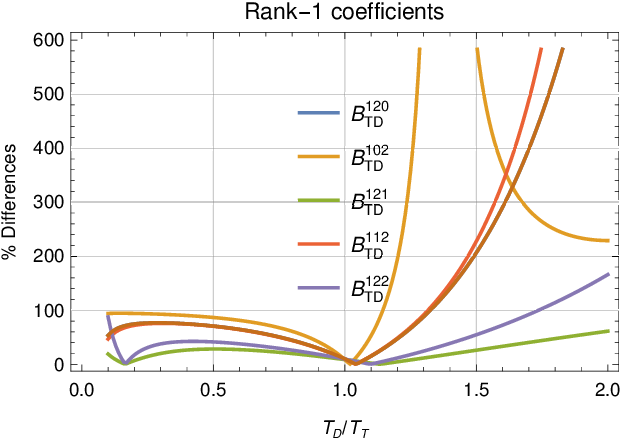}
 \includegraphics[width=0.9\columnwidth]{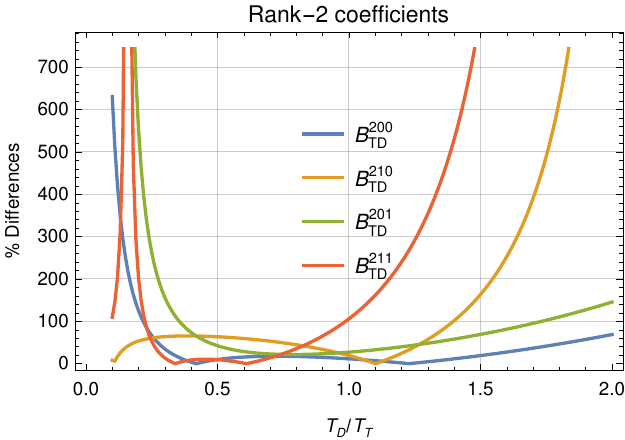}
 \caption{Plots of percentage differences in multi-temperature coefficients $B^{mnl}_{TD}$ calculated by Eqs.\,(\ref{eq:zhdanov21_forms}) and ours calculated from Eq.\,\ref{eq:collision_second_form}. { The top and the middle plots showcase the differences for the rank-1 coefficients, and the bottom plot shows the differences for the rank-2 coefficients.}}
 \label{fig:zhdanov21_errorsB}
\end{figure}

We show the differences in the single-temperature $21N$-moment single-temperature coefficients for the D-T case in Figs.\,(\ref{fig:zhdanov21_errors0}), ((\ref{fig:zhdanov21_errorsA})) and (\ref{fig:zhdanov21_errorsB}).
One can notice in these figures that the differences in the coefficients are significantly higher than those of the $13N$-moment multi-temperature case. Furthermore, the percentage differences in higher order moments, e.g. $A/B_{\i\j}^{11n}$, $A/B_{\i\j}^{12n}$ and $A/B_{\i\j}^{21n}$ are significantly higher than those of the lower order moment ones. The same trends are observed in the cases of carbon, argon and tungsten as well. In order to compare the differences, in Table (\ref{table:21n}), we show the maximum difference in the temperature ratio range of $0.8-1.2$. We notice the same trend as earlier for the lower order coefficients, in the decrease in the percentage differences the heavier the impurity species gets. However, we also notice that the difference for the higher-order $A/B_{\i\j}^{11n}$, $A/B_{\i\j}^{12n}$ and $A/B_{\i\j}^{21n}$ moments can be up to three orders of magnitude higher than the lower order ones. Thus, the $21N$-moment single-temperature coefficients cannot be recommended for simulation purposes with significant temperature differences, compared to the $13N$-moment multi-temperature ones. 

\begin{table}
\centering
 \begin{tabular}{|c||c|c|c|c|}
 \hline
 \rule{0pt}{2ex}
 \text{Coefficient} & \text{D-T} & \text{C-H} & \text{Ar-H} & \text{W-H} \\
 \hline\hline
\rule{0pt}{3ex}$C^{010}$& 16.44 & 18.62 & 12.46 & 10.57 \\
$A^{100}$& 16.44 & 18.62 & 12.46 & 10.57 \\
$B^{100}$& 16.44 & 18.62 & 12.46 & 10.57 \\
$A^{101}$& 14.08 & 12.93 & 8.22 & 10.51 \\
$B^{101}$& 14.08 & 12.93 & 8.22 & 10.51 \\
$A^{102}$& 28.40 & 39.39 & 34.35 & 36.97 \\
$B^{102}$& 28.40 & 39.39 & 34.35 & 36.97 \\
$A^{110}$& 38.98 & 81.66 & 106.88 & 115.30 \\
$B^{110}$& 176.17 & 452.64 & 191.55 & 99.35 \\
$A^{111}$& 27.77 & 47.18 & 62.00 & 67.48 \\
$B^{111}$& 11.53 & 16.38 & 14.66 & 15.20 \\
$A^{112}$& 25.76 & 42.10 & 55.27 & 60.30 \\
$B^{112}$& 10.34 & 25.43 & 23.92 & 27.69 \\
$A^{120}$& 197.05 & 1004.96 & 1491.43 & 1790.87 \\
$B^{120}$& 336.60 & 282.77 & 98.32 & 99.65 \\
$A^{121}$& 40.53 & 102.39 & 134.03 & 144.63 \\
$B^{121}$& 24.87 & 207734.00 & 752.74 & 98.05 \\
$A^{122}$& 45.89 & 110.41 & 137.11 & 146.93 \\
$B^{122}$& 13.20 & 14.10 & 14.06 & 14.76 \\
$A^{200}$& 16.27 & 27.66 & 25.55 & 29.26 \\
$B^{200}$& 19.07 & 25.95 & 22.54 & 22.50 \\
$A^{201}$& 41.50 & 1180.76 & 4493.82 & 21616.10 \\
$B^{201}$& 10.41 & 14.15 & 16.71 & 19.93 \\
$A^{210}$& 41.79 & 1953.74 & 8207.40 & 40552.00 \\
$B^{210}$& 54.69 & 3883.28 & 192.60 & 98.86 \\
$A^{211}$& 234.01 & 4226.10 & 18208.20 & 189903.00 \\
$B^{211}$& 11.93 & 15.65 & 14.30 & 15.18 \\
\hline
\end{tabular}
\caption{Table of maximum percentage differences in the $21N$-moment single-temperature coefficients in the range $T_\i/T_\j=0.8-1.2$.}
\label{table:21n}
\end{table}

\section{Effect of Coefficients on Viscosity and Friction Force Calculations}
\label{sec:intuitive}

We have up until now clearly noted the difference in coefficients numerically. However, it may also be instructive to study some intuitive physical quantities such as viscosity and heat flux. 
The process of obtaining values of viscosity and heat-flux would essentially close the $13N$-moment system of equations given by (\ref{eq:transport}), by eliminating the higher-order moments, $\mathbf{h}$ and $\pi$ in this case, in terms of lower order ones.  

In order to obtain the closed set of equations,  we neglect any electric and magnetic fields, and follow the procedure in Ref.\,\onlinecite{zhdanov_effect_1962}, restricting the collisional terms to the $13N$-moment multi-temperature approximation. Clearly, this approximation will yield viscosity and heat-flux purely inertial in origin. In addition to neglecting the fields, we also assume that the higher order moments evolve much slower in time and have much smaller gradients than the lower order plasma dynamical moments. In terms of the Knudsen number $Kn\sim\lambda/L\sim\tau/T$ where $\lambda$ refers to the mean free path and $L$ the scale length of the system in question, and equivalently $\tau$ is the collision frequency and $T$ the timescale of the evolution of the system, we see, therefore that the plasma dynamical moments are of the order of the $Kn$, and ignore every term that is of a higher order than $Kn$. This implies that we neglect all time and space gradients of higher order moments in the LHS of the moment-averaged Boltzmann equation (\ref{eq:transport}). Hence, the only terms that survive are the ones proportional to $\langle dG^{mn}_\i/dt\rangle$ (through the $T_\i$ dependence of $\gamma_\i$) for rank-1 quantities, and the ones proportional to  $\partial u_r/\partial x_s$ for the rank-2 quantities (as the $c_{\alpha s}$ derivative reduces the order of the moment).

Thus, the reduced evolution equation, now the steady-state equation, for the stress tensor $\pi_\i$ for the species $\i$ becomes 
\begin{align}
 2p_{\i} \varepsilon = \sum_{\beta\neq\alpha}\left( \frac{A_{\i\j}^{200}}{n_\i}\pi_\i + \frac{B_{\i\j}^{200}}{n_\j}\pi_\j   \right),
  \label{eq:reduced_stress}
\end{align}
where $\varepsilon$ is given by 
\begin{equation}
 \varepsilon_{rs}=\left\{\frac{\partial u_r}{\partial x_s} \right\}=\frac{1}{2}\left(\frac{\partial u_r}{\partial x_s}+\frac{\partial u_s}{\partial x_r} \right)-\frac{1}{3}\delta_{rs}\frac{\partial u_l}{\partial x_l}.\nonumber
\end{equation}
This is the usual form by which one represents viscous forces as the viscosity $\eta$ multiplied to a traceless strain rate tensor $\varepsilon$\cite{kulsrud_plasma_2020}. We can re-write this equation in the following form
\begin{equation}
 \sum_\gamma \frac{B^{*200}_{\alpha\gamma}}{n_\gamma}\pi_\gamma=2p_\i \varepsilon,
\end{equation}
where the sum $\gamma$ is over all species including $\i$, and where
\begin{equation}
 B^{*200}_{\alpha\gamma}=\left\{\begin{tabular}{ll}
             $\sum_{\j\neq\i}A_{\i\j}^{200}$ & $,\ \i=\gamma$\\
             $B_{\i\gamma}^{200}$ & $,\ \i\neq\gamma$
             \end{tabular}\right. .
\end{equation}
For $N$ species in the plasma, the set of $N$ equations corresponding to Eq.\,(\ref{eq:reduced_stress}) for all species can then be compactly written in a matricial form
\begin{equation}
 B^{*200}\Pi=P\varepsilon,
\end{equation}
where the $B^{*200}$ is an $N\times N$ matrix whose elements are given by 
\begin{equation}
 B^{*200}=\left(\begin{tabular}{cccc}
          $\frac{\sum_{\j\neq\i}A_{\i\j}^{200}}{n_\i}$ & $\frac{B_{\i\gamma}^{200}}{n_\gamma}$ & $\ldots$  & $\frac{B_{\i\omega}^{200}}{n_\omega}$\\
          $\frac{B_{\gamma\i}^{200}}{n_\i}$ & $\ddots$ & $ $ & $\vdots$\\
          $\vdots$ & $ $ & $\ddots$ & $\vdots$\\
          $\frac{B_{\omega\i}^{200}}{n_\i}$ & $\ldots$ & $\ldots$ & $\frac{\sum_{\j\neq\omega}A_{\omega\j}^{200}}{n_\omega}$
         \end{tabular}\right), \label{eq:bstar200}
\end{equation}
where $\omega$ is an arbitrarily chosen $n^{th}$ species, and where $\Pi$ and $P$ is the column matrix with all the values of $\pi_\gamma$ and $2p_\gamma$ respectively. On inverting the equation, and comparing it to the classical form of stress tensor $\pi_\gamma=-2\eta_\gamma\varepsilon$, we can find the column matrix of the viscosity $E$ given by
\begin{equation}
 E=-\frac{1}{2}(B^{*200})^{-1}P,
\end{equation}
where the elements of $E$ are the partial viscosity of each species $\eta_\gamma$. One can find the total viscosity $\eta=\sum_\gamma \eta_\gamma$ by the formula $\eta=Tr(E^T U)$.

Similarly, the reduced heat-flux $\mathbf{h}_\i$ equation is given by
\begin{align}
 \frac{5}{2}\frac{1}{\gamma_\i}\frac{p_\i}{T_\i}\nabla T_\i &= \sum_{\beta\neq\alpha}\left( \frac{A_{\i\j}^{111}}{n_\i}\mathbf{h}_\i + \frac{B_{\i\j}^{111}}{n_\j}\mathbf{h}_\j\right)\nonumber\\
 &+\sum_{\beta\neq\alpha}\left( {A_{\i\j}^{110}}{m_\i}\mathbf{w}_\i + {B_{\i\j}^{110}}{m_\j}\mathbf{w}_\j\right),
\end{align}
which can then be written in a matrix form as
\begin{equation}
 B^{*111}H+B^{*110}W=T,
\end{equation}
where $B^{*111}$ has the same form as $B^{*200}$ with the $200$-index coefficients replaced with $111$-index coefficients in Eq.\,(\ref{eq:bstar200}).
$B^{*110}$ also has a similar form as Eq.\,(\ref{eq:bstar200}), with the $200$-index coefficients replaced by $110$-index coefficients, and which multiply by $m$ instead of $(1/n)$. 
The column matrices $H$, $W$ and $T$ are matrices containing $\mathbf{h}_\gamma$, $\mathbf{w}_\gamma$, and $\frac{5}{2}\frac{1}{\gamma_\gamma}\frac{p_\gamma}{T_\gamma}\nabla T_\gamma$ respectively. The heat-flux for all species can then be similarly written as
\begin{equation}
 H=(B^{*111})^{-1}T-(B^{*111})^{-1}B^{*110}W.
\end{equation}
One can notice that approximating the heat-fluxes in this manner, reduces them to a linear combination of temperature gradients and flows. On substituting this value of the heat flux, the RHS of the momentum balance equation will depend solely on only two terms, one proportional to the flows and the other proportional to the temperature gradients, hence recovering the familiar form of the collision term, one with friction force dependent on the flow difference and the other a thermal gradient force, dependent on the difference of temperature gradients\cite{braginskii_transport_1965,stangeby_plasma_2000}. The heat-flux term then adds to the existing friction force term and augments it in the following manner
\begin{multline}
 R^{1}_{\i\j,fric}=\left[ \left(A_{\i\j}^{100}m_\i \mathbf{w}_\i-\frac{A_{\i\j}^{101}}{n_\i}[(B^{*111})^{-1}B^{*110}W]_\i)\right)\right.\\
 +\left.\left(B_{\i\j}^{100}m_\j \mathbf{w}_\j-\frac{B_{\i\j}^{101}}{n_\j}[(B^{*111})^{-1}B^{*110}W]_\j\right)\right]
\end{multline}
and the thermal gradient force remains
\begin{multline}
 R^{1}_{\i\j,therm}=\left[ \frac{A_{\i\j}^{101}}{n_\i}[(B^{*111})^{-1}T]_\i +\frac{B_{\i\j}^{101}}{n_\j}[(B^{*111})^{-1}T]_\j\right],
\end{multline}
where $[\dots]_\gamma$ indicates the element of the column matrix corresponding to the species $\gamma$. 
To estimate the augmentation of the friction force, we need to compare the additional term to the original coefficients $A_{\i\j}^{100}m_\i$ and $B_{\i\j}^{100}m_\j$.
One can notice, however, that the friction force term $R^{1}_{\i\j,fric}$ now becomes quite complex, with friction among any two pairs of species depending on the flow velocity of all species. This makes any straightforward comparisons of the coefficients of the friction force cumbersome and prone to overinformation.  Therefore, we only choose the coefficient of $\mathbf{w}_\i$ and $\mathbf{w}_\j$ from the heat-flux contribution $\mathbf{h}_\i$ and $\mathbf{h}_\j$ and compare them to the original coefficients corresponding to the original friction force.

Furthermore, one can repeat the same procedure with a higher-order number of moments as has been done in Ref.\,\onlinecite{yushmanov_diffusion_1980}, often referred to as the ``Zhdanov closure'' when applied to close a $21N$-moment set of equations with the moments of the collision operator given by Eq.\,(\ref{eq:collision_first_form}). It must be mentioned that the higher approximations to the viscosity, which would depend on the electric field and the magnetic field, can be decomposed into a form which is a linear combination of different viscosity contributions\cite{braginskii_transport_1965,alievskii_1963_transport,alievskii_viscous_1964}. Comparison of such approximations with more nuanced field effects and higher number of moments is out of the scope of the current article, since a larger number of moments would imply needing the use of block matrices instead of regular matrices used here, and will be a part of our planned future work.

\begin{figure}
 \centering
 \includegraphics[width=\columnwidth]{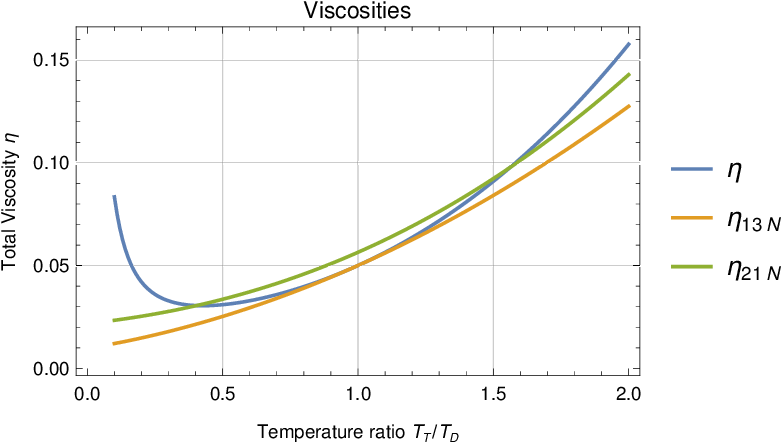}
 \caption{Plot of total viscosities for the Deuterium-Tritium plasma calculated using our coefficients, and Zhdanov's $13N$-moment multi-temperature and $21N$-moment single-temperature coefficients, plotted against the temperature ratio $T_D/T_T$. { Notice the virtual overlap between the viscosities calculated from Zhdanov's $13N$-moment coefficients and ours in the vicinity of equal temperature. In the legends for the plots, first $\eta$ refers to our coefficients, the subscript ``$13N$'' indicates the viscosity calculated Zhdanov's multi-temperature collisional coefficients and similarly the subscript ``$21N$'' for Zhdanov's single-temperature collisional coefficients.}}
 \label{fig:DT_viscosity}
\end{figure} 

\begin{figure}
 \centering
 \includegraphics[width=\columnwidth]{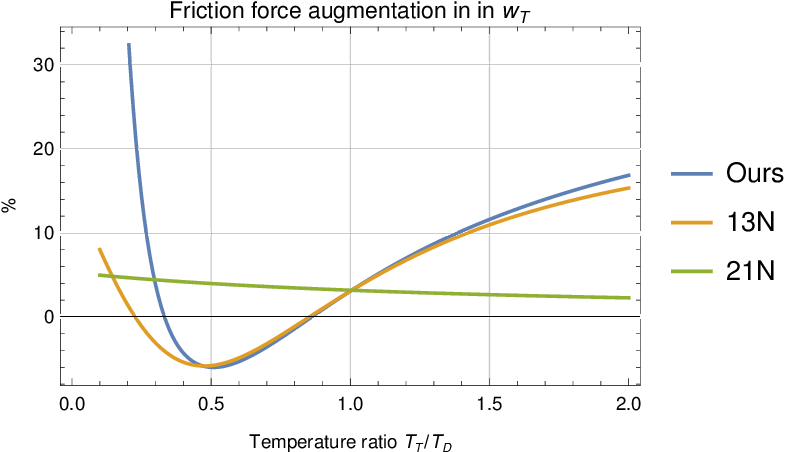}
  \includegraphics[width=\columnwidth]{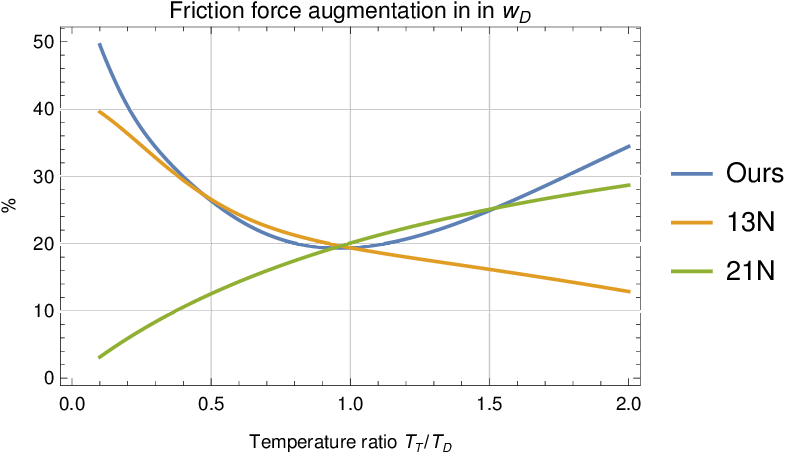}
  \caption{The augmentation of the friction force between Deuterium and Tritium by the contribution of the heat fluxes (in \%), for ours and Zhdanov's coefficients, plotted against the temperature ratio $T_D/T_T$. { In the legends for the plots, the ``$13N$'' indicates the value of the friction force augmentation for Zhdanov's multi-temperature collisional coefficients and similarly the ``$21N$'' refers to the friction force augmentation Zhdanov's single-temperature collisional coefficients.}}
  \label{fig:DT_friction_augmentation}
\end{figure}
For the particular case of a Deuterium-Tritium plasma, one can notice the total viscosity $\eta$ in Fig.\,(\ref{fig:DT_viscosity}) in cases of all three coefficients follow each other quite closely in the vicinity of equal temperatures, with the $13N$-moment multi-temperature ones practically overlapping with our exact values.
One can see in Tables (\ref{table:vis13n}) and (\ref{table:vis21n}) that the differences in viscosity for the $13N$-moment multi-temperature case are significantly lower than those of the $21N$-moment single-temperature case, and also that differences in viscosity generally seem to decrease with increasing mass ratio.
{ Furthermore, we can compare the viscosity values obtained with the prescription for parallel viscosity given by Braginskii (Ref.\,\onlinecite{braginskii_transport_1965}, page 229), where Braginskii provides the magnitude of viscosity by the expression $\eta\sim nkT\tau_{\i\j}$, where $\tau_{\i\j}$ is the mean time between collisions between species $\alpha$ and $\beta$, and where $n=\sum_i n_i$. Values of $\tau_{\i\j}$ are provided by Braginskii for ions and electrons, but not for impurities. However, since Braginskii follows the Chapman-Enskog solution, we can use the original formula provided by Chapman and Cowling (Ref.\,\onlinecite{chapman_mathematical_1952}, Sec.\,9.81) and Zhdanov (Ref.\,\onlinecite{zhdanov_transport_2002}, Sec.\,3.1), for estimating the mean time between collisions $\tau_{\i\j}$
\begin{equation}
 \tau_{\i\j}=\frac{n\mu_{\i\j}[\mathcal{D}_{\i\j}]_1}{n_\beta kT},
\end{equation}
where the first approximation to the diffusion coefficient $[\mathcal{D}_{\i\j}]_1$ is given by
\begin{equation}
 [\mathcal{D}_{\i\j}]_1=\frac{3kT}{16n\mu_{\i\j}\Omega_{\i\j}^{11}}.
\end{equation}
On substituting the physical values from Table I, and the value of $\Omega_{\i\j}^{11}$ from Eq.\,(\ref{eq:kihara_formula}), one can find that the viscosity prescribed by Braginskii, $\eta\sim nkT\tau_{\i\j}$, falls within the same range of values as found in Fig.\,(\ref{fig:DT_viscosity}). }

A similar trend can be partly observed in the augmentation of the friction force coefficients. In Fig.\,(\ref{fig:DT_friction_augmentation}), the increase in friction force is indicated as a percentage of the original value. In general, with increasing mass ratio and decreasing densities, the augmentation in the term proportional to $\mathbf{w}_\i$ decreases and increases in the term proportional to $\mathbf{w}_\j$. This is expected, as the dominant part of the  heat-flux contribution to the friction force is bound to arise from the species with the larger density. In case of the $21N$-moment single-temperature cases for the friction force augmentation for the term proportional to the background flow contribution, the differences seem to rise again for high mass ratios, as evidenced by the W-H case. The percentage differences on these computed physical quantities seem to have a minimum in between the Argon and Tungsten cases, which indicates that even the relatively smaller differences in the $21N$-moment single-temperature case may contribute to significant difference in physical quantities of interest. On the basis of this, we recommend caution in using the $21N$-moment single-temperature coefficients even for heavy impurities when temperature differences may be significant.

\begin{table}
 \centering
 \begin{tabular}{|l||c|c|c|c|}
 \hline
   &D-T&C-H&Ar-H&W-H \\
   \hline\hline
    \rule{0pt}{3ex}Viscosity $\eta$         &1.84&1.12 &0.86 &2.49 \\
  Friction, $\mathbf{w}_\i$&163.03&69.21 &3.88 &4.34 \\
  Friction, $\mathbf{w}_\j$&19.06&23.27 &22.27 & 15.24 \\
  \hline
 \end{tabular}
\caption{Table of percentage differences in viscosity, friction force augmentations in $\mathbf{w}_\i$ and $\mathbf{w}_\beta$ in the $13N$-moment multi-temperature coefficients in the $0.8-1.2$ temperature ratio range.}
\label{table:vis13n}
\end{table}

\begin{table}
 \centering
 \begin{tabular}{|l||c|c|c|c|}
 \hline
   &D-T&C-H&Ar-H&W-H \\
   \hline\hline
  \rule{0pt}{3ex}Viscosity    $\eta$         &16.94&15.14 &6.14 &36.99\\
  Friction, $\mathbf{w}_\i$&1509.65&336.12 &245.95 &119.45\\
  Friction, $\mathbf{w}_\j$&11.20&10.13 & 3.80&231.76\\
  \hline
 \end{tabular}
\caption{Table of percentage differences in viscosity, friction force augmentations in $\mathbf{w}_\i$ and $\mathbf{w}_\beta$ in the $21N$-moment single temperature coefficients in the $0.8-1.2$ temperature ratio range.}
\label{table:vis21n}
\end{table}

Thus, from these numerical results we can conclude 1.\,that the coefficients calculated by our coefficients, Zhdanov's $13N$-moment multi-temperature and $21N$-moment single-temperature expressions tend to follow each other quite closely in the vicinity of equal temperatures, 2.\,that the differences for the $21N$-moment single-temperature coefficients are higher than those of the $13N$-moment multi-temperature coefficients, 3.\,that the differences in coefficients become quite significant outside the vicinity of equal temperatures, however, that in general again, the differences in $13N$-moment multi-temperature coefficients are less than those of the $21N$-moment single-temperature case, 4.\,that the differences in higher-order moments become quite significant in case of the $21N$-moment single-temperature coefficients, 5.\,that the differences in coefficients in general decrease with increasing mass ratio, and 6.\,that despite the agreement and trends being similar in physical quantities such as viscosity and augmentation in friction force, computed from the coefficients in the vicinity of equal temperatures, the values computed from the $21N$-moment single-temperature coefficients may present significant differences as the mass ratio increases, especially in the augmentation of the friction force term.

%%%%%%%%%%%%%%%%%%%%%%%%%%%%%%%%%%

\section{Summary, Conclusions and Future Work}

In this article, we have generalized the calculation of moment averaged collisional coefficients for a multi-component, multi-temperature plasma without making any assumptions on the masses or temperatures of the colliding species with the linearized Boltzmann collision operator for up to rank-2 tensorial moments. We started by taking an ansatz for the distribution function in terms of the Sonine polynomials and the irreducible tensorial Hermite monomials Eq.\,(\ref{eq:ansatz}), and then expressing the moment in terms of these polynomials of different orders Eq.\,(\ref{eq:sonine-hermite}). On taking moments defined in this manner, and then averaging over the Boltzmann collision operator, we obtain a generalized moment-averaged collision term, which expresses itself in terms of partial bracket integrals in Eq.\,(\ref{eq:collision_second_form}). We evaluate these partial bracket integrals analytically and derive general expressions up to rank-2. The collision operator found in this manner is valid for any range of masses and temperatures, but is restricted to the flow differences being much smaller than the order of the thermal velocities of the plasma. Furthermore, the collision term automatically preserves energy and momentum due to symmetry properties of the Boltzmann collision operator. These expressions for the collisional coefficients were found to be very amenable to being implemented in computer algebra systems, and in our case were implemented in Mathematica\cite{mathematica}, and were found to conserve mass, energy and momentum. 

{ The collision coefficients are found to essentially be linear combinations of product of a term $A^{pqrl}_{\i\j/\i\i}$ depending purely on masses and temperatures, and another purely depending on the collisional cross-section $\Omega^{lr}_{\i\j}$, which in turn purely depends on the potential of interaction between the two colliding species and a factor $d_{\i\j}$. For the cross-sections, we chose a formula for the cross-section derived from asymptotic values of the form of the cross-section integral for the shielded Coulomb potential from Eq.\,(\ref{eq:kihara_formula}). We calculated our set of coefficients choosing $d_{\i\j}=\mu_{\i\j}/(2kT)$, as choosing this factor agrees with the calculations of effective cross-section integral for the shielded Coulomb potential from previous literature.}

Previously, Zhdanov et al had derived two sets of expressions for the collision coefficients and cross-sections, the first being derived for a multi-temperature plasma with terms provided up to $13N$-moments, and the second being derived at the plasma common temperature, with terms provided up to $21N$-moments. The procedure of derivation for the former is not explicitly provided for all moments, and no expressions for the coefficients were provided for higher-order collisional moments. { However, with our calculation procedure for the bracket integrals, with the suitable factor $d_{\i\j}=\gamma_{\i\j}/2$, the provided moments were found to be accurately derived. } The procedure for the latter $21N$-moment single-temperature set of coefficients was explicitly provided, but comes at the cost of temperature differences being much smaller than the order of the plasma common temperature. 

We then compared our expressions to the ones provided by Zhdanov for multiple cases of colliding species relevant to fusion. We find that in the vicinity of the temperatures being equal, all sets of coefficients agree very well, but they diverge away from equal temperatures. We further find that the differences in the $13N$-moment multi-temperature coefficients are much smaller than those of the $21N$-moment single-temperature ones (See Appendices \ref{sec:zhdanov13} and \ref{sec:zhdanov21}). We also find that the differences in the coefficients decrease as the mass-ratio of the species increases. Furthermore, we use certain approximations in Knudsen number to obtain reduced equations for the stress-tensor and heat-flux, and use it to calculate the first inertial approximation to total viscosity and the augmentation of the friction force as contributed by the heat-flux. We find that while the differences mostly follow the same patterns as the coefficients, for high mass ratio and low density, as in the case of tungsten impurity in tokamak plasmas, the differences in these physical quantities become significant for the $21N$-moment single-temperature case. On the basis of this, we caution against using the $21N$-moment single-temperature coefficients where the temperature difference between species is significant. In the same vein, we find the $13N$-moment multi-temperature coefficients are agree better than the $21N$-moment single-temperature ones for small temperature differences, especially for modelling heavy-impurity transport, as for tungsten. { For any significant temperature differences between species, we recommend using multi-temperature coefficients.}

In the future, we plan to generalize the expressions for the linearized collision operator up to arbitrary rank-$m$ tensorial moments. We also plan to study in detail linear parallel closure schemes associated with fluid equations derived from the moments under the low Knudsen number approximation, particularly the Zhdanov closure. Furthermore, the study of the closure can help prescribe the appropriate choice of the factor $d_{\i\j}$ through studying its effect on the transport coefficients calculated by the Zhdanov closure scheme. The ones with the aforementioned $21N$-moment single-temperature scheme, derived by Zhdanov et al, often termed the Zhdanov closure, is one such scheme of closure particularly of interest to us, since it has already been implemented in many fluid codes of interest to the fusion community. 
 
\section*{Acknowledgments}

The authors would like to thank Prof.\,V.\,Rat (Université de Limoges, France) for his generous help in obtaining the relevant texts and with explanation of technical details of his work in Ref.\,\onlinecite{rat_transport_2001}. Furthermore, the authors would also like to thank D.\,Brunetti (CCFE, UKAEA, UK), and, O.\,Sauter and S.\,Brunner (SPC, EPFL, Switzerland) for their valuable comments on the paper. We also thank the anonymous referees for their comments which helped improve the article significantly.

The projects leading to this publication have received funding from Excellence Initiative of Aix-Marseille Université - A*MIDEX, a French Investissement d’Avenir Programme, project TOP \& AMX-19-IET-013. This work has also been partly carried out within the framework of the EUROfusion Consortium and has received funding from the Euratom research and training programme 2014-2018 and 2019-2020 under grant agreement No.\,633053. The views and opinions expressed herein do not necessarily reflect those of the European Commission.

\bibliography{bibliography}{}
\bibliographystyle{vancouver}

\appendix

\section{Derivation of the bracket integrals}
\label{sec:bracket_integral_derivation}

We follow the method illustrated in Refs.\,\onlinecite{chapman_mathematical_1952,rat_transport_2001} in order to derive these bracket integrals in their most general form. Certain transforms in Ref.\,\onlinecite{rat_transport_2001} are taken from R.S.\,Devoto's thesis\cite{devoto_thesis}. We proceed in a detailed manner to show all the calculations necessary to calculate the brackets.

There are several transforms one must make to render terms in the bracket integral tractable. The first transform is to move to the center-of-mass frame, where the center-of-mass velocity relative to the common flow $\mathbf{U}$ and relative velocity $\mathbf{g}$ are defined as
\begin{align}
 &\mathbf{U}=\mathbf{G}-\mathbf{u} =\frac{m_\alpha \mathbf{c}^{(\prime)}_\alpha + m_\beta \mathbf{c}^{(\prime)}_\beta}{m_\alpha + m_\beta},\nonumber\\
 &\mathbf{g}^{(\prime)}=\mathbf{c}^{(\prime)}_\beta - \mathbf{c}^{(\prime)}_\alpha,
\end{align}
where the superscript ${(\prime)}$ represents the same corresponding expression for the primed variables, and consequently
\begin{align}
  \mathbf{c}^{(\prime)}_\alpha = \mathbf{U} - \frac{\mu_{\alpha\beta}}{m_\alpha}\mathbf{g}^{(\prime)},\  \mathbf{c}^{(\prime)}_\beta = \mathbf{U} + \frac{\mu_{\alpha\beta}}{m_\beta}\mathbf{g}^{(\prime)},
\label{eq:xgntransform}
 \end{align}
where $\mu_{\alpha\beta}$ is the reduced mass given by $\mu_{\alpha\beta}=m_\alpha m_\beta/(m_\alpha+m_\beta)$. Now we define variable changes $\mathbf{X}$ and $\mathbf{\gn}$
\begin{align}
 \mathbf{X}&=\left( \frac{\g_\ag}{2}+\frac{\g_\bg}{2} \right)^{1/2} \mathbf{U},\nonumber\\
  \mathbf{\gn}^{(\prime)}&=\mu_{\ag\bg}\left( \frac{\g_\ag}{2m_\ag^2}+\frac{\g_\bg}{2m_\bg^2} \right)^{1/2} \mathbf{g}^{(\prime)}
 \label{eq:wxgntransform}
\end{align}
such that
\begin{align}
 \mathbf{W}_\i^{(\prime)}&= M^{1/2}_{\i 1} \mathbf{X} - M^{1/2}_{\i 2} \mathbf{\gn}^{(\prime)},\nonumber\\ \mathbf{W}_\j^{(\prime)}&= M^{1/2}_{\j 1} \mathbf{X} + M^{1/2}_{\j 2} \mathbf{\gn}^{(\prime)},
\end{align}
where 
\begin{align}
 M_{\i 1}&=\frac{\g_\i}{\g_\i+\g_\j},\ M_{\i 2} = \frac{m_\j^2\g_\i}{m_\j^2\g_\i+m_\i^2\g_\j},\nonumber\\
 M_{\j 1}&=\frac{\g_\j}{\g_\i+\g_\j},\ M_{\j 2} = \frac{m_\i^2\g_\j}{m_\j^2\g_\i+m_\i^2\g_\j}.
\end{align}
The Jacobian of the transform is given by
\begin{align}
 \mathcal{J}&=\frac{\partial(\mathbf{c_\i},\mathbf{c_\j})}{\partial(\mathbf{X},\mathbf{\gn})}=\frac{\partial(\mathbf{U},\mathbf{g})}{\partial(\mathbf{X},\mathbf{\gn})}.\frac{\partial(\mathbf{c_\i},\mathbf{c_\j})}{\partial(\mathbf{U},\mathbf{g})}\nonumber\\
 &=\left[\mu_{\i\j}^2 \left( \frac{\g_\i+\g_\j}{2}\right)\left( \frac{\g_\i}{2m_\i^2}+\frac{\g_\j}{2m_\j^2}\right)\right]^{-3/2}.
\end{align}
These transforms will help transform the velocities into quantities that are easier to integrate. 

In order to deal with the Sonine polynomials, it is worth noting that the Sonine polynomial $S^n_m(x)$ is defined to be coefficient of $s^n$ in the expansion of
\begin{equation}
 (1-s)^{-m-1}\exp\left(-\frac{sx}{1-s}\right)=\sum_n s^n S^n_m(x). \label{eq:sonineproperty}
\end{equation}
With this relation in mind, we now proceed to derive some bracket integrals. We describe the derivation of $\left[S^p_{3/2}(W_\beta^2)\mathbf{W}_\beta,S^q_{3/2}(W_\alpha^2)\mathbf{W}_\alpha\right]$ in full detail, after which we provide condensed derivations for the other brackets with only the differing steps.

\subsection{Derivation of $\left[S^p_{3/2}(W_\beta^2)\mathbf{W}_\beta,S^q_{3/2}(W_\alpha^2)\mathbf{W}_\alpha\right]$}

We have the bracket integral defined as follows
\begin{multline}
 \left[S^p_{3/2}(W_\beta^2)\mathbf{W}_\beta,S^q_{3/2}(W_\alpha^2)\mathbf{W}_\alpha\right] =\frac{1}{n_\i n_\j} \int  f^{(0)}_\alpha f^{(0)}_\beta\\
 \times (S^p_{3/2}(W_\beta^{\prime2})\mathbf{W}_\beta^\prime-S^p_{3/2}(W_\beta^2)\mathbf{W}_\beta)\cdot S^q_{3/2}(W_\alpha^2)\mathbf{W}_\alpha \\
 \times g\sigma_{\alpha\beta}(g,\chi)d\Omega d\mathbf{c_{\alpha}}d\mathbf{c_{1\beta}}.
\end{multline}
In order to recover collision coefficients out of this form of a bracket integrals, the general strategy will be to use the property (\ref{eq:sonineproperty}) and absorb the Sonine polynomial terms into the exponential, creating a generating function $\Pi_{\i\j}$ which contains the values of the bracket integrals for all $(p,q)$. Then, the term left outside of the exponential would just be the term corresponding to the full contraction of the irreducible Hermite polynomials. Certain transformations will be performed to integrate this full contraction term over the exponential.  However, since the Sonine polynomials were absorbed into the exponential, the remaining exponential after the integral will have to be expanded into a series to isolate the $(p,q)$ coefficients. Once the integral is performed, the terms left can be expressed as linear combination of moments of the relative velocity $g$ over the cross section $\sigma_{\i\j}$ and a gaussian, $\Omega^{lr}_{\i\j}$ . 

The generating function $\Pi_{\i\j}$ in found to be in the following form with the help of Eq.\,(\ref{eq:sonineproperty}),
\begin{multline}
 \Pi_{\i\j}=\frac{1}{n_\i n_\j}(1-s)^{-5/2}(1-t)^{-5/2}\int f_\i^{(0)}f_\j^{(0)}\\
 \times (\mathbf{W}_\j^\prime e^{-SW_\j^{\prime 2}}-\mathbf{W}_\j e^{-SW_\j^{2}})\cdot\mathbf{W}_\i e^{-TW_\i^2}\\
 \times g\sigma_{\alpha\beta}(g,\chi)d\Omega d\mathbf{c_{\alpha}}d\mathbf{c_{1\beta}},
\end{multline}
where $S=s/(1-s)$ and $T=t/(1-t)$, and where the bracket integral of order $(p,q)$ is the coefficient of $s^p t^q$ in the expansion of $\Pi_{\i\j}$. Now we substitute in the expressions for the zero functions $f_\i^{(0)}$ and $f_\j^{(0)}$, 
\begin{align}
 f_\i^{(0)} &=n_\i \left(\frac{\g_\i}{2\pi}\right)^{3/2}\exp\left(-\mathbf{W}_\i^2\right),\\
 f_\j^{(0)} &=n_\j \left(\frac{\g_\j}{2\pi}\right)^{3/2}\exp\left(-\mathbf{W}_\j^2\right),
\end{align}
and make the coordinate transform $(\mathbf{c_\i},\mathbf{c_\j})\rightarrow(\mathbf{X},\mathbf{\gn})$, and we get
\begin{multline}
  \Pi_{\i\j}=(1-s)^{-5/2}(1-t)^{-5/2} \pi^{-3}{\mathcal{K}_{\i\j}}\\
  \times \int \exp\left(-W_\i^2-W_\j^2 \right)\\
  \times(\mathbf{W}_\j^\prime e^{-SW_\j^{\prime 2}}-\mathbf{W}_\j e^{-SW_\j^{2}})\cdot\mathbf{W}_\i e^{-TW_\i^2}\\
  \times g\sigma_{\alpha\beta}(g,\chi)d\Omega d\mathbf{X}d\mathbf{\gn}.
\end{multline}
where
\begin{equation}
 \mathcal{K}_{\i\j}=\left[\mu_{\i\j}^2 \left( \frac{\g_\i+\g_\j}{{\g_\i\g_\j}}\right)\left( \frac{\g_\i}{m_\i^2}+\frac{\g_\j}{m_\j^2}\right)\right]^{-3/2}.
\end{equation}
Following Ref.\,\onlinecite{chapman_mathematical_1952}, define a $H_{\i\j}(\chi)$ such that
\begin{multline}
 H_{\i\j}(\mathbf{\gn},\chi)= \int \exp\left\{  -W_\i^2-W_\j^2-SW_\j^{\prime 2}-TW_\i^2 \right\}\\
 \times\mathbf{W}_\j^\prime.\mathbf{W}_\i d\mathbf{X},
\end{multline}
which in the limit of a ``no-collision event'' becomes $(\mathbf{\gn},\chi=0)$ such that 
\begin{multline}
 H_{\i\j}(\mathbf{\gn},0)= \int \exp\left\{  -W_\i^2-W_\j^2-SW_\j^{2}-TW_\i^2 \right\}\\
 \times\mathbf{W}_\j.\mathbf{W}_\i d\mathbf{X}.
\end{multline}
This helps us split the generating function first as an integral over $\mathbf{X}$ leaving $H_{\i\j}$ purely a function of $(\mathbf{\gn},\chi)$.  Thus, $\Pi_{\i\j}$ as
\begin{multline}
  \Pi_{\i\j}=(1-s)^{-5/2}(1-t)^{-5/2} \pi^{-3} {\mathcal{K}_{\i\j}}\\
  \times \int \{H_{\i\j}(\mathbf{\gn},\chi)-H_{\i\j}(\mathbf{\gn},0)\} g\sigma_{\alpha\beta}(g,\chi)d\Omega d\mathbf{\gn}.
\end{multline}
Now defining 
\begin{align}
 a_{\i\j}&=1+SM_{\j 1}+TM_{\i 1}\nonumber\\
 &=(1-s)^{-1}(1-t)^{-1}(1-sM_{\i 1}-tM_{\j 1}),\nonumber\\
 a_{\j\i}&=1+SM_{\j 2}+TM_{\i 2}\nonumber\\
 &=(1-s)^{-1}(1-t)^{-1}(1-sM_{\i 2}-tM_{\j 2}),
\end{align}
and 
\begin{align}
 T^*&=\left(\frac{M_{\j 1}M_{\j 2}}{M_{\i 1}M_{\i 2}} \right)^{1/2}-1-T=\theta_{\i\j}-1-T,\nonumber\\
 S^*&=\left(\frac{M_{\j 1}M_{\j 2}}{M_{\i 1}M_{\i 2}} \right)^{1/2}S=\theta_{\i\j}S,
\end{align}
where $\theta_{\i\j}=T_\i/T_\j$. Then $H_{\i\j}(\mathbf{\gn},\chi)$ can be transformed to
\begin{multline}
 H_{\i\j}(\mathbf{\gn},\chi)=\int \exp\left\{ -a_{\i\j}X^2-a_{\j\i}\gn^2\right.\\
 \left. -2(M_{\i 1}M_{\i 2})^{1/2}\mathbf{X}.(S^*\mathbf{\gn}^\prime+T^*\mathbf{\gn})\right\}\mathbf{W}_\j^\prime.\mathbf{W}_\i d\mathbf{X}.
\end{multline}
We now make another variable transformation as follows, so as to write the exponential term purely in terms of the squares of quantities $\xn$ and $\gn$,
\begin{equation}
 \mathbf{\xn}=\mathbf{X}+\frac{(M_{\i 1}M_{\i 2})^{1/2}}{a_{\i\j}}(S^*\mathbf{\gn}^\prime+T^*\mathbf{\gn}),   \label{eq:xxntransform}
\end{equation}
such that the Jacobian of this transform, $(\mathbf{X},\mathbf{\gn})\rightarrow(\mathbf{\xn},\mathbf{\gn})$, is 1, and further defining
\begin{equation}
 b_{\i\j}=a_{\j\i}-\frac{M_{\i 1}M_{\i 2}}{a_{\i\j}}(S^{*2}+T^{*2}+2S^*T^*\cos{\chi}), \label{eq:bij}
\end{equation}
we can write $H_{\i\j}(\mathbf{\gn},\chi)$ as
\begin{equation}
 H_{\i\j}(\mathbf{\gn},\chi)=\int \exp\left\{ -a_{\i\j}\xn^2-b_{\i\j}\gn^2\right\}\mathbf{W}_\j^\prime.\mathbf{W}_\i d\mathbf{\xn}. \label{eq:simplifiedhij}
\end{equation}
At this point, it now remains to derive and expression for $\mathbf{W}_\j^\prime.\mathbf{W}_\i$. In order to do so, we introduce $\mathbf{V}_\i$ and $\mathbf{V}_\j$ and their primed counterparts
\begin{align}
 &\mathbf{V}_\i^{(\prime)} = \frac{M_{\i 1}}{a_{\i\j}}(S^*\mathbf{\gn}^\prime+T^*\mathbf{\gn})+\mathbf{\gn}^{(\prime)},\nonumber\\
 &\mathbf{V}_\j^{(\prime)}=\left(\frac{M_{\j 1}}{M_{\j 2}}\frac{M_{\i 1}M_{\i 2}}{a_{\i\j}^2} \right)^{1/2}(S^*\mathbf{\gn}^\prime+T^*\mathbf{\gn})-\mathbf{\gn}^{(\prime)},
\end{align}
such that $\mathbf{W}_\i$ and $\mathbf{W}_\j^{\prime}$ can be expressed in terms of $\mathbf{\xn}$ and $\mathbf{\gn}^{(\prime)}$
\begin{equation}
 \mathbf{W}_\i^{(\prime)}=M_{\i 1}^{1/2}\mathbf{\xn}-M_{\i 2}^{1/2}\mathbf{V}_\i^{(\prime)},\ \mathbf{W}_\j^{(\prime)}=M_{\j 1}^{1/2}\mathbf{\xn}-M_{\j 2}^{1/2}\mathbf{V}_\j^{(\prime)}.
\end{equation}
These relations are be obtained by substituting Eqs.\,(\ref{eq:xxntransform}) in Eqs.\,(\ref{eq:wxgntransform}). Therefore, the dot product $\mathbf{W}_\j^\prime.\mathbf{W}_\i$ becomes
\begin{multline}
 \mathbf{W}_\j^\prime\cdot\mathbf{W}_\i = (M_{\i 1}M_{\j 1})^{1/2}\xn^2+(M_{\i 2}M_{\j 2})^{1/2}\mathbf{V}_\j^\prime\cdot\mathbf{V}_\i\\
 -\mathbf{\xn}\cdot\left\{ (M_{\i 1}M_{\j 2})^{1/2}\mathbf{V}_\j^\prime+(M_{\i 2}M_{\j 1})^{1/2}\mathbf{V}_\i \right\}.
 \label{eq:wbdotwainitial}
\end{multline}
Now, odd powers of $\mathbf{\xn}$ will vanish in the integral for $H_{\i\j}(\mathbf{\gn},\chi)$, so we only need to concern ourselves with expressing $\mathbf{V}_\j^\prime.\mathbf{V}_\i$ back in terms of $\mathbf{\gn}$. This calculates to
\begin{multline}
 \mathbf{V}_\j^\prime.\mathbf{V}_\i = \left[ \left(\frac{M_{\i 1}M_{\j 1}}{M_{\i 2}M_{\j 2}} \right)^{1/2}(1-b_{\i\j}+M_{\i 2}(\theta_{\i\j}-1))\right.\\
 +(M_{\i 1}(1-\theta_{\i\j})-1 )\cos{\chi} \big{]}\frac{\gn^2}{a_{\i\j}},
\end{multline}
which on substituting back into $\mathbf{W}_\j^\prime.\mathbf{W}_\i$, and defining
\begin{align} 
 A_{\i\j}&=\left(\frac{M_{\i 2}M_{\j 2}}{M_{\i 1}M_{\j 1}} \right)^{1/2}(M_{\i 1}(1-\theta_{\i\j})-1 ), \nonumber\\
 B_{\i\j}&=M_{\i 2}(\theta_{\i\j}-1),
\end{align}
so as to simplify $\mathbf{W}_\j^\prime.\mathbf{W}_\i$ as
\begin{multline}
 \mathbf{W}_\j^\prime.\mathbf{W}_\i=(M_{\i 1}M_{\j 1})^{1/2}\left\{\xn^2\right.\\
 \left.+\frac{\gn^2}{a_{\i\j}}(1-b_{\i\j}+B_{\i\j}+A_{\i\j}\cos{\chi}) \right\}\\
 + \mathrm{odd\ terms\ of\ }\mathbf{\xn}.
\end{multline}
Therefore, now $H_{\i\j}(\mathbf{\gn},\chi)$, on substituting $\mathbf{W}_\j^\prime.\mathbf{W}_\i$ and integrating over $\mathbf{\xn}$,  becomes
\begin{multline}
 H_{\i\j}(\mathbf{\gn},\chi)=\pi^{3/2}\exp(-b_{\i\j}\gn^2)a_{\i\j}^{-5/2}(M_{\i 1}M_{\j 1})^{1/2}\\
 \times \left\{ \frac{3}{2}+\gn^2 (1-b_{\i\j}+B_{\i\j}+A_{\i\j}\cos{\chi}) \right\}.
\end{multline}

Now that $H_{\i\j}(\mathbf{\gn},\chi)$ has been evaluated, it remains to express $\Pi_{\i\j}$ in terms of $s$ and $t$, so that the bracket integral can manifest itself as the coefficient of $s^pt^q$. For this purpose, a number of series expansions need to be performed. {  We begin by performing a Taylor expansion of $\exp(-b_{\i\j}\gn^2)$ in the following form, by separating out an arbitrary factor of $\exp(-k_{\i\j}\gn^2)$ pre-emptively so that quantities in the integral can be expressed later as moments of a gaussian,  
\begin{align}
 \exp(-b_{\i\j}\gn^2)&=\exp(-k_{\i\j}\gn^2)\exp((k_{\i\j}-b_{\i\j})\gn^2)\nonumber\\
 &=\exp(-k_{\i\j}\gn^2)\sum_{r=0}^{\infty} \frac{(k_{\i\j}-b_{\i\j})^r}{r!}\gn^{2r},
\end{align}
and therefore 
\begin{multline}
 H_{\i\j}(\mathbf{\gn},\chi)=\pi^{3/2}\exp(-\gn^2)a_{\i\j}^{-5/2}(M_{\i 1}M_{\j 1})^{1/2}\\
 \times\sum_{r=0}^{\infty}\left\{r+\frac{3}{2}+\gn^2 (1-k_{\i\j}+B_{\i\j}+A_{\i\j}\cos{\chi}) \right\}\frac{(k_{\i\j}-b_{\i\j})^r}{r!}\gn^{2r}.
\end{multline}
}
Now, on substituting the expression for $b_{\i\j}$, expanding $a_{\i\j}$, $a_{\j\i}$, $S^*$ and $T^*$ from their definitions, and setting $S=s/(1-s)$ and $T=t/(1-t)$, $H_{\i\j}(\mathbf{\gn},\chi)$ can be written as follows { 
\begin{multline}
  H_{\i\j}(\mathbf{\gn},\chi)=(1-s)^{5/2}(1-t)^{5/2}\pi^{3/2}\exp(-k_{\i\j}\gn^2)\\
  \times(M_{\i 1}M_{\j 1})^{1/2}\sum_{r=0}^{\infty}\frac{\gn^{2r}}{r!}\left\{r+\frac{3}{2}+\gn^2 (1-k_{\i\j}+B_{\i\j}+A_{\i\j}\cos{\chi}) \right\} \\
  \times \frac{( N_s s+N_t t+N_{st}st+N_1)^r}{(1+D_s s+D_t t)^{r+5/2}}, \label{eq:hij171}
\end{multline}
}where the $N$'s and $D$'s are known coefficients of the form { 
\begin{align}
 N_s =& (1-k_{\i\j})M_{\i1}-M_{\i1}M_{\i2}(\theta_{\i\j}-1)^2\nonumber\\
 &+2M_{\i1}M_{\i2}\theta_{\i\j}(\theta_{\i\j}-1)\cos{\chi}-M_{\j2}\nonumber\\
 N_t =& (1-k_{\i\j})M_{\j1}-M_{\i1}M_{\i2}(\theta_{\i\j}-1)^2\nonumber\\
 &-2M_{\i1}M_{\i2}(\theta_{\i\j}-1)-M_{\i2}= -k_{\i\j}M_{\j1}\nonumber\\
 N_{st} =& M_{\i1}M_{\i2}(\theta_{\i\j}^2-1)+1\nonumber\\
 &-M_{\i1}M_{\j2}-M_{\i2}M_{\j1}-2M_{\i1}M_{\i2}\theta_{\i\j}^2\cos{\chi}\nonumber\\
 N_1 =&(k_{\i\j}-1) M_{\i1}M_{\i2}(\theta_{\i\j}-1)^2\nonumber\\
 D_s =& -M_{\i1}, \ D_t = -M_{\j1}. 
\end{align}
}
Now, we need to express the numerator and denominator as series expansions in order explicitly obtain and group together the powers of $s$ and $t$. We proceed to expand the denominator term as follows
\begin{equation}
 (1+D_s s+D_t t)^{-(r+5/2)}=\sum_{l=0}^{\infty} \frac{(r+\frac{3}{2}+l)_l}{l!}(-D_s s-D_t t)^l,
\end{equation}
where $(a)_b$ is the falling factorial $(a)_b=a(a-1)(a-2)\ldots(a-b+1)$. This expression can be further broken down by the binomial expansion of the term in the brackets into
\begin{multline}
 (1+D_s s+D_t t)^{-(r+5/2)}=\\
 \sum_{l=0}^{\infty}\sum_{l_2=0}^{l} \frac{(r+\frac{3}{2}+l)_l}{(l-l_2)!l_2!}(-D_s s)^{l_2}(-D_t t)^{l-l_2}.
\end{multline}
Similarly, the term in the numerator can be binomially expanded to
\begin{multline}
 ( N_s s+N_t t+N_{st}st+N_1)^r=\\
 \sum_{r_2=0}^{r} \sum_{r_3=0}^{r-r_2}\sum_{r_4=0}^{r-r_2-r_3} \frac{r!}{(r-r_2-r_3-r_4)!r_2!r_3!r_4!}\\
 (N_s s)^{r-r_2-r_3-r_4} (N_t t)^{r_2} (N_{st}st)^{r_3} (N_1)^{r_4}.
\end{multline}
Substituting the expressions for the numerator and the denominator in the expression for $H_{\i\j}(\mathbf{\gn},\chi)$ gives { 
\begin{multline}
  H_{\i\j}(\mathbf{\gn},\chi)=(1-s)^{5/2}(1-t)^{5/2}\pi^{3/2}\exp(-k_{\i\j}\gn^2)\\
  \times(M_{\i 1}M_{\j 1})^{1/2}\sum_{r=0}^{\infty}\sum_{r_2=0}^{r} \sum_{r_3=0}^{r-r_2}\sum_{r_4=0}^{r-r_2-r_3}\sum_{l=0}^{\infty}\sum_{l_2=0}^{l}\\
  \gn^{2r}\left\{r+\frac{3}{2}+\gn^2 (1-k_{\i\j}+B_{\i\j}+A_{\i\j}\cos{\chi}) \right\} \times\\
   \frac{(r+\frac{3}{2}+l)_l}{(r-r_2-r_3-r_4)!r_2!r_3!r_4!(l-l_2)!l_2!}(N_s s)^{r-r_2-r_3-r_4}\times\\
   (N_t t)^{r_2} (N_{st}st)^{r_3} (N_1)^{r_4}(-D_s s)^{l_2}(-D_t t)^{l-l_2},
  \label{eq:hijfinal}
\end{multline}}
One can now notice that $(1-s)^{-5/2}(1-t)^{-5/2}\pi^{-3/2}(M_{\i 1}M_{\j 1})^{-1/2}H_{\i\j}(\mathbf{\gn},\chi)$ would just be a function of $(\gn,\chi,s,t)$ and we can choose to represent it as { 
\begin{multline}
 (1-s)^{-5/2}(1-t)^{-5/2}\pi^{-3/2}(M_{\i 1}M_{\j 1})^{-1/2}H_{\i\j}(\mathbf{\gn},\chi)\\
 =\exp(-k_{\i\j}\gn^2)\sum_{pq\bar{r}\bar{l}} A^{pq\bar{r}\bar{l},1}_{\i\j}  s^p t^q \gn^{2\bar{r}}\cos^{\bar{l}}{\chi},
\end{multline}}
where $A^{pq\bar{r}\bar{l},1}_{\i\j}$ is a factor that is only purely a function of the masses and temperatures of the species. On comparing the exponents of $\gn^2$ and $\cos\chi$ on both the sides of this expression with these limits, one can also find the summation limits on the the maximum value of $r,l$ in the LHS summation, $r\leq p+q+1$ and $l\leq p+q+1$. Now, on substituting the value of $H_{\i\j}(\mathbf{\gn},\chi)$ and $H_{\i\j}(\mathbf{\gn},\chi=0)$ becomes { 
\begin{multline}
 \Pi_{\i\j}=- \pi^{-3/2} {\mathcal{K}_{\i\j}}  (M_{\i 1}M_{\j 1})^{1/2}\sum_{pq\bar{r}\bar{l}} A^{pq\bar{r}\bar{l},1}_{\i\j}s^p t^q \\
  \times \int \exp(-k_{\i\j}\gn^2)   \gn^{2\bar{r}}(1-\cos^{\bar{l}}{\chi}) g\sigma_{\alpha\beta}(g,\chi)d\Omega d\mathbf{\gn}.
\end{multline}}
Now transforming $\mathbf{\gn}$ into spherical coordinates and integrate over the angles. We also write $d\Omega=\sin{\chi}\cos{\phi}$ and integrate over $\phi$.  We also define the Chapman-Cowling cross-section moment integral, in a generalized form, as follows { 
\begin{align}
 \Omega_{\i\j}^{\bar{l}\bar{r}} = &\left(\frac{\pi}{d_{\i\j}}\right)^{1/2}\int_0^\infty \exp(-\zeta^2)   \zeta^{2\bar{r}+3} \phi^{(\bar{l})}_{\i\j} d\zeta,\\
 \phi^{(\bar{l})}_{\i\j}=&\int^\infty_0 (1-\cos^{\bar{l}}{\chi}) \sigma_{\alpha\beta}(g,\chi)\sin{\chi}d\chi,
\end{align}
where $\zeta=k_{\i\j}^{1/2}\gn=d_{\i\j}^{1/2}g$, where 
\begin{equation}
 d_{\i\j}=k_{\i\j}\left\{\mu_{\i\j}^2\left(\frac{\gamma_\i}{2m_\i^2}+\frac{\gamma_\j}{2m_\j^2}\right)\right\},
\end{equation}
The quantity $d_{\i\j}$ is an arbitrarily chosen factor that enters the Chapman-Cowling integral, and in general is purely a function of the masses and temperatures of the colliding species, such that $d_{\i\j}>0$. The coefficients in Zhdanov et al presented in Appendix \ref{sec:zhdanov13} are calculated by choosing $d_{\i\j}=(1/2)\gamma_\i \gamma_\j/(\gamma_\i+\gamma_\j)=\gamma_{\i\j}/2$ (and consequently $k_{\i\j}=1-M_{\i1}M_{\i2}(1-\theta_{\i\j})^2$). A straightforward extension of Chapman and Cowling's original integrals can be performed just by choosing $d_{\i\j}=\mu_{\i\j}/(2kT)$, where $T$ is the plasma common temperature. One can also choose to follow the original calculation by Rat et al by retaining $k_{\i\j}=1$. It must be noted that in principle, the evaluation of the Chapman-Cowling integral and the bracket integrals should not depend on the choice of $k_{\i\j}$, however the choice of $d_{\i\j}$ may lead to slightly different evaluations once the potential of interaction is chosen.} The term $\phi^{(l)}_{\i\j}$ represents the effective cross-section, and we obtain { 
\begin{equation}
 \Pi_{\i\j}=-8\mathcal{K}_{\i\j}  (M_{\i 1}M_{\j 1})^{1/2}  \sum_{pq\bar{r}\bar{l}} s^p t^q \frac{A^{pq\bar{r}\bar{l},1}_{\i\j}}{k_{\i\j}^{\bar{r}+3/2}}\Omega_{\i\j}^{\bar{l}\bar{r}} .
\end{equation}}
Now, the coefficient of $s^pt^q$ in $\Pi_{\i\j}$ is the bracket integral, given by { 
\begin{multline}
 \left[S^p_{3/2}(W_\beta^2)\mathbf{W}_\beta,S^q_{3/2}(W_\alpha^2)\mathbf{W}_\alpha\right]\\
 =-8  \left[\mu_{\i\j}^2 \left( \frac{\g_\i+\g_\j}{{\g_\i\g_\j}}\right)\left( \frac{\g_\i}{m_\i^2}+\frac{\g_\j}{m_\j^2}\right)\right]^{-3/2}\\
 \times (M_{\i 1}M_{\j 1})^{1/2}  \sum_{\bar{r}\bar{l}} \frac{A^{pq\bar{r}\bar{l},1}_{\i\j}}{k_{\i\j}^{\bar{r}+3/2}}\Omega_{\i\j}^{\bar{l}\bar{r}}.
\end{multline}}
The coefficients $A^{pqrl,1}_{\i\j}$ can be obtained by comparing the binomial decomposition of $H_{\i\j}(\mathbf{\gn},\chi)$ with the one solely expressed as a function of $(\gn,\chi,s,t)$, and we obtain { 
\begin{multline}
 \sum_{pq\bar{r}\bar{l}} A^{pq\bar{r}\bar{l},1}_{\i\j}  s^p t^q \gn^{2\bar{r}}\cos^{\bar{l}}{\chi}= \sum_{r=0}^{\infty}\sum_{r_2=0}^{r} \sum_{r_3=0}^{r-r_2}\sum_{r_4=0}^{r-r_2-r_3}\\
 \sum_{l=0}^{\infty}\sum_{l_2=0}^{l}\gn^{2r}\left\{r+\frac{3}{2}+\gn^2 (1-k_{\i\j}+B_{\i\j}+A_{\i\j}\cos{\chi}) \right\} \\
  \times \frac{(r+\frac{3}{2}+l)_l}{(r-r_2-r_3-r_4)!r_2!r_3!r_4!(l-l_2)!l_2!} s^{r-r_2-r_4+l_2} t^{r_2+r_3+l-l_2}\\
  \times (N_s)^{r-r_2-r_3-r_4}\times (N_t)^{r_2} (N_{st})^{r_3} (N_1)^{r_4}(-D_s s)^{l_2}(-D_t t)^{l-l_2},
\end{multline}}
where we have used $\bar{r}$ and $\bar{l}$ to indicate that the $(r,l)$ summation on the LHS is different from that in the RHS. Now setting $p=r-r_2-r_4+l_2$ and  $q=r_2+r_3+l-l_2$, and in doing so selecting the $s^pt^q$ term which fixes $l=p+q+r_4-r_3-r$ and $l_2=p+r_2+r_4-r$, we can write
\begin{widetext}
{ 
\begin{multline}
 \sum_{\bar{r}\bar{l}} A^{pq\bar{r}\bar{l},1}_{\i\j} \gn^{2\bar{r}}\cos^{\bar{l}}{\chi}= \sum_{r=0}^{\infty}\sum_{r_2=0}^{r} \sum_{r_3=0}^{r-r_2}\sum_{r_4=0}^{r-r_2-r_3} \gn^{2r}\left\{r+\frac{3}{2}+\gn^2 (1-k_{\i\j}+B_{\i\j}+A_{\i\j}\cos{\chi}) \right\} \\
  \times \frac{(p+q+r_4-r_3+\frac{3}{2})_{p+q+r_4-r_3-r}}{(r-r_2-r_3-r_4)!r_2!r_3!r_4!(q-r_2-r_3)!(p+r_2+r_4-r)!}\\
  \times (N_s)^{r-r_2-r_3-r_4} (N_t)^{r_2} (N_{st})^{r_3} (N_1)^{r_4}(-D_s)^{p+r_2+r_4-r}(-D_t)^{q-r_2-r_3}.
\end{multline}}
\end{widetext}
Comparing the limits on both sides, one can show that instead of infinite sum over $r$, one only needs to sum until $r\leq p+q$.

\subsection{Derivation of $\left[S^p_{3/2}(W_\alpha^2)\mathbf{W}_\alpha,S^q_{3/2}(W_\alpha^2)\mathbf{W}_\alpha\right]$}

We have
\begin{multline}
 \left[S^p_{3/2}(W_\i^2)\mathbf{W}_\i,S^q_{3/2}(W_\alpha^2)\mathbf{W}_\alpha\right]= \frac{1}{n_\i n_\j} \int  f^{(0)}_\alpha f^{(0)}_\beta\\
 \times(S^p_{3/2}(W_\i^{\prime2})\mathbf{W}_\i^\prime-S^p_{3/2}(W_\i^2)\mathbf{W}_\i)\cdot S^q_{3/2}(W_\alpha^2)\mathbf{W}_\alpha\\
 \times g\sigma_{\alpha\beta}(g,\chi)d\Omega d\mathbf{c_{\alpha}}d\mathbf{c_{1\beta}}.
\end{multline}
Having calculated $\left[S^p_{3/2}(W_\beta^2)\mathbf{W}_\beta,S^q_{3/2}(W_\alpha^2)\mathbf{W}_\alpha\right]$ in the previous section, it becomes much easier to perform the calculation for $\left[S^l_{3/2}(W_\alpha^2)\mathbf{W}_\alpha,S^n_{3/2}(W_\alpha^2)\mathbf{W}_\alpha\right]$. 
Following the previous calculations, we have the bracket integral being the coefficient of $s^pt^q$ in the generating function $\Pi_{\i\i}$
which can then be expressed as
\begin{multline}
  \Pi_{\i\i}=(1-s)^{-5/2}(1-t)^{-5/2} \pi^{-3} {\mathcal{K}_{\i\j}}\\
  \int \{H_{\i\i}(\mathbf{\gn},\chi)-H_{\i\i}(\mathbf{\gn},0)\} g\sigma_{\alpha\beta}(g,\chi)d\Omega d\mathbf{\gn},
\end{multline}
where $H_{\i\i}(\chi)$ is
\begin{equation}
 H_{\i\i}(\mathbf{\gn},\chi)= \int \exp\left\{  -W_\i^2-W_\j^2-SW_\i^{\prime 2}-TW_\i^2 \right\}\mathbf{W}_\i^\prime.\mathbf{W}_\i d\mathbf{X}.
\end{equation}
Now, we again define $a_{\i\j}$ and $a_{\j\i}$ in the following form
\begin{align}
 a_{\i\j} &= 1+(S+T)M_{\i 1}\nonumber\\
 &=(1-s)^{-1}(1-t)^{-1}(1-(s+t)M_{\j1}+st(M_{\j1}-M_{\i1})),\nonumber\\
 a_{\j\i} &= 1+(S+T)M_{\i 2}\nonumber\\
 &=(1-s)^{-1}(1-t)^{-1}(1-(s+t)M_{\j2}+st(M_{\j2}-M_{\i2})),\nonumber\\
 T^* &= -(1-\theta_{\i\j}+T), \ S^*=-S,
\end{align}
such that $H_{\i\i}(\mathbf{\gn},\chi)$, with the help of the Eqs.\,(\ref{eq:xxntransform}) and (\ref{eq:bij}), becomes
\begin{equation}
 H_{\i\i}(\mathbf{\gn},\chi)=\int \exp\left\{ -a_{\i\j}\xn^2-b_{\i\j}\gn^2\right\}\mathbf{W}_\i^\prime.\mathbf{W}_\i d\mathbf{\xn}, \label{eq:hij22}
\end{equation}
which leaves us with needing to derive an expression for $\mathbf{W}_\i^\prime.\mathbf{W}_\i$. In order to do so, we again transform into the variables $\mathbf{V}_\i$ and $\mathbf{V}_\i^\prime$ such that
\begin{equation}
 \mathbf{W}_\i^\prime.\mathbf{W}_\i =M_{\i1}\xn^2+M_{\i2}\mathbf{V}_\i^\prime.\mathbf{V}_\i-(M_{\i1}M_{\i2})^{1/2}\mathbf{\xn}.(\mathbf{V}_\i^\prime+\mathbf{V}_\i)
\end{equation}
where the odd powers of $\mathbf{\xn}$ will integrate out in $H_{\i\i}$. Calculating $\mathbf{V}_\i^\prime.\mathbf{V}_\i$, we get
\begin{multline}
 \mathbf{V}_\i^\prime.\mathbf{V}_\i = \left[ \frac{M_{\i 1}}{M_{\i 2}}(1-b_{\i\j})+M_{\i 1}(\theta_{\i\j}-1)\right.\\
 \left.+(M_{\i 1}(\theta_{\i\j}-1)+1 )\cos{\chi} \frac{}{}\right]\frac{\gn^2}{a_{\i\j}},
\end{multline}
which on substituting back into $\mathbf{W}_\i^\prime.\mathbf{W}_\i$ and defining
\begin{equation}
 A_{\i\j}=\frac{M_{\i 2}}{M_{\i 1}}(M_{\i 1}(\theta_{\i\j}-1)+1 ),\ B_{\i\j}= M_{\i 2}(\theta_{\i\j}-1),
\end{equation}
and on integrating over $\mathbf{\xn}$, we find
\begin{multline}
 H_{\i\i}(\mathbf{\gn},\chi)=\pi^{3/2}\exp(-b_{\i\j}\gn^2)a_{\i\j}^{-5/2}M_{\i 1}\times\\
 \left\{ \frac{3}{2}+\gn^2 (1-b_{\i\j}+B_{\i\j}+A_{\i\j}\cos{\chi}) \right\}.
\end{multline}
So far, this expression agrees with that of Chapman and Cowling. The remaining steps are very similar to the previous derivation, and hence one can find, similar to Eq.\,(\ref{eq:hij171}) { 
\begin{multline}
  H_{\i\i}(\mathbf{\gn},\chi)=(1-s)^{5/2}(1-t)^{5/2}\pi^{3/2}\exp(-k_{\i\j}\gn^2)M_{\i 1}\\
  \times\sum_{r=0}^{\infty}\left\{r+\frac{3}{2}+\gn^2 (1-k_{\i\j}+B_{\i\j}+A_{\i\j}\cos{\chi}) \right\} \\
  \times \frac{( N_s s+N_t t+N_{st}st+N_1)^r}{(1+D_s s+D_t t +D_{st}st)^{r+5/2}}\frac{\gn^{2r}}{r!}, \label{eq:hij271}
\end{multline}}
where the $N$'s and $D$'s are  of the form { 
\begin{align}
 N_s =& (1-k_{\i\j})M_{\j1}-M_{\i2}-2M_{\i1}M_{\i2}(\theta_{\i\j}-1)\cos{\chi}\nonumber\\
 &-M_{\i1}M_{\i2}(\theta_{\i\j}-1)^2\nonumber\\
 N_t =& (1-k_{\i\j})M_{\j1}-M_{\i2}-M_{\i1}M_{\i2}(\theta_{\i\j}^2-1)=-kM_{\j1}\nonumber\\
 N_{st} =& (1-k_{\i\j})(1-2M_{\i1})+2M_{\i2}+2M_{\i1}M_{\i2}\theta_{\i\j}\cos{\chi}\nonumber\\
 &+M_{\i1}M_{\i2}(\theta_{\i\j}^2-3)\nonumber\\
 N_1 =& (k_{\i\j}-1)+M_{\i1}M_{\i2}(\theta_{\i\j}-1)^2\nonumber\\
 D_s =& -M_{\j1}, \ D_t = -M_{\j1},\ D_{st}= M_{\j1}-M_{\i1}. 
\end{align}}
Now, following the binomial expansion for the denominator as in the previous section, one would obtain
\begin{multline}
 (1+D_s s+D_t t+D_{st}st)^{-(r+5/2)}\\
 =\sum_{l=0}^{\infty}\sum_{l_2=0}^{l}\sum_{l_3=0}^{l-l_2} \frac{(r+\frac{3}{2}+l)_l}{(l-l_2-l_3)!l_2!l_3!}\\
 \times(-D_s s)^{l-l_2-l_3}(-D_t t)^{l_2}(-D_{st}st )^{l_3}.
\end{multline}
The binomial expansion for the numerator term remains the same. Combining them, we find the form for $H_{\i\i}(\mathbf{\gn},\chi)$, { 
\begin{multline}
  H_{\i\i}(\mathbf{\gn},\chi)=(1-s)^{5/2}(1-t)^{5/2}\pi^{3/2}\exp(-k_{\i\j}\gn^2)M_{\i 1}\times\\
  \sum_{r=0}^{\infty}\sum_{r_2=0}^{r} \sum_{r_3=0}^{r-r_2}\sum_{r_4=0}^{r-r_2-r_3}\sum_{l=0}^{\infty}\sum_{l_2=0}^{l}\sum_{l_3=0}^{l-l_2} \gn^{2r}\times\\
  \left\{r+\frac{3}{2}+\gn^2 (1-k_{\i\j}+B_{\i\j}+A_{\i\j}\cos{\chi}) \right\} \times\\
   \frac{(r+\frac{3}{2}+l)_l}{(r-r_2-r_3-r_4)!r_2!r_3!r_4!(l-l_2-l_3)!l_2!l_3!}\times\\
  (N_s s)^{r-r_2-r_3-r_4} (N_t t)^{r_2} (N_{st}st)^{r_3} (N_1)^{r_4}\times\\
  (-D_s s)^{l-l_2-l_3}(-D_t t)^{l_2}(-D_{st}st )^{l_3},
\end{multline}}
Looking at this equation, again it can be expressed as a function in terms of $\gn^2$ and $\cos{\chi}$. Therefore, the most general form of $\Pi_{\i\i}$ is going to be of the form { 
\begin{multline}
 \Pi_{\i\i}=-8\left[\mu_{\i\j}^2 \left( \frac{\g_\i+\g_\j}{{\g_\i\g_\j}}\right)\left( \frac{\g_\i}{m_\i^2}+\frac{\g_\j}{m_\j^2}\right)\right]^{-3/2}  M_{\i 1}\times\\
 \sum_{pq\bar{r}\bar{l}} s^p t^q \frac{A^{pq\bar{r}\bar{l},1}_{\i\i}}{k_{\i\j}^{\bar{r}+3/2}}\Omega_{\i\j}^{\bar{l}\bar{r}},
\end{multline}}
and on comparing the powers of $s$ and $t$ in both these expressions and selecting the $s^pt^q$ term, we find that $p=r-r_2-r_4+l-l_2$ and $q=r_2+r_3+l_2+l_3$, and in doing so, we inadvertently set $l=p+q-r-r_3+r_4-l_3$ and $l_2=q-r_2-r_3-l_3$, because of which the expression for calculating  $A^{pqrl,1}_{\i\i}$ becomes
\begin{widetext}
{ 
\begin{multline}
 \sum_{\bar{r}\bar{l}} A^{pq\bar{r}\bar{l},1}_{\i\i} \gn^{2\bar{r}}\cos^{\bar{l}}{\chi}= \sum_{r=0}^{\infty}\sum_{r_2=0}^{r} \sum_{r_3=0}^{r-r_2}\sum_{r_4=0}^{r-r_2-r_3}\sum_{l_3=0}^{p-r+r_2+r_4} \gn^{2r}\left\{r+\frac{3}{2}+\gn^2 (1-k_{\i\j}+B_{\i\j}+A_{\i\j}\cos{\chi}) \right\} \times\\
  \times \frac{(p+q+r_4-r_3-l_3+\frac{3}{2})_{p+q+r_4-r_3-r-l_3}}{(r-r_2-r_3-r_4)!r_2!r_3!r_4!(p-r+r_2+r_4-l_3)!(q-r_2-r_3-l_3)!l_3!} \times\\
  \times (N_s)^{r-r_2-r_3-r_4} (N_t)^{r_2} (N_{st})^{r_3} (N_1)^{r_4}(-D_s)^{p-r+r_2+r_4-l_3}(-D_t)^{q-r_2-r_3-l_3}(-D_{st})^{l_3},
  \label{eq:hiifinal}
\end{multline}}
\end{widetext}
where the sum over $r$ need only be computed such that $r\leq p+q$, and correspondingly $0\leq\bar{r}\leq p+q+1$ and $0\leq\bar{l}\leq p+q+1$. The general solution is given by { 
\begin{multline}
 \left[S^p_{3/2}(W_\alpha^2)\mathbf{W}_\alpha,S^q_{3/2}(W_\alpha^2)\mathbf{W}_\alpha\right]\\
 =-8\left[\mu_{\i\j}^2 \left( \frac{\g_\i+\g_\j}{{\g_\i\g_\j}}\right)\left( \frac{\g_\i}{m_\i^2}+\frac{\g_\j}{m_\j^2}\right)\right]^{-3/2}  M_{\i 1}\times\\
 \sum_{\bar{r}\bar{l}} \frac{A^{pq\bar{r}\bar{l},1}_{\i\i}}{k_{\i\j}^{\bar{r}+3/2}}\Omega_{\i\j}^{\bar{l}\bar{r}}
\end{multline}}
This completes the derivation of the second rank-1 bracket integral.

\subsection{Derivation of $\left[S^p_{5/2}(W_\beta^2)\left(\mathbf{W}_\beta\mathbf{W}_\beta-\frac{1}{3}U W_\beta^2\right),S^q_{5/2}(W_\alpha^2)\left(\mathbf{W}_\alpha\mathbf{W}_\alpha-\frac{1}{3}U W_\alpha^2\right)\right]$}

In order to calculate a higher-rank bracket integral, we first notice that fundamentally, only the terms in the curly brackets in Eq.\,(\ref{eq:hijfinal}), and  in the falling factorial change. Every other term remains the same. Therefore, in the definition of $L_{\i\j}(\mathbf{\gn},\chi)$ (defined analogously to $H_{\i\j}(\mathbf{\gn},\chi)$, we have
\begin{multline}
 L_{\i\j}(\mathbf{\gn},\chi)= \int \exp\left\{ -a_{\i\j}\xn^2-b_{\i\j}\gn^2\right\}\times\\
 (\mathbf{W}_\beta^\prime\mathbf{W}_\beta^\prime-\frac{1}{3}UW_\beta^{\prime2}):(\mathbf{W}_\alpha\mathbf{W}_\alpha-\frac{1}{3}U W_\alpha^2) d\mathbf{\xn}.
\end{multline}
Therefore, we need to calculate only the inner product $(\mathbf{W}_\beta^\prime\mathbf{W}_\beta^\prime-\frac{1}{3}UW_\beta^{\prime2}):(\mathbf{W}_\alpha\mathbf{W}_\alpha-\frac{1}{3}U W_\alpha^2)$, which can be expressed as
\begin{multline}
 \left(\mathbf{W}_\beta^\prime\mathbf{W}_\beta^\prime-\frac{1}{3}U W_\beta^{\prime2}\right):\left(\mathbf{W}_\alpha\mathbf{W}_\alpha-\frac{1}{3}U W_\alpha^2\right)\\
 =(\mathbf{W}_\beta^\prime.\mathbf{W}_\alpha)^2-\frac{1}{3}{W}_\beta^{\prime2}{W}_\alpha^2.
\end{multline}
Now, $\mathbf{W}_\beta^\prime.\mathbf{W}_\alpha$ is given by Eq.\,(\ref{eq:wbdotwainitial}), the square of which would be given by
\begin{multline}
 (\mathbf{W}_\j^\prime.\mathbf{W}_\i)^2 = M_{\i 1}M_{\j 1}\xn^4+M_{\i 2}M_{\j 2}(\mathbf{V}_\j^\prime.\mathbf{V}_\i)^2\\
 +\left[\mathbf{\xn}.\left\{ (M_{\i 1}M_{\j 2})^{1/2}\mathbf{V}_\j^\prime+(M_{\i 2}M_{\j 1})^{1/2}\mathbf{V}_\i \right\}\right]^2+\\
 +2(M_{\i 1}M_{\j 1}M_{\i 2}M_{\j 2})^{1/2}\xn^2(\mathbf{V}_\j^\prime.\mathbf{V}_\i)+\mathrm{\ odd\ terms\ of\ \mathbf{\xn}}.
\end{multline}
And we write down the expression for the product of $W_\j^{\prime2}$ and $W_\i^2$
\begin{multline}
W_\j^{\prime2} W_\i^2 = (M_{\i1}\xn^2+M_{\i2}{V}_\i^2)(M_{\j1}\xn^2+M_{\j2}{V}_\j^{\prime2})\\+4(M_{\i1}M_{\i2}M_{\j1}M_{\j2})^{1/2}(\mathbf{\xn}.\mathbf{V}_\j^{\prime})(\mathbf{\xn}.\mathbf{V}_\i)+\mathrm{\ odd\ terms\ of\ \mathbf{\xn}}.
\end{multline}
Both the terms involving product of dot products of the form $(\mathbf{\xn}.\mathbf{A})(\mathbf{\xn}.\mathbf{B})$ can be written as $\mathbf{A}\mathbf{B}:\mathbf{\xn}\mathbf{\xn}$. This is useful, because in the expression for $L_{\i\j}(\mathbf{\gn},\chi)$, we can take the term $\mathbf{A}\mathbf{B}$ out of the integral with the symmetry rule
\begin{equation}
 \int (\mathbf{A}\mathbf{B}:\mathbf{\xn}\mathbf{\xn})G({\xn})d\mathbf{\xn}=\frac{1}{3}(\mathbf{A}.\mathbf{B})\int \xn^2G({\xn})d\mathbf{\xn}.
 \label{eq:getxout}
\end{equation}
$L_{\i\j}(\mathbf{\gn},\chi)$, on integrating over $\mathbf{\xn}$, therefore becomes
\begin{multline}
 L_{\i\j}(\mathbf{\gn},\chi)=\pi^{3/2}\exp\left\{-b_{\i\j}\gn^2\right\}a_{\i\j}^{-7/2}\left(\frac{5}{2}M_{\i1}M_{\j1}\right.\\
 +\frac{10}{3}a_{\i\j}(M_{\i1}M_{\i2}M_{\j1}M_{\j2})^{1/2}(\mathbf{V}_\beta^\prime.\mathbf{V}_\alpha)+\\
 \left.+a_{\i\j}^2M_{\i2}M_{\j2}(\mathbf{V}_\beta^\prime.\mathbf{V}_\alpha)^2-\frac{1}{3}a_{\i\j}^2M_{\i2}M_{\j2}{V}_\beta^{\prime2}{V}_\alpha^2 \right)
\end{multline}

Following Chapman and Cowling, we can see that $(\mathbf{V}_\beta^\prime\times\mathbf{V}_\alpha)^2={V}_\beta^{\prime2}{V}_\alpha^2-(\mathbf{V}_\beta^\prime.\mathbf{V}_\alpha)^2$, which therefore is
\begin{align}
 \mathbf{V}_\beta^\prime\times\mathbf{V}_\alpha&=\{M_{\i1}(1-\theta_{\i\j})-1 \}\frac{\mathbf{\gn}^\prime\times\mathbf{\gn}}{a_{\i\j}}\\
 \implies 
 (\mathbf{V}_\beta^\prime\times\mathbf{V}_\alpha)^2&=\{M_{\i1}(1-\theta_{\i\j})-1 \}^2 (1-\cos^2{\chi})\frac{\gn^4}{a_{\i\j}^2},
\end{align}
and therefore
\begin{multline}
  (\mathbf{V}_\beta^\prime.\mathbf{V}_\alpha)^2-\frac{1}{3}{V}_\beta^{\prime2}{V}_\alpha^2\\
  =\frac{2}{3}\left[ \frac{M_{\i1}M_{\j1}}{M_{\i2}M_{\j2}}(1-b_{\i\j})^2+  2\frac{M_{\j1}}{M_{\j2}}\left[ M_{\i1}(\theta_{\i\j}-1)\right.\right.\\
   \left.+\eta_{\i\j} \{M_{\i1}(1-\theta_{\i\j})-1\}\cos{\chi}\right](1-b_{\i\j})+\\
 + M_{\i1}M_{\i2}\frac{M_{\j1}}{M_{\j2}}(\theta_{\i\j}-1)^2 +\{M_{\i1}(1-\theta_{\i\j})-1\}^2\cos^2{\chi}\\
 +2M_{\i2}\left(\frac{M_{\i1}M_{\j1}}{M_{\i2}M_{\j2}}\right)^{1/2}(\theta_{\i\j}-1)\{M_{\i1}(1-\theta_{\i\j})-1\}\cos{\chi}\\
 \left.-\frac{1}{2}\{M_{\i1}(1-\theta_{\i\j})-1 \}^2 (1-\cos^2{\chi})\right]\frac{\gn^4}{a_{\i\j}^2},
\end{multline}
where $\eta_{\i\j}=m_\i/m_\j$.
{ We now take a factor of $\exp{(k_{\i\j}-b_{\i\j})\gn^2}$ in $L_{\i\j}(\mathbf{\gn},\chi)$, expand it, shift forward the sum wherever there is are factors of $\exp{(1-b_{\i\j})\gn^2}$, to obtain the following expression
\begin{multline}
 L_{\i\j}(\mathbf{\gn},\chi)=\pi^{3/2}\exp\left\{-k_{\i\j}\gn^2\right\}a_{\i\j}^{-7/2}(M_{\i1}M_{\j1})\times\\
 \sum_{r=0}^{\infty}\left[\frac{5}{2}+ rA_{\i\j}+r(r-1)D_{\i\j}+\right.\\
 [(B_{\i\j}+rE_{\i\j})+(C_{\i\j}+rN_{\i\j})\cos{\chi}]\gn^2+ \\
  \left.+ \left\{O_{\i\j}+P_{\i\j}\cos^2{\chi}+Q_{\i\j}\cos{\chi}\right\}\gn^4 \right]\frac{(k_{\i\j}-b_{\i\j})^r}{r!}\gn^{2r},
\end{multline}
where the coefficients are given as follows
\begin{multline}
A_{\i\j}=\frac{10}{3}, \ B_{\i\j}=\frac{10}{3}[1-k_{\i\j}+M_{\i 2}(\theta_{\i\j}-1)],\\
C_{\i\j}=\frac{10}{3}\left(\frac{M_{\i 2}M_{\j 2}}{M_{\i 1}M_{\j 1}} \right)^{1/2}\{M_{\i 1}(1-\theta_{\i\j})-1 \},\\
D_{\i\j}=\frac{2}{3}, \ E_{\i\j}=\frac{4}{3}[1-k_{\i\j}+M_{\i 2}(\theta_{\i\j}-1)], \\ 
 N_{\i\j}= \frac{4}{3}\frac{M_{\i2}}{M_{\i1}}\eta_{\i\j}\{M_{\i1}(1-\theta_{\i\j})-1\}, \\
 O_{\i\j}=\frac{2}{3}[1-k_{\i\j}+ M_{\i2}(\theta_{\i\j}-1)]^2-\frac{1}{3}\frac{M_{\i2}M_{\j2}}{M_{\i1}M_{\j1}}\{M_{\i1}(1-\theta_{\i\j})-1\}^2,\\
 P_{\i\j}=\frac{M_{\i2}M_{\j2}}{M_{\i1}M_{\j1}}\{M_{\i1}(1-\theta_{\i\j})-1\}^2, \\
 Q_{\i\j}=\frac{4}{3}\frac{M_{\i2}}{M_{\i1}}\eta_{\i\j} \{M_{\i1}(1-\theta_{\i\j})-1\}(1-k_{\i\j})\\
 +\frac{4}{3}M_{\i2}\left(\frac{M_{\i2}M_{\j2}}{M_{\i1}M_{\j1}}\right)^{1/2}(\theta_{\i\j}-1)\{M_{\i1}(1-\theta_{\i\j})-1\}.
\end{multline}}
The term in the square brackets is fundamentally what we require; it will replace the one in the curly bracket in the expansion. Additionally, the falling factorial has a $5/2$ instead of a $3/2$ because we expand $a_{\i\j}^{-(r+7/2)}$ from the denominator instead of $a_{\i\j}^{-(r+5/2)}$, with the final expression for the bracket integral given by
\begin{widetext}
{ 
\begin{multline}
 \left[S^p_{m+1/2}(W_\beta^2)(\mathbf{W}_\j\mathbf{W}_\j-\frac{1}{3}\mathbf{U} W_\j^2),S^q_{m+1/2}(W_\alpha^2)(\mathbf{W}_\alpha\mathbf{W}_\alpha-\frac{1}{3}\mathbf{U} W_\alpha^2)\right]=\\
 -8  \left[\mu_{\i\j}^2 \left( \frac{\g_\i+\g_\j}{{\g_\i\g_\j}}\right)\left( \frac{\g_\i}{m_\i^2}+\frac{\g_\j}{m_\j^2}\right)\right]^{-3/2}  (M_{\i 1}M_{\j 1})  \sum_{\bar{r}\bar{l}} \frac{A^{pq\bar{r}\bar{l},2}_{\i\j}}{k_{\i\j}^{\bar{r}+3/2}}\Omega_{\i\j}^{\bar{l}\bar{r}}.
\end{multline}
\begin{multline}
 \sum_{\bar{r}\bar{l}} A^{pq\bar{r}\bar{l},2}_{\i\j} \gn^{2\bar{r}}\cos^{\bar{l}}{\chi}= \sum_{r=0}^{\infty}\sum_{r_2=0}^{r} \sum_{r_3=0}^{r-r_2}\sum_{r_4=0}^{r-r_2-r_3} \gn^{2r}\left\{\frac{5}{2}+ rA_{\i\j}+r(r-1)D_{\i\j}+\right.\\ [(B_{\i\j}+rE_{\i\j})+(C_{\i\j}+rN_{\i\j})\cos{\chi}]\gn^2+
  \left.+ (O_{\i\j}+P_{\i\j}\cos^2{\chi}+Q_{\i\j}\cos{\chi})\gn^4 \bigg\}\right. \times\\
  \times \frac{(p+q+r_4-r_3+\frac{5}{2})_{p+q+r_4-r_3-r}}{(r-r_2-r_3-r_4)!r_2!r_3!r_4!(q-r_2-r_3)!(p+r_2+r_4-r)!}  (N_s)^{r-r_2-r_3-r_4} (N_t)^{r_2} (N_{st})^{r_3} (N_1)^{r_4}(-D_s)^{p+r_2+r_4-r}(-D_t)^{q-r_2-r_3},
\end{multline} 
}
\end{widetext}
where the sum over $r$ needs to be only computed till $r\leq p+q$, and correspondingly $0\leq\bar{r}\leq p+q+2$ and $0\leq\bar{l}\leq p+q+2$.

\subsection{Derivation of $\left[S^p_{5/2}(W_\alpha^2)(\mathbf{W}_\alpha\mathbf{W}_\alpha-\frac{1}{3}\mathbf{U} W_\alpha^2),S^q_{5/2}(W_\alpha^2)(\mathbf{W}_\alpha\mathbf{W}_\alpha-\frac{1}{3}\mathbf{U} W_\alpha^2)\right]$}

Again, as in the previous section,  we notice that fundamentally, only the terms in the curly brackets in Eq.\,(\ref{eq:hiifinal}) and the index in the falling factorial change. Every other term remains the same. Therefore, in the definition of $L_{\i\i}(\mathbf{\gn},\chi)$ (defined analogously to $L_{\i\j}(\chi)$, we have
\begin{multline}
L_{\i\i}(\mathbf{\gn},\chi)= \int \exp\left\{ -a_{\i\j}\xn^2-b_{\i\j}\gn^2\right\}\times\\
 (\mathbf{W}_\i^\prime\mathbf{W}_\i^\prime-\frac{1}{3}UW_\i^{\prime2}):(\mathbf{W}_\alpha\mathbf{W}_\alpha-\frac{1}{3}U W_\alpha^2) d\mathbf{\xn},
\end{multline}
where the inner product $(\mathbf{W}_\i^\prime\mathbf{W}_\i^\prime-\frac{1}{3}UW_\i^{\prime2}):(\mathbf{W}_\alpha\mathbf{W}_\alpha-\frac{1}{3}U W_\alpha^2)$ can be expressed as
\begin{multline}
 \left(\mathbf{W}_\i^\prime\mathbf{W}_\i^\prime-\frac{1}{3}U W_\i^{\prime2}\right):\left(\mathbf{W}_\alpha\mathbf{W}_\alpha-\frac{1}{3}U W_\alpha^2\right)\\
 =(\mathbf{W}_\i^\prime.\mathbf{W}_\alpha)^2-\frac{1}{3}{W}_\i^{\prime2}{W}_\alpha^2.
\end{multline}
The product $(\mathbf{W}_\i^\prime.\mathbf{W}_\alpha)^2$
\begin{multline}
 (\mathbf{W}_\i^\prime.\mathbf{W}_\i)^2 = M_{\i 1}^2\xn^4+M_{\i 2}^2(\mathbf{V}_\i^\prime.\mathbf{V}_\i)^2\\
 +M_{\i 1}M_{\i 2}\left[\mathbf{\xn}.\left\{ \mathbf{V}_\i^\prime+\mathbf{V}_\i \right\}\right]^2+
 2M_{\i 1}M_{\i 2}\xn^2(\mathbf{V}_\i^\prime.\mathbf{V}_\i)\\+\mathrm{\ odd\ terms\ of\ \mathbf{\xn}}.
\end{multline}
And we write down the expression for $W_\i^{\prime2} W_\i^2$
\begin{multline}
W_\i^{\prime2} W_\i^2 = (M_{\i1}\xn^2+M_{\i2}{V}_\i^2)(M_{\i1}\xn^2+M_{\i2}{V}_\i^{\prime2})\\
+4M_{\i1}M_{\i2}(\mathbf{\xn}.\mathbf{V}_\i^{\prime})(\mathbf{\xn}.\mathbf{V}_\i)\\
+\mathrm{\ odd\ terms\ of\ \mathbf{\xn}}.
\end{multline}
Using identity Eq.\,(\ref{eq:getxout}), we have
Therefore, on integrating over $\mathbf{\xn}$, $L_{\i\i}(\mathbf{\gn},\chi)$ becomes
\begin{multline}
 L_{\i\i}(\mathbf{\gn},\chi)=\pi^{3/2}\exp\left\{-b_{\i\j}\gn^2\right\}a_{\i\j}^{-7/2}\left(\frac{5}{2}M_{\i1}^2\right.\\
 +\frac{10}{3}a_{\i\j}M_{\i1}M_{\i2}(\mathbf{V}_\i^\prime.\mathbf{V}_\alpha)+a_{\i\j}^2M_{\i2}^2(\mathbf{V}_\i^\prime.\mathbf{V}_\alpha)^2\\
 \left.-\frac{1}{3}a_{\i\j}^2M_{\i2}^2{V}_\i^{\prime2}{V}_\alpha^2 \right).
\end{multline}
Following the same procedure as in the previous section, we find
\begin{multline}
 (\mathbf{V}_\i^\prime.\mathbf{V}_\alpha)^2-\frac{1}{3}{V}_\i^{\prime2}{V}_\alpha^2=\frac{2}{3}\left[ \frac{M_{\i1}^2}{M_{\i2}^2}(1-b_{\i\j})^2+\right.\\
 2\frac{M_{\i1}}{M_{\i2}}\left[ M_{\i1}(\theta_{\i\j}-1)+\{M_{\i1}(\theta_{\i\j}-1)+1\}\cos{\chi}\right](1-b_{\i\j}) \\
 + M_{\i1}^2(\theta_{\i\j}-1)^2 +\{M_{\i1}(\theta_{\i\j}-1)+1\}^2\cos^2{\chi}\\
 +2M_{\i1}(\theta_{\i\j}-1)\{M_{\i1}(\theta_{\i\j}-1)+1\}\cos{\chi}\\
\left. -\frac{1}{2}\{M_{\i1}(\theta_{\i\j}-1)+1 \}^2 (1-\cos^2{\chi})\frac{}{}\right]\frac{\gn^4}{a_{\i\j}^2}.
\end{multline}
{ Substituting the values of $\mathbf{V}_\i^\prime.\mathbf{V}_\alpha$ and $(\mathbf{V}_\i^\prime.\mathbf{V}_\alpha)^2-(1/3){V}_\i^{\prime2}{V}_\alpha^2$ in the expression for $L_{\i\i}(\mathbf{\gn},\chi)$, then expanding $\exp{(k_{\i\j}-b_{\i\j})\gn^2}$ term and shifting the sum for the factors of $(k_{\i\j}-b_{\i\j})\gn^2$,  we get
\begin{multline}
 L_{\i\i}(\mathbf{\gn},\chi)=\pi^{3/2}\exp\left\{-k_{\i\j}\gn^2\right\}a_{\i\j}^{-7/2}(M_{\i1}^2)\\
  \sum_{r=0}^{\infty}\left[\frac{5}{2}+ rA_{\i\i}+r(r-1)D_{\i\i}+\right.\\
 [(B_{\i\i}+rE_{\i\i})+(C_{\i\i}+rN_{\i\i})\cos{\chi}]\gn^2+ \\
  \left.+ \left\{O_{\i\i}+P_{\i\i}\cos^2{\chi}+Q_{\i\i}\cos{\chi}\right\}\gn^4 \frac{}{}\right]\frac{(k_{\i\j}-b_{\i\j})^r}{r!}\gn^{2r},
\end{multline}
where the coefficients are given by
\begin{multline}
A_{\i\j}=\frac{10}{3}, \ B_{\i\j}=\frac{10}{3}[(1-k_{\i\j})+M_{\i 2}(\theta_{\i\j}-1)],\\
C_{\i\j}=\frac{10}{3}\frac{M_{\i 2}}{M_{\i 1}}\{M_{\i 1}(\theta_{\i\j}-1)+1 \}, \\
D_{\i\j}=\frac{2}{3}, \ E_{\i\j}=\frac{4}{3}[(1-k_{\i\j})+M_{\i 2}(\theta_{\i\j}-1)],\\
N_{\i\j}= \frac{4}{3}\frac{M_{\i2}}{M_{\i1}}\{M_{\i1}(\theta_{\i\j}-1)+1\},\\
  O_{\i\j}=\frac{2}{3}[1-k_{\i\j}+  M_{\i2}(\theta_{\i\j}-1)]^2-\frac{1}{3}\frac{M_{\i2}^2}{M_{\i1}^2}\{M_{\i1}(\theta_{\i\j}-1)+1\}^2,\\
 P_{\i\j}=\frac{M_{\i2}^2}{M_{\i1}^2}\{M_{\i1}(\theta_{\i\j}-1)+1\}^2, \\
 Q_{\i\j}=\frac{4}{3}\frac{M_{\i2}}{M_{\i1}}\{M_{\i1}(\theta_{\i\j}-1)+1\}(1-k_{\i\j})\\
 +\frac{4}{3}\frac{M_{\i2}^2}{M_{\i1}}(\theta_{\i\j}-1)\{M_{\i1}(\theta_{\i\j}-1)+1\}.
\end{multline}}
The term in the square brackets will replace the one in the curly bracket in the expansion for computing $A^{pqrl,2}_{\i\i}$, accompanied by the falling factorial term having a $5/2$ instead of a $3/2$ because we expand $a_{\i\j}^{-(r+7/2)}$ from the denominator instead of $a_{\i\j}^{-(r+5/2)}$. Thus, the general form of the bracket integral and the coefficients $A^{pqrl,2}_{\i\i}$ are now given by
\begin{widetext}
{ 
\begin{multline}
 \left[S^p_{5/2}(W_\alpha^2)(\mathbf{W}_\alpha\mathbf{W}_\alpha-\frac{1}{3}\mathbf{U} W_\alpha^2),S^q_{5/2}(W_\alpha^2)(\mathbf{W}_\alpha\mathbf{W}_\alpha-\frac{1}{3}\mathbf{U} W_\alpha^2)\right] = \\
 -8\left[\mu_{\i\j}^2 \left( \frac{\g_\i+\g_\j}{{\g_\i\g_\j}}\right)\left( \frac{\g_\i}{m_\i^2}+\frac{\g_\j}{m_\j^2}\right)\right]^{-3/2}  M_{\i 1}^2 \sum_{\bar{r}\bar{l}} \frac{A^{pq\bar{r}\bar{l},2}_{\i\i}}{k_{\i\j}^{\bar{r}+3/2}}\Omega_{\i\j}^{\bar{l}\bar{r}}.
\end{multline}
\begin{multline}
 \sum_{\bar{r}\bar{l}} A^{pq\bar{r}\bar{l},2}_{\i\i} \gn^{2\bar{r}}\cos^{\bar{l}}{\chi}= \sum_{r=0}^{\infty}\sum_{r_2=0}^{r} \sum_{r_3=0}^{r-r_2}\sum_{r_4=0}^{r-r_2-r_3} \gn^{2r}\left\{\frac{5}{2}+ rA_{\i\j}+r(r-1)D_{\i\j}+\right.\\ [(B_{\i\j}+rE_{\i\j})+(C_{\i\j}+rN_{\i\j})\cos{\chi}]\gn^2+
  \left.+ (O_{\i\j}+P_{\i\j}\cos^2{\chi}+Q_{\i\j}\cos{\chi})\gn^4 \bigg\}\right. \times\\
  \times \frac{(p+q+r_4-r_3-l_3+\frac{5}{2})_{p+q+r_4-r_3-r-l_3}}{(r-r_2-r_3-r_4)!r_2!r_3!r_4!(p-r+r_2+r_4-l_3)!(q-r_2-r_3-l_3)!l_3!} \times\\
  \times (N_s)^{r-r_2-r_3-r_4} (N_t)^{r_2} (N_{st})^{r_3} (N_1)^{r_4}(-D_s)^{p-r+r_2+r_4-l_3}(-D_t)^{q-r_2-r_3-l_3}(-D_{st})^{l_3},
\end{multline}
}
\end{widetext}
where the sum over $r$ needs to be computed for $r\leq p+q$, and correspondingly $0\leq\bar{r}\leq p+q+2$ and $0\leq\bar{l}\leq p+q+2$.

\subsection{Derivation of $\left[S^p_{1/2}(W_\beta^2),S^q_{1/2}(W_\alpha^2)\right]$ and $\left[S^p_{1/2}(W_\alpha^2),S^q_{1/2}(W_\alpha^2)\right]$}

Following the previous sections, we notice that, in order to derive the brackets  $\left[S^p_{1/2}(W_\beta^2),S^q_{1/2}(W_\alpha^2)\right]$ and $\left[S^p_{1/2}(W_\alpha^2),S^q_{1/2}(W_\alpha^2)\right]$, we just need to drop the term that arises from the curly brackets in Eq.\,(\ref{eq:hijfinal})and  Eq.\,(\ref{eq:hiifinal}) respectively, and  and change the index in the falling factorial to $1/2$. Thus, the general solutions for the brackets and  and the coefficients $A^{pqrl,0}_{\i\j},A^{pqrl,0}_{\i\i}$ are given by
\begin{widetext}
{ 
\begin{equation}
 \left[S^p_{3/2}(W_\beta^2),S^q_{3/2}(W_\alpha^2)\right]=-8  \left[\mu_{\i\j}^2 \left( \frac{\g_\i+\g_\j}{{\g_\i\g_\j}}\right)\left( \frac{\g_\i}{m_\i^2}+\frac{\g_\j}{m_\j^2}\right)\right]^{-3/2}    \sum_{\bar{r}\bar{l}} \frac{A^{pq\bar{r}\bar{l},0}_{\i\j}}{k^{\bar{r}+3/2}}\Omega_{\i\j}^{\bar{l}\bar{r}},
\end{equation}
\begin{multline}
 \sum_{\bar{r}\bar{l}} A^{pq\bar{r}\bar{l},0}_{\i\j} \gn^{2\bar{r}}\cos^{\bar{l}}{\chi}= \sum_{r=0}^{\infty}\sum_{r_2=0}^{r} \sum_{r_3=0}^{r-r_2}\sum_{r_4=0}^{r-r_2-r_3} \gn^{2r} \frac{(p+q+r_4-r_3+\frac{1}{2})_{p+q+r_4-r_3-r}}{(r-r_2-r_3-r_4)!r_2!r_3!r_4!(q-r_2-r_3)!(p+r_2+r_4-r)!}\times\\
 \times (N_s)^{r_2} (N_t)^{r_3} (N_{st})^{r_4} (N_1)^{r-r_2-r_3-r_4}(-D_s)^{q-r_2-r_3}(-D_t)^{p+r_2+r_4-r}.
\end{multline}
\begin{equation}
 \left[S^p_{3/2}(W_\alpha^2),S^q_{3/2}(W_\alpha^2)\right]=-8\left[\mu_{\i\j}^2 \left( \frac{\g_\i+\g_\j}{{\g_\i\g_\j}}\right)\left( \frac{\g_\i}{m_\i^2}+\frac{\g_\j}{m_\j^2}\right)\right]^{-3/2}  \sum_{\bar{r}\bar{l}} \frac{A^{pq\bar{r}\bar{l},0}_{\i\i}}{k^{\bar{r}+3/2}}\Omega_{\i\j}^{\bar{l}\bar{r}},
\end{equation}
and
\begin{multline} 
 \sum_{\bar{r}\bar{l}} A^{pq\bar{r}\bar{l},0}_{\i\i} \gn^{2\bar{r}}\cos^{\bar{l}}{\chi}= \sum_{r=0}^{\infty}\sum_{r_2=0}^{r} \sum_{r_3=0}^{r-r_2}\sum_{r_4=0}^{r-r_2-r_3}\sum_{l_3=0}^{p-r+r_2+r_4} \gn^{2r}\times\\
 \frac{(p+q+r_4-r_3-l_3+\frac{1}{2})_{p+q+r_4-r_3-r-l_3}}{(r-r_2-r_3-r_4)!r_2!r_3!r_4!(p-r+r_2+r_4-l_3)!(q-r_2-r_3-l_3)!l_3!} \times\\
  \times (N_s)^{r_2} (N_t)^{r_3} (N_{st})^{r_4} (N_1)^{r-r_2-r_3-r_4}(-D_s)^{p-r+r_2+r_4-l_3}(-D_t)^{q-r_2-r_3-l_3}(-D_{st})^{l_3}.
\end{multline}
}
\end{widetext}
where the sum over $r$ goes till $r\leq p+q$, and correspondingly $0\leq\bar{r}\leq p+q$ and $0\leq\bar{l}\leq p+q$.

\section{Bracket integral values for verifying conservation properties}
\label{sec:bracket_values}

In this section, $\eta_{\i\j}=m_\i/m_\j$ and $\theta_{\i\j}=T_\i/T_\j$ and vice versa. One can use Eq.\,(\ref{eq:collision_second_form}) for $m=1,p=0$ and $m=0,p=1$ to verify the momentum and energy conservation respectively with the tables below for $q=0,1,2$.

\subsection{$\left[S^p_{3/2}(W_\j^2)\mathbf{W}_\j,S^q_{3/2}(W_\alpha^2)\mathbf{W}_\alpha\right]$}

\begin{multline}
 p=0,q=0:\  A^{0011,1}_{\i\j} \Omega_{\i\j}^{11}\\
 A^{0011,1}_{\i\j}=\frac{8 \sqrt{\eta _{\alpha \beta }} \left(\eta _{\alpha \beta }+1\right){}^4 \theta _{\alpha \beta }^2}{\left(k_{\alpha \beta } \left(\eta _{\alpha \beta }+\theta _{\alpha \beta }\right) \left(\eta _{\alpha
   \beta } \theta _{\alpha \beta }+1\right)\right){}^{5/2}} 
\end{multline}

\begin{multline}
p=0,q=1:\  A^{0111,1}_{\i\j} \Omega_{\i\j}^{11}+A^{0121,1}_{\i\j} \Omega_{\i\j}^{12}\\
 A^{0111,1}_{\i\j}=  \frac{20 \eta _{\alpha \beta }^{3/2} \left(\eta _{\alpha \beta }+1\right){}^4 \theta _{\alpha \beta }^2}{\left(\eta _{\alpha \beta }+\theta _{\alpha \beta }\right){}^{7/2} \left(k_{\alpha \beta } \left(\eta
   _{\alpha \beta } \theta _{\alpha \beta }+1\right)\right){}^{5/2}}\\
A^{0121,1}_{\i\j}=\frac{4 \eta _{\alpha \beta }^{3/2} \left(\eta _{\alpha \beta }+1\right){}^4 \theta _{\alpha \beta }^2 }{\left(\eta _{\alpha \beta }+\theta _{\alpha \beta
   }\right){}^{9/2} \left(k_{\alpha \beta } \left(\eta _{\alpha \beta } \theta _{\alpha \beta }+1\right)\right){}^{7/2}}\\
   \times\left(5 k_{\alpha \beta } \left(\eta _{\alpha \beta }^2 \theta _{\alpha \beta }+\eta _{\alpha \beta }
   \left(\theta _{\alpha \beta }^2+1\right)+\theta _{\alpha \beta }\right)-7 \left(\eta _{\alpha \beta }+1\right){}^2 \theta _{\alpha \beta }\right)
\end{multline}

\begin{multline}
p=0,q=2:\   A^{0211,1}_{\i\j} \Omega_{\i\j}^{11}+A^{0221,1}_{\i\j} \Omega_{\i\j}^{12}+A^{0231,1}_{\i\j} \Omega_{\i\j}^{13}\\
A^{0211,1}_{\i\j}=\frac{35 \left(\eta _{\alpha \beta }+1\right){}^4 \theta _{\alpha \beta }^2}{\left(\eta _{\alpha \beta }+\theta _{\alpha \beta }\right){}^{9/2} \left(k_{\alpha \beta } \left(\frac{1}{\eta _{\alpha \beta }}+\theta
   _{\alpha \beta }\right)\right){}^{5/2}}\\
A^{0221,1}_{\i\j}=\frac{7 \eta _{\alpha \beta }^{5/2} \left(\eta _{\alpha \beta }+1\right){}^4 \theta _{\alpha \beta }^2 }{\left(\eta _{\alpha \beta }+\theta _{\alpha \beta }\right){}^{11/2} \left(k_{\alpha \beta }
   \left(\eta _{\alpha \beta } \theta _{\alpha \beta }+1\right)\right){}^{7/2}}\\
   \times\left(5 k_{\alpha \beta } \left(\eta _{\alpha \beta }+\theta _{\alpha \beta }\right) \left(\eta _{\alpha
   \beta } \theta _{\alpha \beta }+1\right)-9 \left(\eta _{\alpha \beta }+1\right){}^2 \theta _{\alpha \beta }\right)\\
A^{0231,1}_{\i\j}=\frac{\eta _{\alpha \beta }^{5/2} \left(\eta _{\alpha \beta }+1\right){}^4 \theta _{\alpha \beta }^{3/2} }{2 \sqrt{\frac{\eta _{\alpha \beta }}{\theta _{\alpha \beta }}+1}
   \left(\eta _{\alpha \beta }+\theta _{\alpha \beta }\right){}^6 \left(k_{\alpha \beta } \left(\eta _{\alpha \beta } \theta _{\alpha \beta }+1\right)\right){}^{9/2}}\\
   \times \left(99 \left(\eta _{\alpha \beta }+1\right){}^4 \theta _{\alpha \beta }^2\right.\\
   -126 \left(\eta _{\alpha \beta
   }+1\right){}^2 \theta _{\alpha \beta } k_{\alpha \beta } \left(\eta _{\alpha \beta }+\theta _{\alpha \beta }\right) \left(\eta _{\alpha \beta } \theta _{\alpha \beta }+1\right)\\
   \left.+35 k_{\alpha \beta }^2
   \left(\eta _{\alpha \beta }+\theta _{\alpha \beta }\right){}^2 \left(\eta _{\alpha \beta } \theta _{\alpha \beta }+1\right){}^2\right)
\end{multline}

\subsection{$\left[S^p_{3/2}(W_\alpha^2)\mathbf{W}_\alpha,S^q_{3/2}(W_\alpha^2)\mathbf{W}_\alpha\right]$}

\begin{multline}
p=0,q=0:\   A^{0011,1}_{\i\i}\Omega_{\i\j}^{11}\\
A^{0011,1}_{\i\i}=-\frac{8 \left(\eta _{\alpha \beta }+1\right){}^4 \theta
   _{\alpha \beta }{}^{5/2}}{\left(k_{\alpha \beta } \left(\eta _{\alpha \beta }+\theta _{\alpha \beta }\right) \left(\eta _{\alpha \beta } \theta _{\alpha \beta }+1\right)\right){}^{5/2}}
\end{multline}

\begin{multline}
p=0,q=1:\  A^{0111,1}_{\i\i}\Omega_{\i\j}^{11}+A^{0121,1}_{\i\i}\Omega_{\i\j}^{12}\\
A^{0111,1}_{\i\i}=-\frac{20 \left(\eta _{\alpha \beta }+1\right){}^4 \left(\frac{\theta _{\alpha \beta }}{\eta _{\alpha \beta }+\theta _{\alpha \beta }}\right){}^{7/2}}{\left(k_{\alpha \beta } \left(\eta _{\alpha \beta } \theta
   _{\alpha \beta }+1\right)\right){}^{5/2}}\\
A^{0121,1}_{\i\i}=-\frac{4 \left(\eta _{\alpha \beta }+1\right){}^4 }{\left(\eta _{\alpha \beta }+\theta _{\alpha \beta }\right){}^{9/2} \left(k_{\alpha \beta } \left(\eta _{\alpha \beta }+\frac{1}{\theta _{\alpha
   \beta }}\right)\right){}^{7/2}}\\
   \times\left(5 k_{\alpha \beta } \left(\eta _{\alpha \beta }+\theta _{\alpha \beta }\right) \left(\eta _{\alpha \beta } \theta _{\alpha \beta }+1\right)-7 \left(\eta
   _{\alpha \beta }+1\right){}^2 \theta _{\alpha \beta }\right)
\end{multline}

\begin{multline}
p=0,q=2:\   A^{0211,1}_{\i\i}\Omega_{\i\j}^{11}+A^{0221,1}_{\i\i}\Omega_{\i\j}^{12}+A^{0231,1}_{\i\i}\Omega_{\i\j}^{13}\\
A^{0211,1}_{\i\i}=-\frac{35 \left(\eta _{\alpha \beta }+1\right){}^4 \left(\frac{\theta _{\alpha \beta }}{\eta _{\alpha \beta }+\theta _{\alpha \beta }}\right){}^{9/2}}{\left(k_{\alpha \beta } \left(\eta _{\alpha \beta } \theta
   _{\alpha \beta }+1\right)\right){}^{5/2}}\\
A^{0221,1}_{\i\i}=-\frac{7 \left(\eta _{\alpha \beta }+1\right){}^4 \theta _{\alpha \beta }^{9/2} }{\left(\eta _{\alpha \beta }+\theta _{\alpha \beta }\right){}^{11/2} \left(k_{\alpha \beta } \left(\eta _{\alpha \beta
   } \theta _{\alpha \beta }+1\right)\right){}^{7/2}}\\
   \times\left(5 k_{\alpha \beta } \left(\eta _{\alpha \beta }+\theta _{\alpha \beta }\right) \left(\eta _{\alpha \beta } \theta _{\alpha
   \beta }+1\right)-9 \left(\eta _{\alpha \beta }+1\right){}^2 \theta _{\alpha \beta }\right)\\
A^{0231,1}_{\i\i}=-\frac{\left(\eta _{\alpha \beta }+1\right){}^4 \theta _{\alpha \beta }^4 }{2 \sqrt{\frac{\eta _{\alpha \beta }}{\theta _{\alpha \beta }}+1} \left(\eta _{\alpha \beta }+\theta _{\alpha
   \beta }\right){}^6 \left(k_{\alpha \beta } \left(\eta _{\alpha \beta } \theta _{\alpha \beta }+1\right)\right){}^{9/2}}\\
   \times\left(99 \left(\eta _{\alpha \beta }+1\right){}^4 \theta _{\alpha \beta }^2\right.\\
   -126 \left(\eta _{\alpha \beta }+1\right){}^2 \theta _{\alpha
   \beta } k_{\alpha \beta } \left(\eta _{\alpha \beta }+\theta _{\alpha \beta }\right) \left(\eta _{\alpha \beta } \theta _{\alpha \beta }+1\right)\\
   \left.+35 k_{\alpha \beta }^2 \left(\eta _{\alpha \beta }+\theta
   _{\alpha \beta }\right){}^2 \left(\eta _{\alpha \beta } \theta _{\alpha \beta }+1\right){}^2\right)
\end{multline}

\subsection{$\left[S^p_{3/2}(W_\j^2),S^q_{3/2}(W_\alpha^2)\right]$}

\begin{multline}
p=1,q=0:\  A^{1011,0}_{\i\j} \Omega_{\i\j}^{11}\\
A^{1011,0}_{\i\j}=\frac{16 \eta _{\alpha \beta } \left(\eta _{\alpha \beta }+1\right){}^3 \left(\theta _{\alpha \beta }-1\right)\theta _{\alpha \beta }{}^{3/2}}{\left(k_{\alpha \beta } \left(\eta _{\alpha \beta }+\theta _{\alpha
   \beta }\right) \left(\eta _{\alpha \beta } \theta _{\alpha \beta }+1\right)\right){}^{5/2}}
\end{multline}

\begin{multline}
p=1,q=1:\  A^{1111,0}_{\i\j} \Omega_{\i\j}^{11}+A^{1121,0}_{\i\j} \Omega_{\i\j}^{12}\\
A^{1111,0}_{\i\j}=\frac{8 \eta _{\alpha \beta } \left(\eta _{\alpha \beta }+1\right){}^3 \theta _{\alpha \beta }^{3/2} \left(\eta _{\alpha \beta } \left(5 \theta _{\alpha \beta }-3\right)+2 \theta _{\alpha \beta
   }\right)}{\left(\eta _{\alpha \beta }+\theta _{\alpha \beta }\right){}^{7/2} \left(k_{\alpha \beta } \left(\eta _{\alpha \beta } \theta _{\alpha \beta }+1\right)\right){}^{5/2}}\\
A^{1121,0}_{\i\j}=\frac{8 \eta _{\alpha \beta } \left(\eta _{\alpha \beta }+1\right){}^3 \theta _{\alpha \beta }^{3/2} }{\left(\eta _{\alpha \beta }+\theta _{\alpha \beta }\right){}^{9/2} \left(k_{\alpha \beta }
   \left(\eta _{\alpha \beta } \theta _{\alpha \beta }+1\right)\right){}^{7/2}}\\
   \times\left(\left(\eta _{\alpha \beta }+1\right){}^2 \theta _{\alpha \beta } \left(\eta _{\alpha \beta } \left(5-7
   \theta _{\alpha \beta }\right)-2 \theta _{\alpha \beta }\right)\right.\\
   \left.+k_{\alpha \beta } \left(\eta _{\alpha \beta }+\theta _{\alpha \beta }\right) \left(\eta _{\alpha \beta } \theta _{\alpha \beta }+1\right)
   \left(\eta _{\alpha \beta } \left(5 \theta _{\alpha \beta }-3\right)+2 \theta _{\alpha \beta }\right)\right)
\end{multline}

\begin{multline}
 p=1,q=2:\ A^{1211,0}_{\i\j} \Omega_{\i\j}^{11}+A^{1221,0}_{\i\j} \Omega_{\i\j}^{12}+A^{1231,0}_{\i\j} \Omega_{\i\j}^{13}\\
A^{1211,0}_{\i\j}=\frac{10 \eta _{\alpha \beta }^2 \left(\eta _{\alpha \beta }+1\right){}^3 \theta _{\alpha \beta }^{3/2} \left(\eta _{\alpha \beta } \left(7 \theta _{\alpha \beta }-3\right)+4 \theta _{\alpha \beta
   }\right)}{\left(\eta _{\alpha \beta }+\theta _{\alpha \beta }\right){}^{9/2} \left(k_{\alpha \beta } \left(\eta _{\alpha \beta } \theta _{\alpha \beta }+1\right)\right){}^{5/2}}\\
A^{1221,0}_{\i\j}=\frac{2 \eta _{\alpha \beta }^2 \left(\eta _{\alpha \beta }+1\right){}^3 \theta _{\alpha \beta }^{3/2} }{\left(\eta _{\alpha \beta }+\theta _{\alpha \beta }\right){}^{11/2} \left(k_{\alpha \beta }
   \left(\eta _{\alpha \beta } \theta _{\alpha \beta }+1\right)\right){}^{7/2}}\\
   \times\left(5 k_{\alpha \beta } \left(\eta _{\alpha \beta }+\theta _{\alpha \beta }\right) \left(\eta _{\alpha
   \beta } \theta _{\alpha \beta }+1\right) \left(\eta _{\alpha \beta } \left(7 \theta _{\alpha \beta }-3\right)+4 \theta _{\alpha \beta }\right)\right.\\
   \left.-7 \left(\eta _{\alpha \beta }+1\right){}^2 \theta _{\alpha \beta }
   \left(\eta _{\alpha \beta } \left(9 \theta _{\alpha \beta }-5\right)+4 \theta _{\alpha \beta }\right)\right)\\
A^{1231,0}_{\i\j}=\frac{\eta _{\alpha \beta }^2 \left(\eta _{\alpha \beta }+1\right){}^3 \theta _{\alpha \beta } }{\sqrt{\frac{\eta _{\alpha \beta }}{\theta _{\alpha \beta }}+1} \left(\eta _{\alpha \beta }+\theta _{\alpha \beta }\right){}^6 \left(k_{\alpha \beta } \left(\eta _{\alpha \beta } \theta
   _{\alpha \beta }+1\right)\right){}^{9/2}}\\
   \times \left(9 \left(\eta _{\alpha \beta }+1\right){}^4 \theta _{\alpha \beta }^2 \left(\eta _{\alpha \beta } \left(11
   \theta _{\alpha \beta }-7\right)+4 \theta _{\alpha \beta }\right)\right.\\
   \left.-14 \left(\eta _{\alpha \beta }+1\right){}^2 \theta _{\alpha \beta } k_{\alpha \beta } \left(\eta _{\alpha \beta }+\theta _{\alpha \beta
   }\right) \left(\eta _{\alpha \beta } \theta _{\alpha \beta }+1\right)\right.\\
   \left.\times\left(\eta _{\alpha \beta } \left(9 \theta _{\alpha \beta }-5\right)+4 \theta _{\alpha \beta }\right)\right.\\
   \left.+5 k_{\alpha \beta }^2 \left(\eta
   _{\alpha \beta }+\theta _{\alpha \beta }\right){}^2 \left(\eta _{\alpha \beta } \theta _{\alpha \beta }+1\right){}^2 \left(\eta _{\alpha \beta } \left(7 \theta _{\alpha \beta }-3\right)+4 \theta _{\alpha \beta
   }\right)\right)
\end{multline}

\subsection{$\left[S^p_{3/2}(W_\i^2),S^q_{3/2}(W_\alpha^2)\right]$}

\begin{multline}
 p=1,q=0:\  A^{1011,0}_{\i\i} \Omega_{\i\j}^{11}\\
A^{1011,0}_{\i\i}= \frac{16 \eta _{\alpha \beta } \left(\eta _{\alpha \beta }+1\right){}^3 \left(\theta _{\alpha \beta }-1\right)\theta _{\alpha \beta }{}^{3/2}}{ \left(k_{\alpha \beta } \left(\eta _{\alpha \beta }+\theta _{\alpha
   \beta }\right) \left(\eta _{\alpha \beta } \theta _{\alpha \beta }+1\right)\right){}^{5/2}}
\end{multline}

\begin{multline}
 p=1,q=1:\  A^{1111,0}_{\i\i} \Omega_{\i\j}^{11}+A^{1121,0}_{\i\i} \Omega_{\i\j}^{12}\\
A^{1111,0}_{\i\i}=-\frac{8 \eta _{\alpha \beta } \left(\eta _{\alpha \beta }+1\right){}^3 \left(2 \eta _{\alpha \beta }-3 \theta _{\alpha \beta }+5\right)}{\left(\eta _{\alpha \beta }+\theta _{\alpha \beta }\right){}^{7/2}
   \left(k_{\alpha \beta } \left(\eta _{\alpha \beta }+\frac{1}{\theta _{\alpha \beta }}\right)\right){}^{5/2}}\\
A^{1121,0}_{\i\i}=\frac{8 \eta _{\alpha \beta } \left(\eta _{\alpha \beta }+1\right){}^3 \theta _{\alpha \beta }^{5/2} }{\left(\eta _{\alpha \beta }+\theta _{\alpha \beta }\right){}^{9/2} \left(k_{\alpha \beta } \left(\eta _{\alpha \beta } \theta _{\alpha \beta }+1\right)\right){}^{7/2}}\\
\times\left(\left(\eta _{\alpha \beta }+1\right){}^2 \theta _{\alpha \beta } \left(2 \eta _{\alpha \beta }-5 \theta
   _{\alpha \beta }+7\right)\right.\\
   \left.-k_{\alpha \beta } \left(2 \eta _{\alpha \beta }-3 \theta _{\alpha \beta }+5\right) \left(\eta _{\alpha \beta }+\theta _{\alpha \beta }\right) \left(\eta _{\alpha \beta } \theta
   _{\alpha \beta }+1\right)\right)
\end{multline}

\begin{multline}
 p=1,q=2:\  A^{1211,0}_{\i\i} \Omega_{\i\j}^{11}+A^{1221,0}_{\i\i} \Omega_{\i\j}^{12}+A^{1231,0}_{\i\i} \Omega_{\i\j}^{13}\\
A^{1211,0}_{\i\i}=-\frac{10 \eta _{\alpha \beta } \left(\eta _{\alpha \beta }+1\right){}^3 \theta _{\alpha \beta }^{7/2} \left(4 \eta _{\alpha \beta }-3 \theta _{\alpha \beta }+7\right)}{\left(\eta _{\alpha \beta }+\theta _{\alpha
   \beta }\right){}^{9/2} \left(k_{\alpha \beta } \left(\eta _{\alpha \beta } \theta _{\alpha \beta }+1\right)\right){}^{5/2}}\\
A^{1221,0}_{\i\i}= \frac{2 \eta _{\alpha \beta } \left(\eta _{\alpha \beta }+1\right){}^3 \theta _{\alpha \beta }^3 }{k_{\alpha \beta }^{7/2} \sqrt{\eta _{\alpha \beta }+\frac{1}{\theta _{\alpha \beta }}} \left(\eta _{\alpha \beta }+\theta _{\alpha \beta }\right){}^{11/2} \left(\eta _{\alpha
   \beta } \theta _{\alpha \beta }+1\right){}^3}\\
   \times\left(7 \left(\eta _{\alpha \beta }+1\right){}^2 \theta _{\alpha \beta } \left(4 \eta _{\alpha \beta }-5 \theta
   _{\alpha \beta }+9\right)\right.\\
   \left.-5 k_{\alpha \beta } \left(4 \eta _{\alpha \beta }-3 \theta _{\alpha \beta }+7\right) \left(\eta _{\alpha \beta }+\theta _{\alpha \beta }\right) \left(\eta _{\alpha \beta } \theta
   _{\alpha \beta }+1\right)\right)\\
A^{1231,0}_{\i\i}=-\frac{\eta _{\alpha \beta } \left(\eta _{\alpha \beta }+1\right){}^3 \theta _{\alpha \beta }^3 }{\sqrt{\frac{\eta _{\alpha \beta }}{\theta _{\alpha \beta }}+1} \left(\eta _{\alpha \beta }+\theta
   _{\alpha \beta }\right){}^6 \left(k_{\alpha \beta } \left(\eta _{\alpha \beta } \theta _{\alpha \beta }+1\right)\right){}^{9/2}}\\
   \times \left(9 \left(\eta _{\alpha \beta }+1\right){}^4 \theta _{\alpha \beta }^2 \left(4 \eta _{\alpha \beta }-7 \theta
   _{\alpha \beta }+11\right)\right.\\
   \left.-14 \left(\eta _{\alpha \beta }+1\right){}^2 \theta _{\alpha \beta } k_{\alpha \beta }\right.\\
   \left.\times\left(4 \eta _{\alpha \beta }-5 \theta _{\alpha \beta }+9\right) \left(\eta _{\alpha \beta
   }+\theta _{\alpha \beta }\right) \left(\eta _{\alpha \beta } \theta _{\alpha \beta }+1\right)\right.\\
   \left.+5 k_{\alpha \beta }^2 \left(4 \eta _{\alpha \beta }-3 \theta _{\alpha \beta }+7\right) \left(\eta _{\alpha \beta
   }+\theta _{\alpha \beta }\right){}^2 \left(\eta _{\alpha \beta } \theta _{\alpha \beta }+1\right){}^2\right)
\end{multline}

\section{$13N$-moment equations and collisional coefficients under Zhdanov's approximations}
\label{sec:zhdanov13}

We provide here the $13N$-moment multi-temperature coefficients given in Chapter 4 of Ref.\,\onlinecite{zhdanov_transport_2002}, originally in Ref.\,\onlinecite{alievskii_1963_transport}, 
where there are the three primary plasmadynamical quantities $(\rho_\i,T_\i,\mathbf{w}_\i)$ and in addition, the pressure-stress tensor $\pi_\i$ and the heat flux $h_\i$.

The collisional moments are given by $R^{mn}$. They are generally the form of sum of terms, each being coefficients multiplied to the moments. There is no explicit derivation provided in previous literature by Zhdanov et al for the provided coefficients. We reproduce them below for the sake of comparison.

\begin{align}
 \mathbf{R}_\i^{10}&= \sum_\j G^{(1)}_{\i\j}(\mathbf{w}_\i-\mathbf{w}_\j)+\sum_\j \gamma_{\i\j}G^{(2)}_{\i\j}\left(\frac{\mathbf{h}_\i}{\gamma_\i p_\i}-\frac{\mathbf{h}_\j}{\gamma_\j p_\j}\right)\\
 R_\i^{01}&=\sum_\j G^{(1)}_{\i\j}\frac{3k}{m_\i+m_\j}(T_\i-T_\j)\\
 R_\i^{20}&= \sum_\j \frac{1}{\gamma_\i+\gamma_\j}\left(G^{(3)}_{\i\j}\frac{\pi_\i}{p_\i}+G^{(4)}_{\i\j}\frac{\pi_\j}{ p_\j}\right)\\
 \mathbf{R}_\i^{11}&=\frac{1}{\gamma_\i}\sum_\j \left( G^{(5)}_{\i\j}\frac{\mathbf{h}_\i}{p_\i}+G^{(6)}_{\i\j}\frac{\mathbf{h}_\j}{p_\j}\right.\nonumber\\
 &\left.+\frac{5}{2}\frac{\gamma_{\i\j}}{\gamma_\i}G^{(7)}_{\i\j}(\mathbf{w}_\i-\mathbf{w}_\j) +5\Theta_{\i\j}G^{(1)}_{\i\j}\mathbf{w}_\i \right).
\end{align}
Here, $\Theta_{\i\j}=(1-T_\j/T_\i)/(1+m_\j/m_\i)$. The $G$-coefficients are given by
\begin{align}
 G^{(1)}_{\i\j}&=B_{\i\j}^{(1)}\\
 G^{(2)}_{\i\j}&=B_{\i\j}^{(1)}\\ 
 G^{(3)}_{\i\j}&=B_{\i\j}^{(3)}-\frac{\gamma_\j}{\gamma_\i}\Theta_{\i\j}C_{\i\j}^{(1)}\\
 G^{(4)}_{\i\j}&=B_{\i\j}^{(4)}-\Theta_{\i\j}C_{\i\j}^{(1)}
 \end{align}
and
 \begin{align}
 G^{(5)}_{\i\j}&=B_{\i\j}^{(5)}+\frac{1}{2}\frac{\gamma_\j}{\gamma_\i}\Theta_{\i\j}\left(C_{\i\j}^{(2)}-C_{\i\j}^{(3)}-2\frac{\gamma_\i}{\gamma_\j}C_{\i\j}^{(4)}\right)\\
 &-\frac{\gamma_\j}{\gamma_\i}\Theta_{\i\j}^2\left(C_{\i\j}^{(2)}-\frac{1}{2}C_{\i\j}^{(3)}\right)\\
 G^{(6)}_{\i\j}&=B_{\i\j}^{(6)}-\frac{1}{2}\Theta_{\i\j}\left(C_{\i\j}^{(2)}-C_{\i\j}^{(3)}+2C_{\i\j}^{(4)}\right)\\
 &+\Theta_{\i\j}^2\left(C_{\i\j}^{(2)}-\frac{1}{2}C_{\i\j}^{(3)}\right)\\
 G^{(7)}_{\i\j}&=B_{\i\j}^{(2)}+\Theta_{\i\j}C_{\i\j}^{(5)}-\Theta_{\i\j}^2C_{\i\j}^{(6)}.
\end{align}
In this representation, the $B$-coefficients refer to the terms that remain in the equal-temperature case, and the $C$-coefficients refer to terms that come into play in the case of non-equal temperatures. The $B$-coefficients are given as follows
\begin{align}
 B_{\i\j}^{(1)}&=-\frac{16}{3}\mu_{\i\j}n_\i n_\j \Omega_{\i\j}^{11}\\
 B_{\i\j}^{(2)}&=-\frac{16}{3}\mu_{\i\j}n_\i n_\j\left(\frac{2}{5}\Omega_{\i\j}^{12}-\Omega_{\i\j}^{11}\right)\\
 B_{\i\j}^{(3)}&=-\frac{16}{5}\mu_{\i\j}n_\i n_\j\left(\frac{\gamma_\j}{\gamma_\i}\Omega_{\i\j}^{22}+\frac{10}{3}\Omega_{\i\j}^{11}\right)\\
 B_{\i\j}^{(4)}&=-\frac{16}{5}\mu_{\i\j}n_\i n_\j\left(\Omega_{\i\j}^{22}-\frac{10}{3}\Omega_{\i\j}^{11}\right)\\
 B_{\i\j}^{(5)}&=-\frac{64}{15}\kappa_{\i\j}\mu_{\i\j}n_\i n_\j\left\{\Omega_{\i\j}^{22}+\left(\frac{15}{4}\frac{\gamma_\i}{\gamma_\j}+\right.\right.\\
 &\left.\left.\frac{25}{8}\frac{\gamma_\j}{\gamma_\i} \right)\Omega_{\i\j}^{11}-\frac{1}{2}\frac{\gamma_\j}{\gamma_\i}(5\Omega_{\i\j}^{12}-\Omega_{\i\j}^{13})  \right\}\\
 B_{\i\j}^{(6)}&=-\frac{64}{15}\kappa_{\i\j}\mu_{\i\j}n_\i n_\j\left\{\Omega_{\i\j}^{22}-\frac{55}{8}\Omega_{\i\j}^{11}\right.\\
 &\left.+\frac{1}{2}(5\Omega_{\i\j}^{12}-\Omega_{\i\j}^{13})  \right\},
\end{align}
and the $C$-coefficients are given as follows
\begin{align}
 C_{\i\j}^{(1)}&=-\frac{16}{5}\mu_{\i\j}n_\i n_\j\left(\Omega_{\i\j}^{22}-\frac{4}{3}\Omega_{\i\j}^{12}\right)\\
 C_{\i\j}^{(2)}&=-\frac{64}{15}\kappa_{\i\j}\mu_{\i\j}n_\i n_\j\left(5\Omega_{\i\j}^{12}-2\Omega_{\i\j}^{13}\right)\\
 C_{\i\j}^{(3)}&=-\frac{64}{15}\kappa_{\i\j}\mu_{\i\j}n_\i n_\j\left(5\Omega_{\i\j}^{22}-2\Omega_{\i\j}^{23}\right)\\
 C_{\i\j}^{(4)}&=-\frac{64}{15}\kappa_{\i\j}\mu_{\i\j}n_\i n_\j\left(\Omega_{\i\j}^{22}-\frac{11}{2}\Omega_{\i\j}^{12}+\frac{25}{4}\Omega_{\i\j}^{11} \right)\\
 C_{\i\j}^{(5)}&=-\frac{64}{15}\mu_{\i\j}n_\i n_\j\left(\Omega_{\i\j}^{22}-\Omega_{\i\j}^{12}-\frac{5}{2}\Omega_{\i\j}^{11} \right)\\
 C_{\i\j}^{(6)}&=-\frac{64}{15}\mu_{\i\j}n_\i n_\j\left(\Omega_{\i\j}^{22}-2\Omega_{\i\j}^{12} \right),
\end{align}
where $\kappa_{\i\j}=\gamma_{\i\j}/(\gamma_\i+\gamma_\j)$. The form of the Chapman-Cowling integral is given by
\begin{multline}
 \Omega^{lr}_{\alpha\beta} = \left( \frac{2\pi}{\gamma_{\alpha\beta}}\right)^{1/2} \int^{\infty}_0 \int^{\pi}_0  \zeta^{2r+3} \exp{(-\zeta^2)}\times\\
 \times(1-\cos^l{\chi})\sin{\chi}\sigma_{\alpha\beta}(\zeta,\chi)d\chi d\zeta,
 \label{eq:zhdanov_omegarl_proper}
\end{multline}
where $\zeta=(\gamma_{\alpha\beta}/2)^{1/2}g$. { One can obtain the same coefficients, and even higher-order ones, as above by choosing $d_{\i\j}=\gamma_{\alpha\beta}/2$ in Appendix \ref{sec:bracket_integral_derivation}.}

\section{$21N$-moment approximate coefficients provided by Zhdanov}
\label{sec:zhdanov21}

In this section, we provide the approximated collisional coefficients used by Zhdanov in Chapter 8 of Ref.\,\onlinecite{zhdanov_transport_2002}, and originally in Ref.\,\onlinecite{yushmanov_diffusion_1980}. They are provided for the case of a fully ionized plasma with multiple species, each species having sub-components at different charge states. Fundamentally, this approximation involves two components. The first is to choose the approximation for the Chapman-Cowling integral $\Omega_{\i\j}^{lr}$, in a fashion similar to the one suggested by Preuss\cite{preuss_bilanzgleichungen_1970},
\begin{equation}
 \Omega_{\i\j}^{lr}=\sqrt{\pi}l(r-1)!\left(\frac{Z_\i Z_\j e^2}{4\pi\epsilon_0}\right)^2 \frac{\ln\Lambda_{\i\j}}{\mu_{\i\j}^{1/2}(2kT)^{3/2}}, \label{eq:zhdanov21Ncrosssections}
\end{equation}
{ which approximates the  Chapman-Cowling integrals  for Coulomb potential up to the order of the logarithmic term.} The Coulomb logarithm, $\ln\Lambda_{\i\j}$ is given by
\begin{equation}
 \ln\Lambda_{\i\j}=\frac{12\pi\epsilon_0^{3/2}kT}{Z_{eff}e^2}\left\{\frac{kT}{n_e e^2(1+Z_{eff})} \right\},
\end{equation}
where the effective charge $Z_{eff}$ is given by
\begin{equation}
 Z_{eff}=\frac{\sum_i n_i Z_i^2}{\sum_i n_i Z_i},
\end{equation}
where $i$ refers to ions of similar kind but at different charge states. One can notice that these expressions contain a common $T$ instead of the exact species (or component) temperature. The general requirement for this to hold seems to be that $|T_\i-T_\j| \ll T_\i$, i.e. that the temperatures of all species are close to each other, in case of which one can use the following expression for the common temperature
\begin{equation}
 nT=\sum_\i n_\i T_\i,\ n=\sum_\i n_\i.
\end{equation}
In case of such an approximation, the only terms that remain in calculation of the collision coefficients are the equal-temperature bracket integral terms, equivalent to the $B$-coefficients of the previous section. { However, the LHS is treated as if it operates at different temperatures for different species and their sub-components. This would be a reasonable assumption if the masses of the species were similar, thus making the relaxation timescales similar, under which both species would have temperatures very close to the plasma common temperature. However, it may diverge with species of masses at different orders, such as with heavy impurities in the plasma.}

The collision coefficients for the rank-0, rank-1 and rank-2 quantities are as follows
\begin{widetext}
\begin{align}
R_\i^{01}&=\sum_\j G^{(1)}_{\i\j}\frac{3k}{m_\i+m_\j}(T_\i-T_\j)\\
 \mathbf{R}_\i^{10}&= \sum_\j G^{(1)}_{\i\j}(\mathbf{w}_\i-\mathbf{w}_\j)+\sum_\j \frac{\mu_{\i\j}}{kT}G^{(2)}_{\i\j}\left(\frac{\mathbf{h}_\i}{\rho_\i}-\frac{\mathbf{h}_\j}{\rho_\j}\right)+\sum_\j \left(\frac{\mu_{\i\j}}{kT}\right)^2 G^{(8)}_{\i\j}\left(\frac{\mathbf{r}_\i}{\rho_\i}-\frac{\mathbf{r}_\j}{\rho_\j}\right)\\
  \mathbf{R}_\i^{11}&=\frac{kT}{m_\i}\sum_\j \left\{ \frac{5}{2}\frac{\mu_{\i\j}}{m_\i}G^{(2)}_{\i\j}(\mathbf{w}_\i-\mathbf{w}_\j)+G^{(5)}_{\i\j}\frac{\mathbf{h}_\i}{p_\i}+G^{(6)}_{\i\j}\frac{\mathbf{h}_\j}{p_\j} +\frac{\mu_{\i\j}}{kT}\left(G^{(9)}_{\i\j}\frac{\mathbf{r}_\i}{p_\i}+G^{(10)}_{\i\j}\frac{\mathbf{r}_\j}{p_\j}\right)  \right\}\\
 \mathbf{R}_\i^{12}&= \left(\frac{kT}{m_\i}\right)^2\sum_\j\left\{ \frac{35}{2}\left(\frac{\mu_{\i\j}}{m_\i}\right)^2G^{(8)}_{\i\j}(\mathbf{w}_\i-\mathbf{w}_\j)+7\frac{\mu_{\i\j}}{m_\i}\left( G^{(9)}_{\i\j}\frac{\mathbf{h}_\i}{p_\i}+G^{(10)}_{\i\j}\frac{\mathbf{h}_\j}{p_\j}\right)+\frac{1}{kT}\left(m_\i G^{(11)}_{\i\j}\frac{\mathbf{r}_\i}{p_\i}+m_\j G^{(12)}_{\i\j}\frac{\mathbf{r}_\j}{p_\j} \right)       \right\},\\
  R_\i^{20}&= \sum_\j \frac{kT}{m_\i+m_\j}\left\{ G^{(3)}_{\i\j}\frac{\pi_\i}{p_\i}+G^{(4)}_{\i\j}\frac{\pi_\j}{ p_\j}  +\frac{\mu_{\i\j}}{kT}\left(G^{(13)}_{\i\j}\frac{\sigma_\i}{p_\i}+G^{(14)}_{\i\j}\frac{\sigma_\j}{ p_\j}\right)                \right\}\\
  R_\i^{21}&=\sum_\j \left(\frac{kT}{m_\i+m_\j}\right)^2\left\{ \frac{7}{2}\frac{\mu_{\i\j}}{m_\i}\left( G^{(13)}_{\i\j}\frac{\pi_\i}{p_\i}+G^{(14)}_{\i\j}\frac{\pi_\j}{ p_\j}\right)  +\frac{\mu_{\i\j}}{kT}\left(G^{(15)}_{\i\j}\frac{\sigma_\i}{p_\i}+G^{(16)}_{\i\j}\frac{\sigma_\j}{ p_\j}\right)\right\}.
\end{align}
\end{widetext}
After applying the approximations for the Chapman-Cowling integral and equal-temperature assumption, the $G$-coefficients are as follows
\begin{align}
 G^{(1)}_{\i\j}&=-\lambda_{\i\j}\\
 G^{(2)}_{\i\j}&=\frac{3}{5}\lambda_{\i\j}\\
 G^{(3)}_{\i\j}&=-2\left(1+\frac{3}{5}\frac{m_\j}{m_\i} \right)\lambda_{\i\j}\\
 G^{(4)}_{\i\j}&=\frac{4}{5}\lambda_{\i\j}\\
 G^{(5)}_{\i\j}&=-\left(\frac{13}{10}\frac{m_\j}{m_\i}+\frac{8}{5}+3\frac{m_\i}{m_\j} \right)\kappa_{\i\j}\lambda_{\i\j}\\
 G^{(6)}_{\i\j}&=\frac{27}{10}\kappa_{\i\j}\lambda_{\i\j}\\
 G^{(8)}_{\i\j}&=-\frac{3}{14}\lambda_{\i\j}\\
 G^{(9)}_{\i\j}&=\frac{3}{5}\left(\frac{23}{28}\frac{m_\j}{m_\i}+\frac{8}{7}+3\frac{m_\i}{m_\j}\right)\kappa_{\i\j}\lambda_{\i\j}\\
 G^{(10)}_{\i\j}&=-\frac{45}{28}\kappa_{\i\j}\lambda_{\i\j}
\end{align}

\begin{align}
 G^{(11)}_{\i\j}&=-\left[\frac{433}{280}\left(\frac{m_\j}{m_\i}\right)^2+\frac{136}{35}\frac{m_\j}{m_\i}+\frac{459}{35}+\frac{32}{5}\frac{m_\i}{m_\j}\right.\\
 &\left.+5\left(\frac{m_\i}{m_\j}\right)^2 \right]\kappa_{\i\j}^2\lambda_{\i\j}\\
 G^{(12)}_{\i\j}&=\frac{75}{8}\kappa_{\i\j}^2\lambda_{\i\j}\\
 G^{(13)}_{\i\j}&=\left(\frac{18}{35}\frac{m_\j}{m_\i}+\frac{6}{5} \right)\lambda_{\i\j}\\
 G^{(14)}_{\i\j}&=-\frac{24}{35}\lambda_{\i\j}\\
 G^{(15)}_{\i\j}&=-\left\{\frac{51}{35}\left(\frac{m_\j}{m_\i}\right)^2 +\frac{37}{7}\frac{m_\j}{m_\i}+\frac{22}{5}+4\frac{m_\i}{m_\j} \right\}\kappa_{\i\j}\lambda_{\i\j}\\
 G^{(16)}_{\i\j}&=\frac{24}{7}\frac{m_\j}{m_\i}\kappa_{\i\j}\lambda_{\i\j},
\end{align}
where here $\kappa_{\i\j}=m_\i m_\j/(m_\i+m_\j)^2$, and $\lambda_{\i\j}$ is given by
\begin{equation}
 \lambda_{\i\j}=\frac{1}{3}(2\pi)^{-3/2}n_\i n_\j e^4 Z_\i^2 Z_\j^2\mu_{\i\j}^{1/2} \frac{\ln\Lambda_{\i\j}}{(kT)^{3/2}\epsilon_0^2},
\end{equation}
{  which is obtained from Eq.\,(\ref{eq:zhdanov21Ncrosssections}).} Notice the difference in the coefficients for $R^{20}$ and $R^{21}$ from Ref.\,\cite{zhdanov_transport_2002}.
\end{document}